\documentclass{article}
\usepackage[left=3cm,right=3cm,top=3cm,bottom=3cm]{geometry} % page settings
\usepackage{authblk}

\usepackage{graphicx,psfrag,epsf}
\usepackage{graphicx,hyperref}
\usepackage{amssymb}
\usepackage{amsmath}
\usepackage{multirow}
\usepackage{float}
\usepackage{mathtools}
\usepackage{xcolor}
\usepackage{lmodern}
\mathtoolsset{showonlyrefs=true}
\usepackage{natbib}
\usepackage{subfig}
\usepackage{lipsum}
\usepackage{graphicx}
\usepackage{epsfig}
\usepackage{epstopdf}
\usepackage[normalem]{ulem}

\usepackage{xr}
\usepackage{fancyhdr,lipsum}

\makeatletter

\newsavebox{\@linebox}
\savebox{\@linebox}[3em][t]{\parbox[t]{3em}{%
		\@tempcnta\@ne\relax
		\loop{\underline{\scriptsize\the\@tempcnta}}\\
		\advance\@tempcnta by \@ne\ifnum\@tempcnta<48\repeat}}

\usepackage[title]{appendix}
\usepackage{amssymb}
\begin{document}

\title{The modeling  of multiple animals that share behavioral features}

\author{{Gianluca} {Mastrantonio}}
\affil{Politecnico di Torino, Dipartimento di Scienze Matematiche}

	\date{}

	\maketitle

			\begin{abstract}

In this work, we propose a model that can be used to infer the behavior of multiple animals. Our proposal is defined as a set of hidden Markov models that are  based on the sticky hierarchical Dirichlet process, with a shared base-measure, and a  STAP emission distribution. The latent classifications are representative of the behavior assumed by the animals, which is  described by the  STAP parameters. Given the latent classifications, the animals are independent.

As a result of  the way we formalize the distribution over the STAP parameters,  the  animals  may share, in different behaviors, the set or a subset of the parameters, thereby allowing us to investigate the similarities between them. The hidden Markov models, based on the Dirichlet process, allow us to estimate   the number of latent behaviors for each animal, as a model parameter.

This proposal  is motivated by a real data problem, where the GPS coordinates of  six Maremma Sheepdogs have been observed.
Among the other results, we show that four   dogs share most of the behavior characteristics, while   two have specific behaviors.

\end{abstract}

\section{Introduction}\label{newsec:intro}

% The statistical models to analyze animal movement data have become increasingly popular and, since the first paper of \cite{dunn77}, very flexible and complex approaches have been proposed. These are  used to understand different aspects of the movement, ranging from the habitat selection \citep{Hebblewhite} to behavior analysis \citep{MERRILL2000,Anderson2003,Maruotti2015a,MASTRANTONIO2018}; for a detailed review, the reader may refer to \cite{hooten2017animal}.

Movement data are often based on  a time-series of 2-dimensional spatial coordinates  recorded using a GPS device attached to  an animal.
Since the first paper by \cite{dunn77},  the statistical models used to analyze such data have become increasingly popular, and  are  used to understand different aspects of the movement of animals, ranging from  habitat selection \citep{Hebblewhite} to behavior analysis \citep{MERRILL2000,Anderson2003,Maruotti2015a,MASTRANTONIO2018}; for a detailed review, the reader may refer to \cite{hooten2017animal}.\\
%and the time-intervals between consecutive observations (called also \textit{fixes}),
% are  set by the researcher.
\indent Three  major categories of   movement-description models can be identified in the  behavior modeling framework: biased random walks (BRWs), correlated random walks (CRWs), and bias and   correlated random walks (BCRWs). In a BRW,  the animal movement  is attracted (or biased) toward a point in space, which is called \textit{center-of-attraction}  \citep[see, for example, ][]{BLACKWELL199787,dunn77}. The center-of-attraction can be interpreted as a proxy of the home-range  \citep{Christ2008} or it can  describe a  movement toward a patch of space \citep{McClintock2012}.  In a CRW, the movement direction, at any given time,  depends  on the previous direction.   This characteristic is called \textit{directional persistence} \citep{Jonsen2005} and it is useful to describe  a constant change in direction between consecutive observations.
If   both  directional persistence and attractors are used to describe a movement, the model is   a BCRW  \citep{Fortin2005,codling2008,McClintock2012}.  A movement-description model is generally   used as an emission distribution of  a mixture-type model, where a latent cluster-membership variable  is used to identify the   behavior assumed by an animal. If the observed time-window is wide enough, the use of a mixture-type model is  justified by the assumption that  an animal  exhibits more than one behavior during the day \citep{Patterson2222222}, e.g., sleeping and  hunting.   The switching between behaviors is  often temporally structured and, if  formulated in a discrete-time framework, the  model is usually the hidden Markov model (HMM) \citep{Michelot2016,Langrock2012}.\\% which is use for its easiness of implementation and interpretation.\\
\indent  The literature on the  modeling of multiple animals is not as extensive as that on  single individuals, even though   coordinates of different animals are  often recorded. Nonetheless,    interest in this topic is increasing   \cite[see, for example, ][]{Westley2018} since, as shown by \citep{Jonsen2016}, the joint  modeling of multiple animals often increases  the precision of the  estimates.
By adopting the classification given by \cite{Scharf2020}, it is possible to model multiple animals using two approaches. In the first approach, called  \textit{indirect}, the parameters that govern the behaviors are seen as random effects across animals, i.e., they come from a common distribution, whose parameters must be estimated, and the animals are  conditionally independent  \citep[see, for example,][]{McClintock2013,Michelot2017,Buderman2018}.
On the other hand, in the  \textit{direct} approach, the dependence between animals is described by an unobserved graph or social network, see, for example, \cite{Calabrese2018} \cite{Hooten2018},  \cite{Milner2021} and \cite{niu2021}.\\
%Although, in our opinion, the indirect  approach is promising, it is not yet widely used.
\indent We here propose a Bayesian model which can be used  to describe multiple animals that  share certain  movement characteristics, observed over different time-windows. The model is based on the hierarchical Dirichlet process (HDP)  \citep{Teh2006} and it is a generalization of the  sticky hierarchical Dirichlet process HMM (sHDP-HMM) of \cite{fox2011}. In the model,
given the latent classification and likelihood parameters, the animals are  independent and the behaviors are  described by the five parameters of  the STAP distribution, which is a BCRW emission-distribution  that has recently been proposed by \cite{mastrantonio2020modeling}.
The main contributions of the present  proposal are  the possibility of estimating the number of latent behaviors of each animal   as model parameters, and of introducing the sharing of parameters between behaviors and animals in HMMs based on Dirichlet processes (DPs).
The former is a by-product of the  DP modeling, which allows us to avoid  the use of  information criteria to select the number of behaviors,  which has been   shown to be problematic in this context \citep{Pohle2017}, or a trans-dimensional Markov chain Monte Carlo (MCMC) algorithm, such as the reversible jump MCMC (RJMCMC),
which presents challenges in its implementation \citep{Hastie2012}.
The sharing of parameters is introduced  in the lower level of the model hierarchy, where  the   distribution over the  STAP parameters is defined  by  combining  five  different  DPs. This distribution is discrete, with
a countable number of   atoms being defined so that they can share some of their multivariate components. This approach is similar   to the one  proposed  by \cite{mastrantonioEnv21}, which was used to model climate data in a change-point framework.   Therefore, the behaviors within or between animals   can have the same value of an STAP parameter, and this  allows us   to investigate  similarities and differences between the analyzed animals.
Other approaches also exist that have the sharing of parameters    as one of their characteristics (see, for example, \cite{Jonsen2016} or \cite{Milner2021}), but they require one to select, a priori, what parameters are allowed to change and  the number of values that a parameter can assume. On the other hand, in our proposal, everything is done during the model fitting and driven by the information within the data.
\\
% , the animal-specific unknown number of behaviors  is estimated as a random quantity, and the STAP allows us to detect how much a movement pattern is characterized by an attractive point or directional persistence. \\
% is this context can be problematic for at least two reason: the tendency of these criteria to select an often unrealistic larger number of states and, more importantly, if we allow the number of behavior to be different between animals, the number of models that has to be tested can de large for moderete number on animal and possible state, e.g., with 6 animal and two different number of state, the models to be tested are $2^6$. A possible solution would be the use of a trans-dimensional Markov chain Monte Carlo algorithm (MCMC), such as the reversible jump MCMC (RJMCMC),  that ahs been used in the context of animal behaviors, see, for example, \cite{McClintock2012}, but it presents challanges in its implementation, see, for example, \citep{Hastie2012}.\\
\indent Our proposal has been  used to model the trajectories of 6 Maremma Sheepdogs,   observed in Australia with recorded coordinates every 30 minutes.
These dogs are used all over Europe and Asia to protect livestock from possible predators and, in recent years, also in Australia \citep[see, for example, ][]{bommel2,Gehring}.
Maremma Sheepdogs are able to work in synergy with the shepherd to keep the stock together but this is not always possible when the  property is too large. For this reason, the  dogs are often left alone and are rarely visited by the shepherd. The owner has no supervision over the dogs and it is therefore   interesting to analyze  and understand their behavior.
The used dataset was taken from the movebank repository (\url{www.movebank.org}) and is described in detail in \cite{SheepdogRep} and \cite{Sheepdog}.  With our model, we have identified many similarities and  some specific features  between dogs,   that are easy to interpret and which give a better insight into the  behavior of the dogs.  Two competitive models have also been estimated on the same data and the results are compared with our proposal. \\
\indent  The paper is organized as follows.  We introduce the STAP density in Section \ref{newsec:stap}, and the hierarchical formalization of our proposal  in Section \ref{newsec:mod}, while Section
\ref{newsec:realdata} contains the results of the real data application. The paper ends with some conclusive remarks in Section
\ref{newsec:rem}.  The Appendix contain details of the MCMC algorithm  and on the results of the competitive models.

\section{The STAP distribution} \label{newsec:stap}
%\section{The model}\label{newsec:model}

% \begin{figure}[t]
% \centering
% \includegraphics[scale=0.40]{TransX}
% \caption{A graphical representation  of the relations between the spatial locations, the movement-metrics  and the displacement-coordinates. The arrow represents    $\protect\vec{\mathbf{F}}_{t_i}$
%      while the ellipse  is an area containing  95\% of the probability mass of the conditional distribution of $\mathbf{s}_{t_{i+1}}$.} \label{newfig:coord}
% \end{figure}

\begin{figure}[t!]
    \centering
		{\subfloat[CRW]{\includegraphics[trim = 10 0 10 30,scale=0.32]{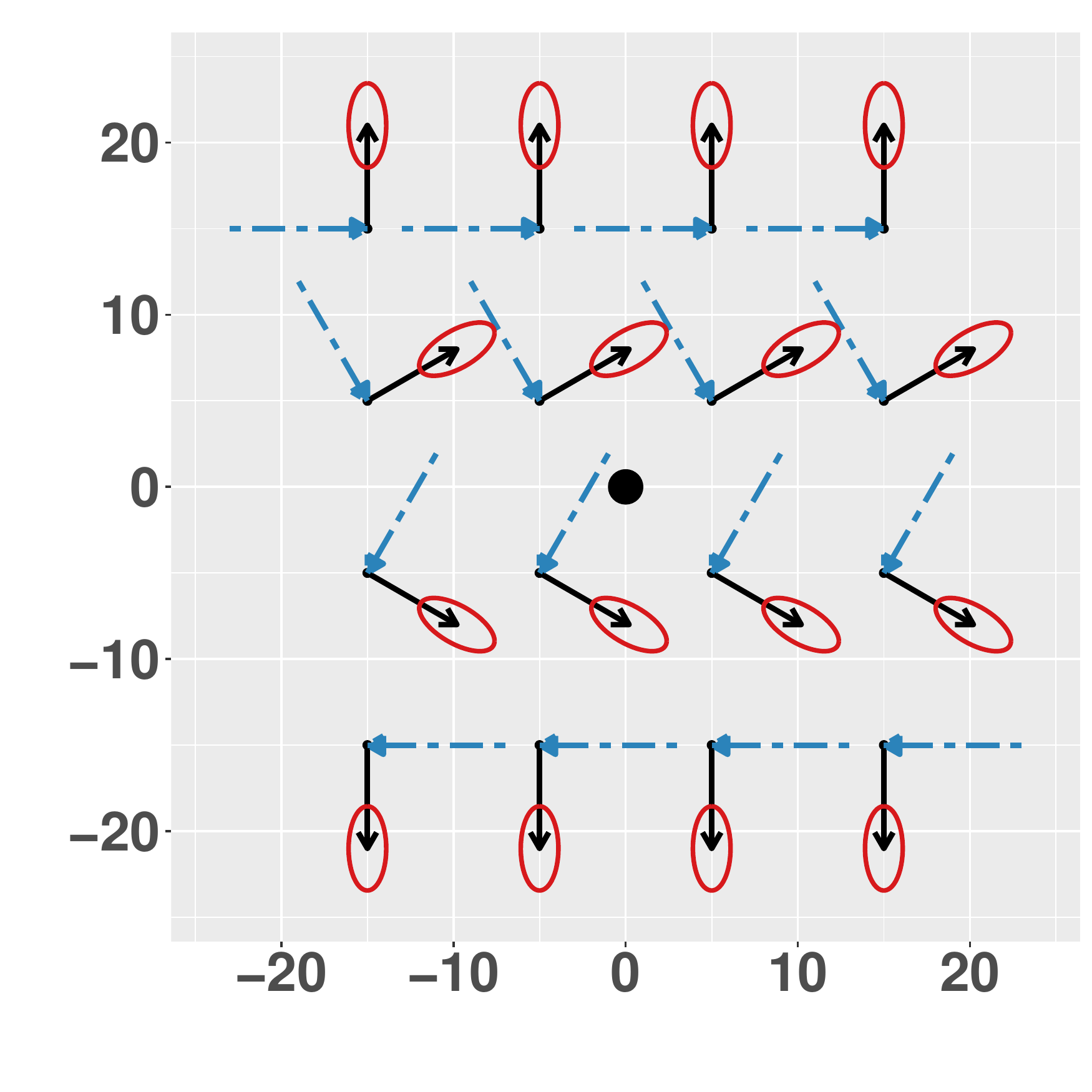}}}
    {\subfloat[BRW]{\includegraphics[trim = 10 0 10 30,scale=0.32]{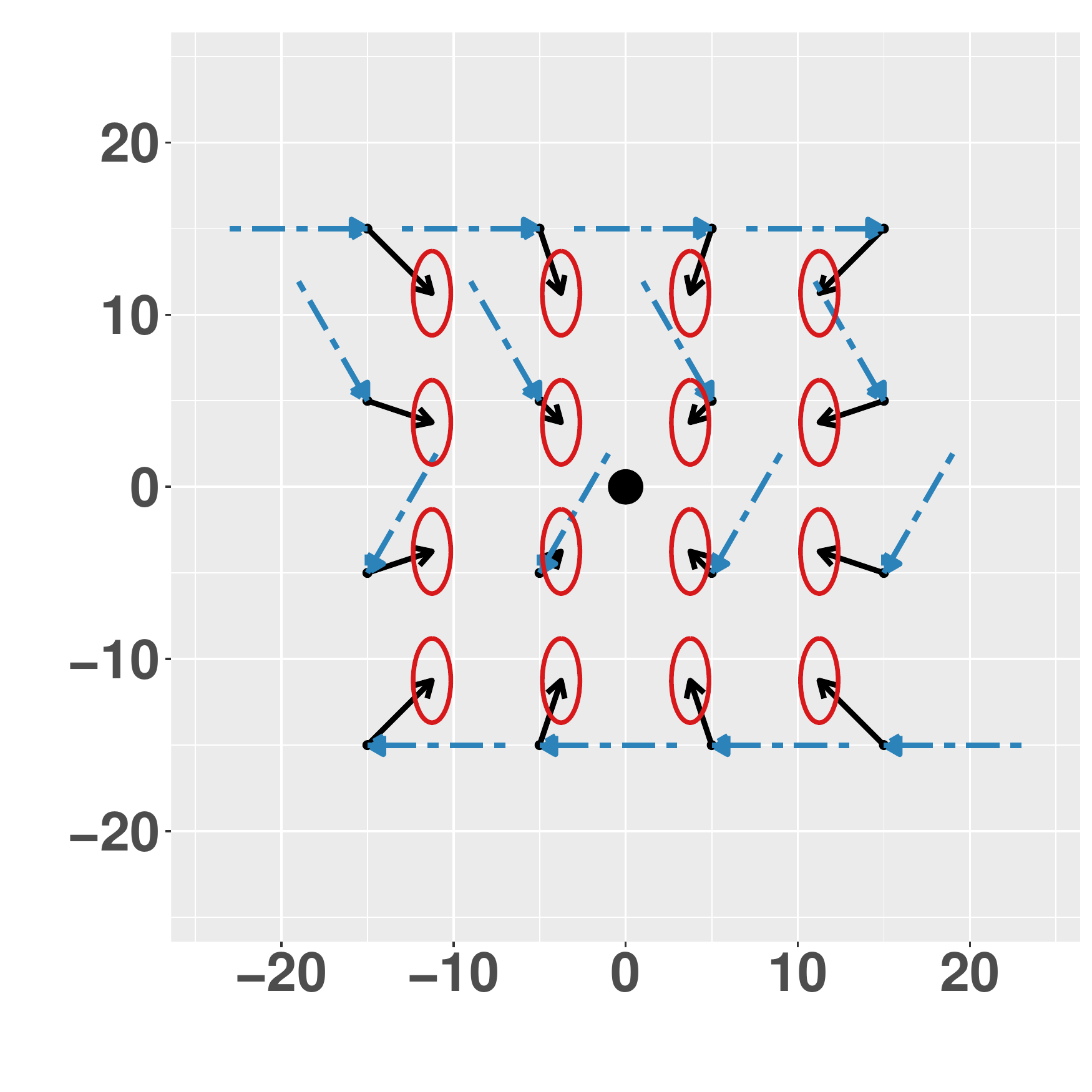}}}\\
    {\subfloat[BCRW with $\rho=2/3$]{\includegraphics[trim = 10 0 10 30,scale=0.32]{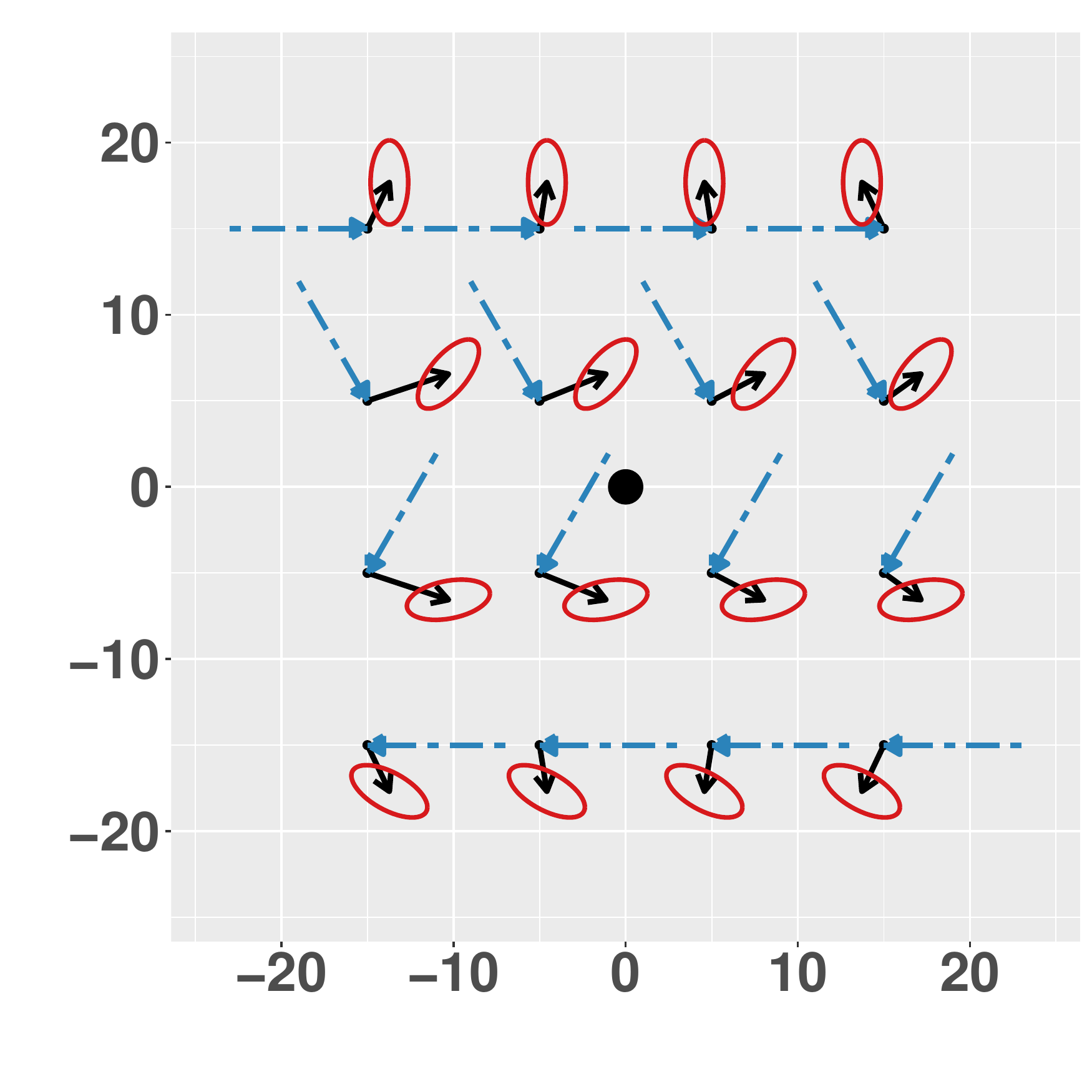}}}
    {\subfloat[BCRW with $\rho=1/3$]{\includegraphics[trim = 10 0 10 30,scale=0.32]{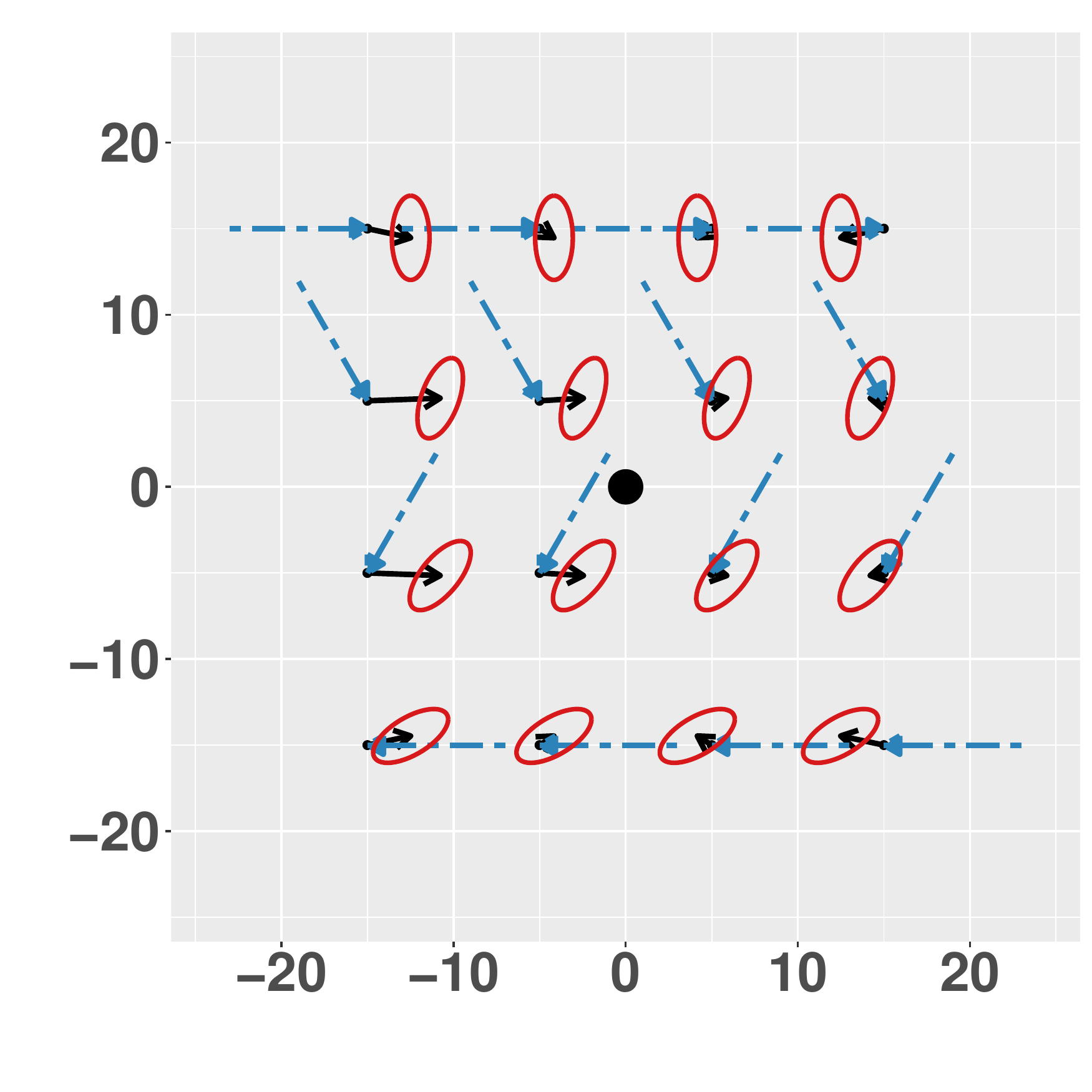}}}
     \caption{
% % 		Graphical representation of the conditional distribution of $\mathbf{s}_{t_{i+1}}$ for different  $\mathbf{s}_{t_{i}}$,  previous direction, and value of $\rho$.
% %     The dashed arrow represents the movement between  $\mathbf{s}_{t_{i-1}}$ and $\mathbf{s}_{t_{i}}$, the solid arrow is $\protect\vec{\mathbf{F}}_{t_i}$,  while the ellipse is an area containing  95\% of the probability mass of  the  conditional  distribution of $\mathbf{s}_{t_{i+1}}$.   In both figures,   $\boldsymbol{\mu} = (0,0)'$, $\boldsymbol{\eta} = (0,5)'$,  $\tau=0.25$,
% %    $ \boldsymbol{\Sigma}= \left(
% % \begin{array}{cc}
% % 0.2& 0\\
% % 0& 1
% % \end{array} \right)$.
 Graphical representation of the conditional distribution of $\mathbf{s}_{{i+1}}$ for different possible values of $\mathbf{s}_{{i}}$ and  the previous directions.
    The dashed arrow represents the movement between  $\mathbf{s}_{{i-1}}$ and $\mathbf{s}_{{i}}$. The solid arrow is $\protect\vec{\mathbf{F}}_{{i}}$, while the ellipse is the area containing  95\% of the probability mass of  the  conditional  distribution of $\mathbf{s}_{{i+1}}$.
		   $\boldsymbol{\mu} = (0,0)'$, $\boldsymbol{\eta} = (0,6)'$,
		$\tau=0.25$,
   $ \boldsymbol{\Sigma}= \left(
\begin{array}{cc}
0.2& 0\\
0& 1
\end{array} \right)$ in all figures, and the central dot is the location $\boldsymbol{\mu}$, which is the attractor  in the BRW and the BCRW.
}\label{newfig:EsModel}
\end{figure}

With the aim of    better understanding  the results of the real data application (Section \ref{newsec:realdata}) and the formalization of our proposal, we   briefly describe the  STAP distribution, which was  introduced in \cite{mastrantonio2020modeling}, and its parameters; for a more   detailed description the reader may refer to \cite{mastrantonio2020modeling}. \\
\indent We assume we have a time-series  of two-dimensional spatial locations
$\mathbf{s} = (\mathbf{s}_{t_1},\dots, \mathbf{s}_{t_T})'$  that represent an animal's path, where $\mathbf{s}_{t_i} = (s_{t_i,1},s_{t_i,2}) \in \mathcal{D} \subset \mathbb{R}^2$, and  $t_i$ is a temporal index. The coordinates are recorded without any measurement error and the time difference between consecutive points is constant.
In order to formalize the STAP,  we  introduce the bearing angle
\begin{equation}
\phi_{t_i} = \text{atan}^*({s}_{t_{i+1},2}-{s}_{t_{i} ,2}, {s}_{t_{i+1},1}-{s}_{t_{i},1}) \in [-\pi, \pi), \label{neweq:Phinew}
\end{equation}
and the rotation matrix
\begin{equation}
\mathbf{R}(x) = \left(
\begin{array}{cc}
\cos(x) & -\sin(x)\\
\sin(x) & \cos(x)
\end{array}
\right),
\end{equation}
where $\text{atan}^*(\cdot)$ is the two-argument tangent function \citep{Jammalamadaka2004}. The bearing-angle measures the direction of the movement between time $t_{i}$ and $t_{i+1}$, and the rotation matrix can be  used to  perform a rotation in a two-dimensional space, so that if it multiplies a two-dimensional vector, the vector  is rotated anti-clockwise by an angle $x$.
The conditional distribution of $\mathbf{s}_{t_{i+1}}$ is assumed to be second-order Markovian,  with the following specification:
\begin{align} \label{neweq:scondnew}
    \mathbf{s}_{t_{i+1}} |\mathbf{s}_{t_{i}},\mathbf{s}_{t_{i-1}} & \sim N(\mathbf{s}_{t_{i}}+\mathbf{M}_{t_i},\mathbf{V}_{t_i} ), \, \, i \in \{1, \dots , T-1\},\\
    \mathbf{M}_{t_i} & =  (1-\rho)\tau \left(\boldsymbol{\mu}-  \mathbf{s}_{t_i} \right)+\rho\mathbf{R}(\phi_{t_{i-1}})\boldsymbol{\eta},\\
   \mathbf{V}_{t_i} & = \mathbf{R}(\rho\phi_{t_{i-1}})\boldsymbol{\Sigma}\mathbf{R}'(\rho\phi_{t_{i-1}}),
\end{align}
where $\boldsymbol{\mu}, \boldsymbol{\eta} \in \mathbb{R}^2$, $\tau \in (0,1)$, $\rho \in [0,1]$, and $\boldsymbol{\Sigma}$ is a 2-dimensional covariance matrix.  The location  $\mathbf{s}_{t_{1}}$ is   fixed, and $  \mathbf{s}_{t_{0}}$
is another parameters that is needed to  compute $\phi_{t_0}$ in the conditional distribution of $\mathbf{s}_{t_{2}}$. If the path follows equation \eqref{neweq:scondnew}, we write
$\mathbf{s}_{t_{i+1}} |\mathbf{s}_{t_{i}},\mathbf{s}_{t_{i-1}} \sim \text{STAP}(\boldsymbol{\theta})$, with $\boldsymbol{\theta} = (\boldsymbol{\mu}, \boldsymbol{\eta},\boldsymbol{\Sigma}, \tau, \rho)$.\\
\indent The movement described by the STAP  can have directional-persistence and attraction to a point in space, therefore, the STAP is a BCRW. To better understand these  two properties and how they are formalized in  equation \eqref{neweq:scondnew},
 we  introduce  the vector $\vec{\mathbf{F}}_{t_i}$, which  is a vector   with initial and terminal points  equal to $\mathbf{s}_{t_i}$  and $\mathbf{s}_{t_{i}}+\mathbf{M}_{t_i}$,
 respectively. This vector represents the expected movement between time $t_{i}$ and $t_{i+1}$, since its initial point is the previously observed location $\mathbf{s}_{t_i}$ and the
 terminal one is equal to $\mathbb{E}(\mathbf{s}_{t_{i+1}}|\mathbf{s}_{t_{i}})$. If $\rho=0$, the STAP reduces to a two-dimensional AR(1) (a  BRW),  and  $\vec{\mathbf{F}}_{t_i}$ points to the spatial location $\boldsymbol{\mu}$, which is therefore the attractor.  The length of $\vec{\mathbf{F}}_{t_i}$ is
 $\tau|| \left(\boldsymbol{\mu}-  \mathbf{s}_{t_i} \right)||$, which shows that $\tau$ measures how much of the total distance between the last observation
 ($\mathbf{s}_{t_i}$) and the attractor ($\boldsymbol{\mu}$) is expected to be covered or, in other words, how strong  the attraction to $\boldsymbol{\mu}$ is.
 If $\rho=1$,  the STAP reduces to a CRW, based on a normal density. In this case, the direction of $\vec{\mathbf{F}}_{t_i}$ is the same as the direction of $\mathbf{R}(\phi_{t_{i-1}})\boldsymbol{\eta}$, which depends on the previous   bearing angle $\phi_{t_{i-1}}$,  and thus induces a   directional-correlation between consecutive points.  If  $\rho \in (0,1)$,  $\vec{\mathbf{F}}_{t_i}$
 is  a  weighted mean between its value in a  BRW  and a CRW, with weights given by $(1-\rho)$ and $\rho$, respectively. The covariance matrix of the conditional distribution of $\mathbf{s}_{t_{i+1}}$ is fixed in a  BRW ($\text{Cov}(\mathbf{s}_{t_{i+1}}|\mathbf{s}_{t_{i}})=\boldsymbol{\Sigma}$), while it rotates with  the bearing-angle for   any
 $\rho>0$ ($\text{Cov}(\mathbf{s}_{t_{i+1}}|\mathbf{s}_{t_{i}})=\mathbf{R}(\rho\phi_{t_{i-1}})\boldsymbol{\Sigma}\mathbf{R}'(\rho\phi_{t_{i-1}})$): for more details, the reader may refer to  \cite{mastrantonio2020modeling}.\\
 \indent  We show examples of STAP densities, and   the associated BRW ($\rho=0$) and CRW ($\rho=1$) in Figure \ref{neweq:scondnew}, to better understand the differences between the BRW, CRW and BCRW.   The dashed arrow in the figure is the movement  between times $t_{i-1}$ and $t_{i}$, the solid arrow is  $\vec{\mathbf{F}}_{t_i}$, and the ellipse  is a contour  of the   conditional distribution of $\mathbf{s}_{t_{i+1}}$ with a constant density containing 95\% of the total probability mass.
 From \ref{neweq:scondnew} (a), we can see that the direction of
  $\vec{\mathbf{F}}_{t_i}$ and the ellipse change according to $\phi_{t_{i-1}}$ in a CRW. On the other hand $\vec{\mathbf{F}}_{t_i}$ and the ellipse are independent from the previous direction in a  BRW (Figure \ref{neweq:scondnew} (b)), and $\vec{\mathbf{F}}_{t_i}$  points to the spatial attractor $\boldsymbol{\mu}=(0,0)'$. When $\rho \in (0,1)$, the ellipse and   $\vec{\mathbf{F}}_{t_i}$
	are dependent on both the previous direction and the spatial attractor, see Figures \ref{neweq:scondnew} (c) and (d).

\section{The proposed model} \label{newsec:mod}

% It is unrealistic to assume that the parameters describing the movement  do not change with time, since different behaviors can be observed in a time-window. For this reason, often
%
%
%
% heterogeneity is introduced with a mixture-type model, where the latent classification is   the behavior exhibited by the animal at a given time-point.  Different approaches have been proposed, see, for example, \cite{Patterson2017}, \cite{Harris201329} or \cite{mastrantonio2019}, but the most commonly used is the HMM, which is the one we use here.
% The way we define the  the distributions  over the STAP parameters allows us to introduced the  main contribution of our proposal, that is the ability to share parameters between behaviors between and within animal, and, as a by-product, we are able to estimate the number of latent behaviors for each animal.

In this section, we introduce the  components of the model and how they are used to introduce the characteristics of our proposal.
 We extend the notation of the previous section to describe the path of $m$  animals, and to allow changes in the behavior to be considered.\\
\indent   We indicate   the path of the $j$-th animal with  $\mathbf{s}_{j}=(\mathbf{s}_{j,t_{j,1}},\mathbf{s}_{j,t_{j,2}},\dots , \mathbf{s}_{j,t_{j,T_j}})$, where $j=1,\dots , m$, and
$\mathcal{T}_{j} \equiv (t_{j,1}, t_{j,2}, \dots ,t_{j,T_j} )  $ is the set of temporal points, equally spaced in time, where the position of the $j$-th dog is recorded. The sets
 $\mathcal{T}_{j}$ and $\mathcal{T}_{j'}$ can contain different time-points, but the time difference must be constant across animals, i.e., $t_{j,i+1}-t_{j,i} = c$ for all $j= 1,\dots , m$ and $i = 1,\dots T_j-1$.
  We  introduce a discrete random variable $z_{j,t_{j,i}} \in \mathbb{N}$ to represent the animal behavior at time $t_{j,i}$,
where $z_{j,t_{j,i}}=k$ indicates that
animal $j$ follows behavior  $k$ at time $t_{j,i}$. Given the behavior assumed by each animal, the paths are independent and
\begin{equation} \label{neweq:stapHMMpar}
\mathbf{s}_{j,t_{j,i+1}}|\mathbf{s}_{j,t_{j,i}},\mathbf{s}_{j,t_{j,i-1}}, z_{j,t_{j,i}}  \sim \text{STAP}(\boldsymbol{\theta}_{z_{j,t_{j,i}}} )
\end{equation}
where $\boldsymbol{\theta}_k = (\boldsymbol{\mu}_k, \boldsymbol{\eta}_k,\boldsymbol{\Sigma}_k, \tau_k, \rho_k)$. In other words, if  the $j$-th animal is following the $k$-th behavior at time $t_{j,i}$ (i.e., $z_{j,t_{j,i}}=k$),  the path is described by the set $\boldsymbol{\theta}_k$ of STAP parameters. It should be noted that   the $k$-th behaviors are represented by the same set of parameters $\boldsymbol{\theta}_k$  for all animals.\\
\indent Let $\mathbf{{s}} = \{\mathbf{{s}}_{j}\}_{j=1}^{m}$, $\mathbf{z}_j= \{  z_{j,t}   \}_{t \in \mathcal{T}_j}$,  $\mathbf{{z}} = \{\mathbf{{z}}_{j}\}_{j=1}^{m}$, and
$\boldsymbol{\theta} = \{\boldsymbol{\theta}_k\}_{k \in \mathbb{N}}$, then
the model we propose  is
\begin{align}
f(\mathbf{{s}}|\boldsymbol{\theta}, \mathbf{z})& = \prod_{j=1}^m \prod_{i=1 }^{T_j-1} f(\mathbf{s}_{j,t_{j,i+1}}|\mathbf{s}_{j,t_{j,i}},\mathbf{s}_{j,t_{j,i-1}},\boldsymbol{\theta}_{z_{j,t_{j,i}}}) ,\label{neweq:s1}\\
\mathbf{s}_{j,t_{j,i+1}}|\mathbf{s}_{j,t_{j,i}},\mathbf{s}_{j,t_{j,i-1}},z_{j,t_{j,i}},\boldsymbol{\theta}_{z_{j,t_{j,i}}} &\sim \text{STAP}(\boldsymbol{\theta}_{z_{j,t_{j,i}}} ),  \quad \mathbf{s}_{j,t_{j,0}}  \sim \text{Unif}(\mathcal{D}),\label{neweq:s2}\\
z_{j,t_{j,i}}|z_{j,t_{j,i-1}},\boldsymbol{\pi}_{j,z_{j,t_{j,i-1}}} &\sim \text{Multinomial}(1,\boldsymbol{\pi}_{j,z_{j,t_{j,i-1}}}),   \label{neweq:z}\\
z_{j,t_{j,0}} &\sim \text{Geom}(\epsilon),  \\
\boldsymbol{\pi}_{j,l}|\alpha, \nu, \boldsymbol{\beta} & \sim \text{DP}\left(\alpha+\nu,   \frac{\alpha\boldsymbol{\beta}+ \nu\delta_{l}}{\alpha+\nu}\right),\label{neweq:pi}\\
\{{\beta}_k \}_{k \in \mathbb{N}} & = C_1(\boldsymbol{\beta}_{\boldsymbol{\mu}}^*, \boldsymbol{\beta}_{\boldsymbol{\eta}}^*,\boldsymbol{\beta}_{\boldsymbol{\Sigma}}^*, \boldsymbol{\beta}_{\tau}^*, \boldsymbol{\beta}_{\rho}^*),  \label{neweq:c1}\\
\{\boldsymbol{\theta}_k\}_{k \in \mathbb{N}} & = C_2(\boldsymbol{\mu}^*, \boldsymbol{\eta}^*,\boldsymbol{\Sigma}^*,  \boldsymbol{\tau}^*,  \boldsymbol{\rho}^*),\label{neweq:c2}\\
\boldsymbol{\beta}_{\boldsymbol{\mu}}^*|\gamma_{\boldsymbol{\mu}} \sim \text{Gem}(\gamma_{\boldsymbol{\mu}}),\, \boldsymbol{\beta}_{\boldsymbol{\eta}}^*|\gamma_{\boldsymbol{\eta}}&\sim \text{Gem}(\gamma_{\boldsymbol{\eta}}),\, \boldsymbol{\beta}_{\boldsymbol{\Sigma}}^*|\gamma_{\boldsymbol{\Sigma}} \sim \text{Gem}(\gamma_{\boldsymbol{\Sigma}}), \label{neweq:betaper}\\
\boldsymbol{\beta}_{\tau}^*|\gamma_{\tau}& \sim \text{Gem}(\gamma_{\tau}), \,
\boldsymbol{\beta}_{\rho}^*|\gamma_{\rho}  \sim \text{Gem}(\gamma_{\rho}), \\
\boldsymbol{\mu}_p^*|H_{\boldsymbol{\mu}}\sim H_{\boldsymbol{\mu}},\,
\boldsymbol{\eta}_p^*|H_{\boldsymbol{\eta}}&\sim H_{\boldsymbol{\eta}}, \,
\boldsymbol{\Sigma}_p^*|H_{\boldsymbol{\Sigma}}\sim H_{\boldsymbol{\Sigma}},
 \label{neweq:distpar} \\
\tau_p^*|H_{\tau} &\sim H_{\tau},\,
\rho_p^*|H_{\rho}\sim H_{\rho},
\end{align}
where  $p \in \mathbb{N}$, $l \in \mathbb{N}$, and $i=1,\dots , T_j-1$. A full description of the  components of the model and how they  are used to introduce the main novelties of our model is given below.

\paragraph*{\textbf{The DPs.}}
In order to simplify the description of the lower levels of the model hierarchy, we use
$\chi$ as a variable that can be $\boldsymbol{\mu}$, $\boldsymbol{\eta}$, $\boldsymbol{\Sigma}$, $\nu$, or $\rho$, and it is used when whatever is described can be applied  to any of the five parameters. In equation \eqref{neweq:distpar}, the  values of the STAP parameter $\chi_p$ are sampled from the distribution $H_{\chi}$, and
the  infinite-dimensional vector of probabilities $\boldsymbol{\beta}_{\chi}^* = \{ \beta_{\chi,p}^*  \}_{p \in \mathbb{N}}$,  associated with parameter $\chi$,  is  Gem distributed \citep{Gnedin01acharacterization} with \textit{scaling parameter} $\gamma_{\chi}$.
The scaling parameter can  easily be interpreted with the stick-breaking representation of
the   GEM distribution,   defined as follows:
\begin{align} \label{eq:sb}
\beta_{\chi,1}^* &\sim \text{B}(1, \gamma_{\chi}),\\
\frac{\beta_{\chi,p}^*}{1-\sum_{h=1}^{p-1} \beta_{\chi,h}^*} &\sim \text{B}(1, \gamma_{\chi}), \, p \neq 1.
\end{align}
From \eqref{eq:sb}, we  see that
the smaller  $\gamma_{\chi}$ is, and the
smaller is the number of elements of $\boldsymbol{\beta}_{\chi}$ that contains most of the probability mass with  $\lim_{\gamma_{\chi} \rightarrow 0}\beta_{\chi,1}^* = 1$ and $\lim_{\gamma_{\chi} \rightarrow 0} \beta_{\chi,p}^* = 0$ for all $p \neq 1$.
The vectors $\boldsymbol{\beta}_{\chi}^* $ and $\boldsymbol{\chi}^* = \{  {\chi}_p^*\}_{p \in \mathbb{N}}$ can be used to define the discrete distribution
 %\citep{Ferguson1973}, one for each   STAP parameter. For example, for parameter $\boldsymbol{\mu}$ we can say  that the discrete distribution
\begin{equation} \label{neweq:g}
G_{\chi} = \sum_{p \in \mathbb{N}} \beta_{\chi,p}^* \delta_{\chi_p^*},
\end{equation}
where
$\delta_{\cdot}$ is the Dirac delta function.  Equation \eqref{neweq:g} is a draw  from a $\text{DP}(\gamma_{\chi},H_{\chi})$, and thus we can  equivalently describe   $\boldsymbol{\beta}_{\chi}^* $ and $\chi_p$
 as the components of a sample from $\text{DP}(\gamma_{\chi},H_{\chi})$,
 or as equations  \eqref{neweq:betaper} and \eqref{neweq:distpar}.  The vectors  $\boldsymbol{\chi}^* $ and $\boldsymbol{\beta}_{\chi}^*$
contain, respectively, the  values that the parameters can assume (${\chi}_p^*$) and the ``base'' probability (${\beta}_{\chi_p}^*$) that a particular value of the parameter is selected in a behavior (see equation  \eqref{neweq:mean2} below).
%Then,  if $\gamma_{\chi} \rightarrow 0 $ the distribution $G_{\chi}$ concentrates the entire mass of probability on a single value,  if $\gamma_{\boldsymbol{\mu}} \rightarrow \infty $ the distribution becomes continuous.\\
%The scaling parameter specifies how strong this discretization is: in the limit of {\displaystyle \alpha \rightarrow 0}\alpha \rightarrow 0, the realizations are all concentrated at a single value, while in the limit of {\displaystyle \alpha \rightarrow \infty }\alpha \rightarrow \infty  the realizations become continuous. Between the two extremes the realizations are discrete distributions with less and less concentration as {\displaystyle \alpha }\alpha  increases
\paragraph*{\textbf{The functions $C_1(\cdot)$ and $C_2(\cdot)$.}}
The function $C_1(\cdot)$ and $C_2(\cdot)$  (equations \eqref{neweq:c1} and \eqref{neweq:c2})   introduce the sharing of parameters between behaviors, which is  one of the novelties of our proposal.
The role of function $C_2(\cdot)$  is to produce the set of STAP parameters $\{\boldsymbol{\theta}_k\}_{k \in \mathbb{N}}$, where we remind the reader that
 $\boldsymbol{\theta}_k=(\boldsymbol{\mu}_k,\boldsymbol{\eta}_k,\boldsymbol{\Sigma}_k,\tau_k, \rho_k)$.  The set $\{\boldsymbol{\theta}_k\}_{k \in \mathbb{N}}$ is
comprised of all the possible combinations of the 5 STAP
parameters, without repetitions.
%
%
%
% Given the sets of parameters  sampled from equation \eqref{neweq:distpar}, the function $C_2(\cdot)$ produces the infinite dimensional vector of STAP parameters $\{\boldsymbol{\theta}_k\}_{k \in \mathbb{N}}$, where we remind the reader that, from equation \eqref{neweq:stapHMMpar},
%  $\boldsymbol{\theta}_k=(\boldsymbol{\mu}_k,\boldsymbol{\eta}_k,\boldsymbol{\Sigma}_k,\tau_k, \rho_k)$ is the set of STAP parameters of the $$k$-th$ behaviors.
%
% The  function $C_2(\cdot)$ creates  all possible combinations, without repetition
% of the elements in the 5 sets $\boldsymbol{\mu}$, $\boldsymbol{\eta}$, $\boldsymbol{\Sigma}$, $\boldsymbol{\tau}$, and $\boldsymbol{\rho}$.
This means that  $\boldsymbol{\theta}_k \neq \boldsymbol{\theta}_{k'}$, if  $k \neq k'$, but  we can have a subset of elements  that has the same value, e.g.,
$\tau_k \equiv \tau_{k'}$. Hence,  since each behavior selects its STAP parameters  in $\boldsymbol{\theta} = \{\boldsymbol{\theta}_k\}_{k \in \mathbb{N}}$, different behaviors can share   parameters, even though they are described by a different $\boldsymbol{\theta}_k$. \\
\indent Function $C_1(\cdot)$ is closely related  to $C_2(\cdot)$ since,  if  we  introduce  the  new variables
 $\lambda_{\boldsymbol{\mu},k}$, $\lambda_{\boldsymbol{\eta},k}$, $\lambda_{\boldsymbol{\Sigma},k}$, $\lambda_{\tau,k}$ and $\lambda_{\rho,k}$ that represent what parameter is in $\boldsymbol{\theta}_k$, i.e.,
\begin{equation}
\boldsymbol{\mu}_k = \boldsymbol{\mu}_{\lambda_{\boldsymbol{\mu},k}}^*, \,\,
\boldsymbol{\eta}_k = \boldsymbol{\eta}_{\lambda_{\boldsymbol{\eta},k}}^*, \,\,
\boldsymbol{\Sigma}_k = \boldsymbol{\Sigma}_{\lambda_{\boldsymbol{\Sigma},k}}^*, \,\,
\boldsymbol{\tau}_k = {\tau}_{\lambda_{ {\tau},k}}^*, \,\,
\boldsymbol{\rho}_k = {\rho}_{\lambda_{ {\rho},k}}^*, \label{neweq:ss}
\end{equation}
we can associate a value  $\beta_k$  to $\boldsymbol{\theta}_k$ which is  computed as
%
% Since probability values ($\beta_{\chi,p}^*$) is associated to each   STAP parameters atom, see equation \ref{neweq:g}, we  can associate to $\boldsymbol{\theta}_k$  a value $\beta_k$ of a vector of probability,  which is computed as the product of the associated $\beta_{\chi,p}^*$:
\begin{equation}\label{neweq:beta1}
 \beta_k =\beta_{\boldsymbol{\mu},{\lambda_{\boldsymbol{\mu},k}}}^* \beta_{\boldsymbol{\eta},{\lambda_{\boldsymbol{\eta},k}}}^* \beta_{\boldsymbol{\Sigma},{\lambda_{\boldsymbol{\Sigma},k}}}^* \beta_{\tau,{\lambda_{\tau,k}}}^* \beta_{\rho,{\lambda_{\rho,k}}}^*.
\end{equation}
The set $\{\beta_k\}_{k \in \mathbb{N}}$ is the output of $C_1(\cdot)$ and it is a probability vector, since $\sum_{k \in \mathbb{N}} \beta_k=1$ and $\beta_k \in (0,1)$.
We can define the discrete distribution
\begin{equation}
G_{0} = \sum_{k \in \mathbb{N}} \beta_{k} \delta_{\boldsymbol{\theta}_k}, \label{neweq:g000}
\end{equation}
where, similarly to \eqref{neweq:g},
its atoms
contain all the possible values that   $\boldsymbol{\theta}_k$ can assume  and  $\beta_{k}$ is connected to the expected value of  the probability of selection $\boldsymbol{\theta}_k$   as the vector of parameter in a behavior (see equation \eqref{neweq:mean} below). This way to define the distribution $G_{0}$ is closely related to the  shared base-distribution of the change-point model of
\cite{mastrantonioEnv21}.
% \begin{align}
% \beta_{k} &\sim  \text{GEM}(1,\gamma),\\
% \boldsymbol{\theta}_k & \sim H
% \end{align}
% instead of \eqref{neweq:betaper} and \eqref{neweq:distpar}, we retrieve  the sHDP-HMM.
% \text{Gem}(\gamma_{\boldsymbol{\Sigma}}), \label{neweq:betaper}\\
% \boldsymbol{\beta}_{\tau}^*|\gamma_{\tau}& \sim \text{Gem}(\gamma_{\tau}), \,
% \boldsymbol{\beta}_{\rho}^*|\gamma_{\rho}  \sim \text{Gem}(\gamma_{\rho}), \\
% \boldsymbol{\mu}_p^*|H_{\boldsymbol{\mu}}\sim H_{\boldsymbol{\mu}},\,
% \boldsymbol{\eta}_p^*|H_{\boldsymbol{\eta}}&\sim H_{\boldsymbol{\eta}}, \,
% \boldsymbol{\Sigma}_p^*|H_{\boldsymbol{\Sigma}}\sim H_{\boldsymbol{\Sigma}},
%  \label{neweq:distpar} \\
% \tau_p^*|H_{\tau} &\sim H_{\tau},\,
% \rho_p^*|H_{\rho}\sim H_{\rho},

\paragraph*{\textbf{Behavior switching.}}

Let   $\boldsymbol{\Pi}_j$  be the matrix that has  $\boldsymbol{\pi}_{j,l} = \{ \pi_{j,l,k}\}_{k \in \mathbb{N}}$ as  $l$-th row. Matrix $\boldsymbol{\Pi}_j$ rules the switching between  the behaviors of animal $j$ (equation \eqref{neweq:z}) and if  the $j$-th animal is following behavior $l$ at time $t_{j,i-1}$,  the probability of switching to behavior $k$ is given by the element of  $\boldsymbol{\Pi}_j$
in row $l$ and column $k$. Hence,  the time evolution of $z_{j,t_{j,i}}$ is modeled  by a discrete first-order Markov process,  which defines an HMM with transition matrix $\boldsymbol{\Pi}_j$ and  initial state $z_{j,t_{j,0}}$. The initial state is drawn from a Geometric distribution with parameter $\epsilon$,  which is defined as   the number of Bernoulli trials needed to have one success.
 The row $\boldsymbol{\pi}_{j,l}$ is DP distributed (see equation \eqref{neweq:pi}) and the  expected value of ${\pi}_{j,l,k} $   is equal to
\begin{equation} \label{neweq:mean}
  \text{E}( {\pi}_{j,l,k}| \alpha, \kappa,\boldsymbol{\beta}) = \frac{\alpha{\beta_k}+ \nu\delta(l,k)}{\alpha+\nu}
\end{equation}
\citep[see][]{fox2011},
where $\delta(l,k)$ is equal to 1 if $l=k$, and 0 otherwise.
From \eqref{neweq:mean}, we can see that the  $k$-th element of  $\boldsymbol{\beta}$  is associated with the  expected value of   ${\pi}_{j,l,k}$, for all the animals ($j=1,\dots , m$) and $l  \in \mathbb{N}$. Hence,  a larger  $\beta_k$ increases the probability of switching from any behavior $l$ to the $k$-th,  described by  $ \boldsymbol{\theta}_k$. On the other hand, equation \eqref{neweq:mean} can also be stated as
\begin{equation} \label{neweq:mean2}
  \text{E}( {\pi}_{j,l,k}| \alpha, \kappa,\boldsymbol{\beta}) = \frac{\alpha \beta_{\boldsymbol{\mu},{\lambda_{\boldsymbol{\mu},k}}}^* \beta_{\boldsymbol{\eta},{\lambda_{\boldsymbol{\eta},k}}}^* \beta_{\boldsymbol{\Sigma},{\lambda_{\boldsymbol{\Sigma},k}}}^* \beta_{\tau,{\lambda_{\tau,k}}}^* \beta_{\rho,{\lambda_{\rho,k}}}^* + \nu\delta(l,k)}{\alpha+\nu},
\end{equation}
which highlights how the value $\beta_{\chi,p}^*$ is connected to all  ${\pi}_{j,l,k}$, with $l , k \in \mathbb{N}$ and $j=1,\dots , m$, so that $\lambda_{\chi,k}=p$. Therefore, a larger $\beta_{\chi,p}^*$ increases the  expected values of all these probabilities, and  for this reason we call  $\beta_{\chi}^*$  the ``base probability'' as $\chi_p^*$.
The variable $\alpha$ is the scaling parameter of the DP of equation \eqref{neweq:pi}, which has the same interpretation of
$\gamma_{\chi}$,
while $\kappa$ is a weight that is added to the self transitions ${\pi}_{j,l,l}$ to increase their expected value, see equation \eqref{neweq:mean},
which in turn is used to  reduce the tendency of the HDP-HMM  to create redundant behaviors, i.e., behaviors with similar parameter vectors. For a  more detailed description  of parameters $\alpha$ and $\nu$, the reader may refer to \cite{fox2011}.\\
%\indent Given the transition matrix $\boldsymbol{\Pi}_j$, the behavior assumed by the $$j$-th$ animal at time $t_{j,i}$, is selected from  a multinomial   with vector of probabilities given by  row  $l$ of   $\boldsymbol{\Pi}_j$ if $z_{j,t_{j,i-1}}=l$.
\indent
It should be noted that, in most applications, see, for example, \cite{McClintock2012} and \cite{LeosBarajas2017},    $z_{j,t_{j,i}} \in \{1,2,\dots, K^*\}$ is assumed,  where $K^*$ indicates the maximum number of behaviors, while  we have  $z_{j,t_{j,i}} \in \mathbb{N}$ in this work, since we  define the HMM using DPs.
Thus, the model assumes an infinite and countable number of possible behaviors  for each animal, but, since we have a finite number of observed time-points, only a finite number
$K_j$ of them can be ``occupied'';
these are generally called   ``non-empty states'' \citep{Schnatte2019},  or, in this context, ``non-empty behaviors''.
The random variable  $K_j$   is used to estimate  the number of latent  behaviors of the $j$-th animal.
% The possibility to share set of parameters between animals and behavior is introduced in the way the distribution over $\boldsymbol{\theta}_k$ is defined.
Parameters $\alpha$, $\kappa$ and  $\gamma$ (through $\boldsymbol{\beta}$) determine the number of non-empty states $K_j$, since they are responsible with the total mass associated with each of the $\boldsymbol{\Pi}_j$ elements.

%\indent If a small number of elements  of $\boldsymbol{\Pi}_j$ have most of the probability mass, then a few number of behaviors can be observed, and this depends on the scaling parameter $\alpha$, $\kappa$, and the values of $\boldsymbol{\beta}$: the latter are themselves dependent on the scaling parameters $\gamma$.

\paragraph*{\textbf{The emission-distribution.}}

The model specification is concluded with the emission distribution, which is given by
\eqref{neweq:s1} and \eqref{neweq:s2}. It should be noted that, given the latent behaviors, we consider the animals independent but, since they share the same set of atoms $\{\boldsymbol{\theta}_k\}_{k \in \mathbb{N}}$, the  behaviors of the  different animals can  be described by the same STAP distribution.   From equation \eqref{neweq:ss}, we know that $\boldsymbol{\theta}_k$ and $\boldsymbol{\theta}_{k'}$ can have common components and therefore, when an animal changes behavior, it is not necessary for all the parameters to change and, more importantly, we can also identify the features that two animals share for different behaviors, e.g., two behaviors can have  the same attractive-point  ($\boldsymbol{\mu}_p^*$), even though the strength of attraction
($ {\tau}_p^*$) is different. This feature is one of the main novelties of our proposal, and, although other approaches have a similar characteristic (see, for example, \cite{Jonsen2016} or \cite{Milner2021}),  they do not allow  the number of latent behaviors  to be estimate, which is the other main novelty of our proposal,  the possibility of evaluating, during model fitting,  what parameters are allowed to change, or the number of values that a parameter can assume. The sharing of a subset of the parameters between states   is new in the context of HMMs based on DPs.

\paragraph{\textbf{Connection to the sHDP-HMM.}} To conclude this section, we would like to show that our proposal can be considered as a generalization of the sHDP-HMM. The model of
 \cite{fox2011}  is defined for a single time-series,  and  $G_{0}$ is a  draw from a DP.
It is easy to see that, if we consider only one animal and  use only one multivariate parameter in equation \eqref{neweq:distpar}, e.g., $\boldsymbol{\theta}_p^* = (\boldsymbol{\mu}_p^*,\boldsymbol{\eta}_p^*,\boldsymbol{\Sigma}_p^*,\tau_p^*, \rho_p^*)$, with the associated vector of probability $\boldsymbol{\beta}_{\boldsymbol{\theta}}^*$, the distribution   $G_{0}$ is a draw from a DP, since
$\beta_{k} = {\beta}_{\boldsymbol{\theta},p}^*$ and $\boldsymbol{\theta}_k = \boldsymbol{\theta}_p^*$. Hence,  the model reduces to a sHDP-HMM.

%
% \paragraph*{}
% Notice that  the number of possible behaviors is infinite,  since $\boldsymbol{\pi}_{j,l}$  is infinite-dimensional. Nonetheless,  in the observed time-window, only a finite  number of behaviors can be observed. The  $K_j$ unique values assumed by $z_{j, t_i}$ is then  a random variable that  we use to  estimate  the number of latent behaviors.

\section{Real data application} \label{newsec:realdata}

\begin{figure}[t]
  \centering
  {\subfloat[Woody]{\includegraphics[scale=0.22]{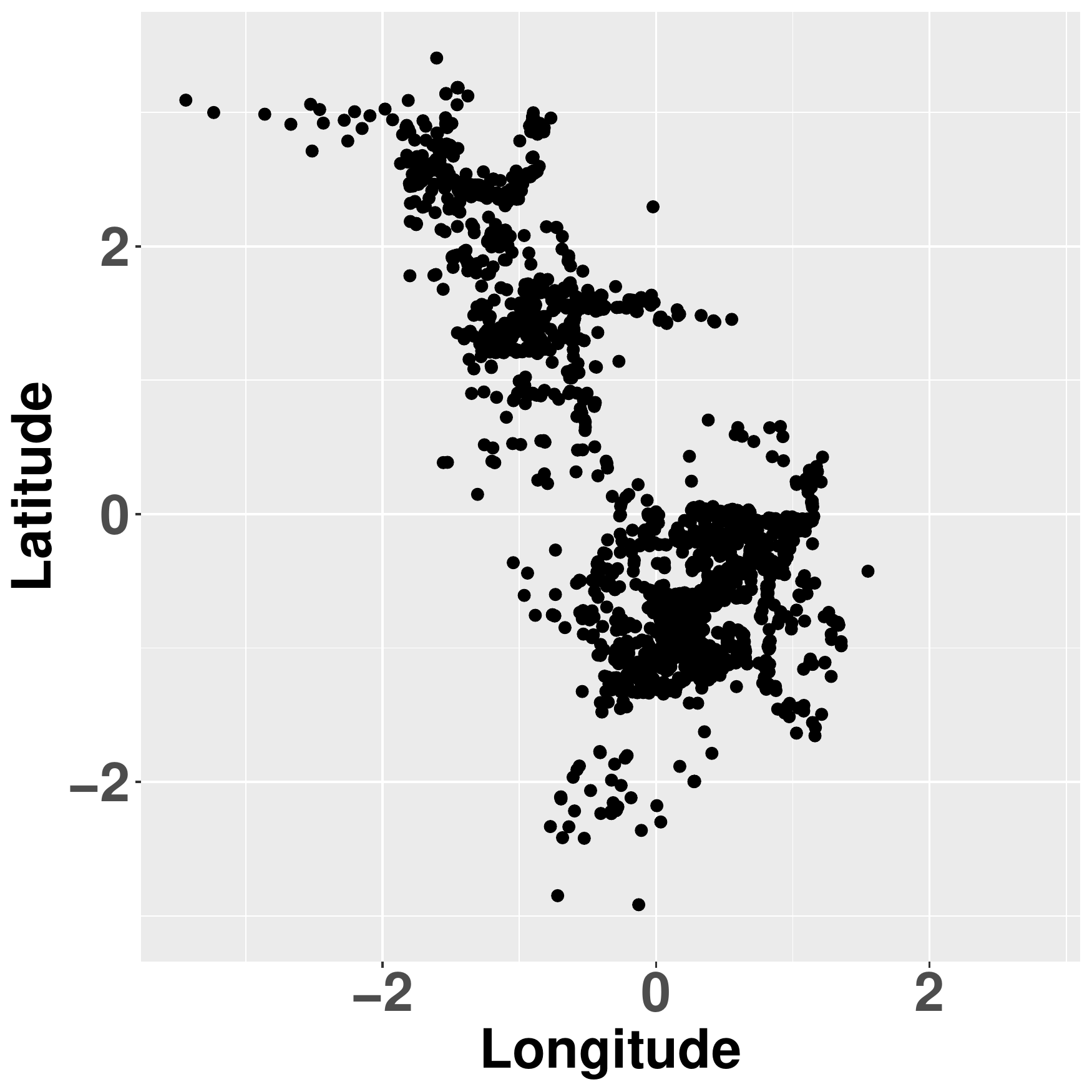}}}
  {\subfloat[Sherlock]{\includegraphics[scale=0.22]{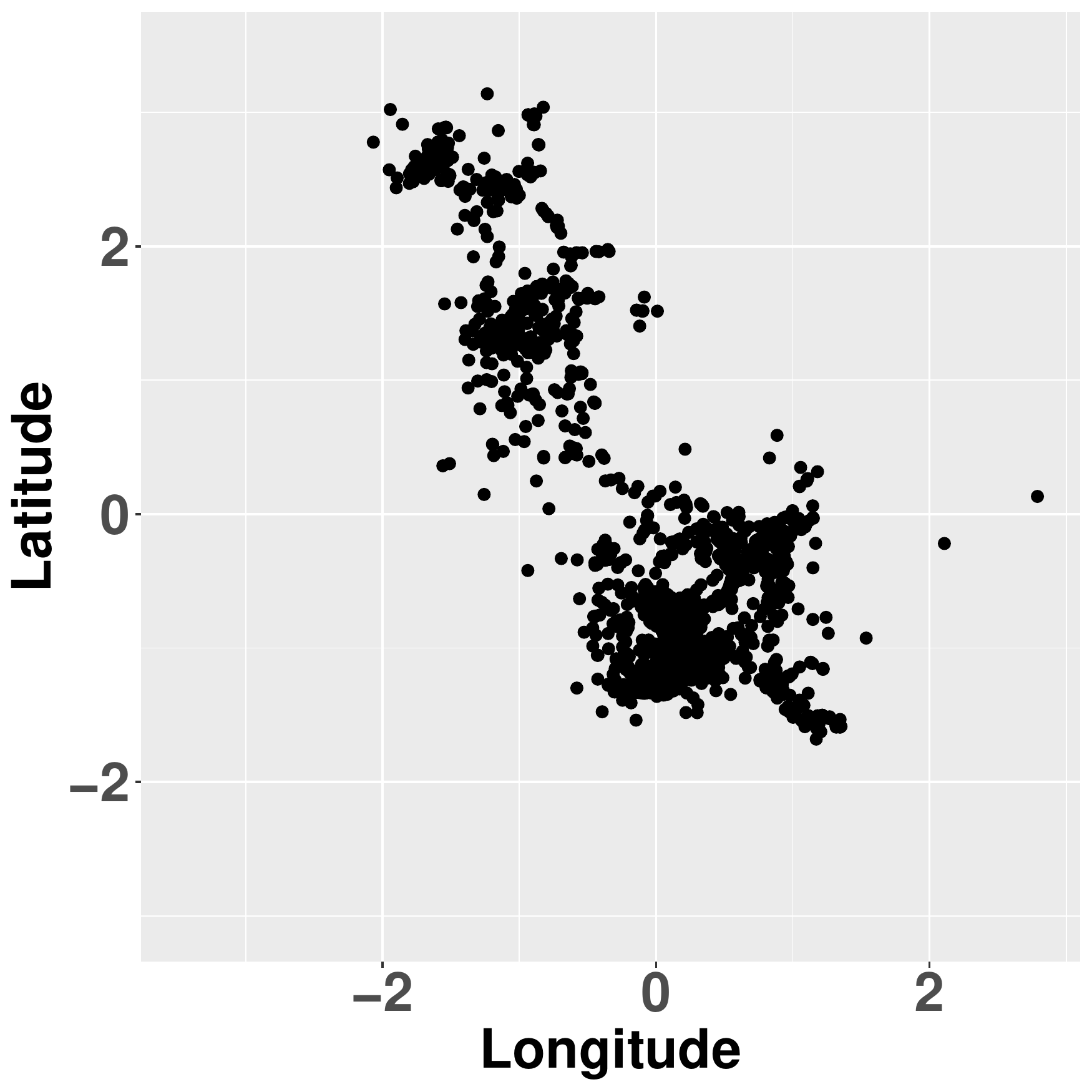}}}
  {\subfloat[Alvin]{\includegraphics[scale=0.22]{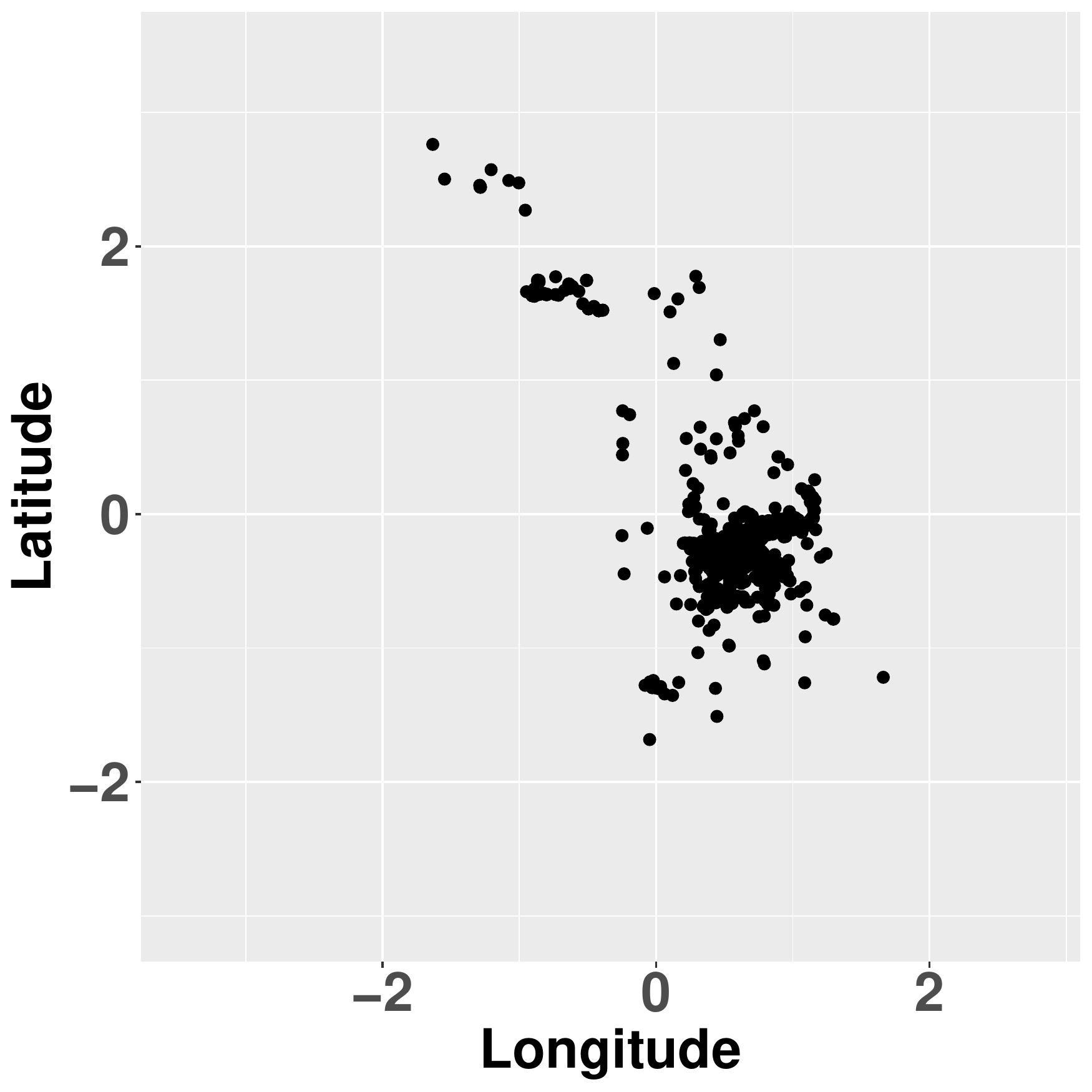}}}\\
  {\subfloat[Rosie]{\includegraphics[scale=0.22]{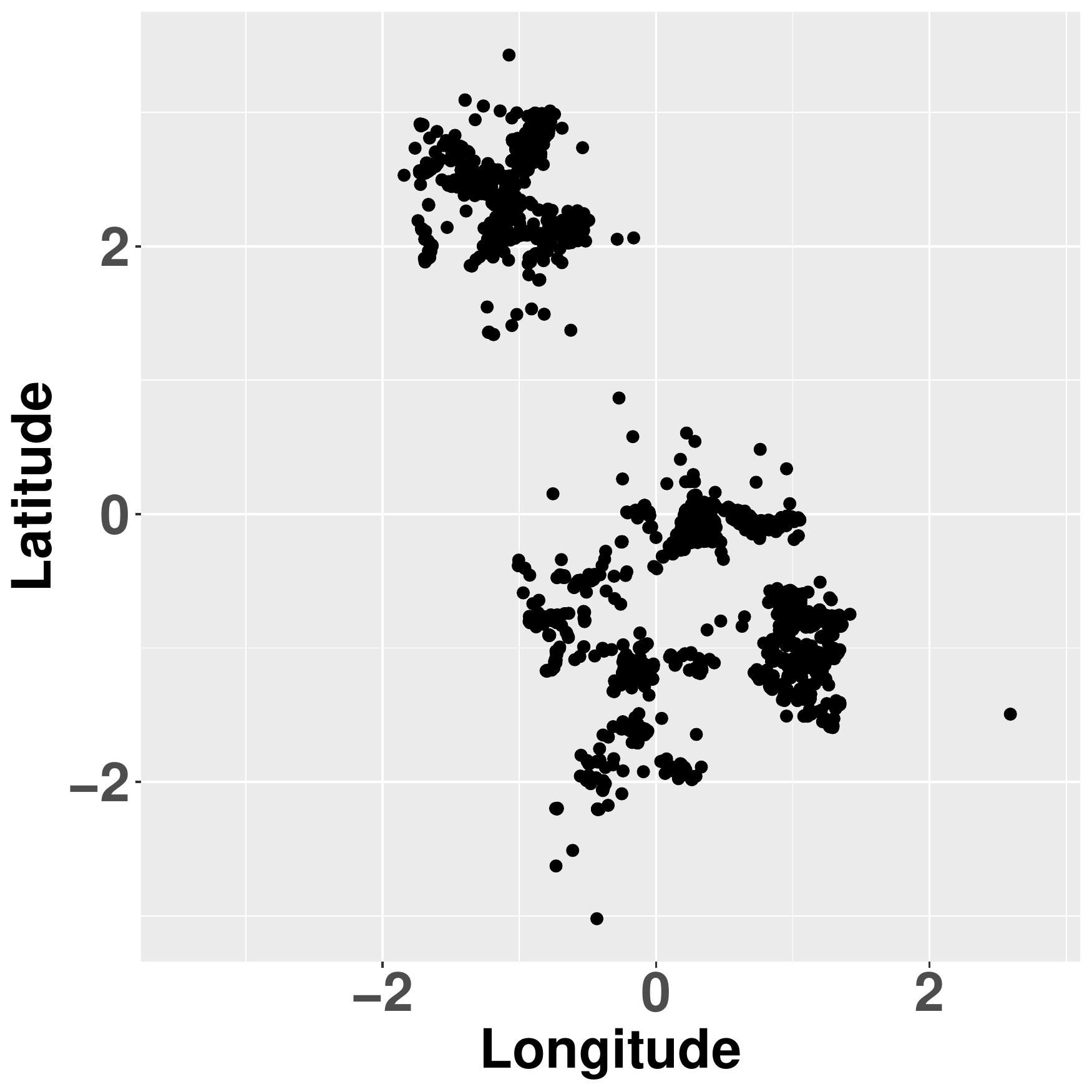}}}
  {\subfloat[Bear]{\includegraphics[scale=0.22]{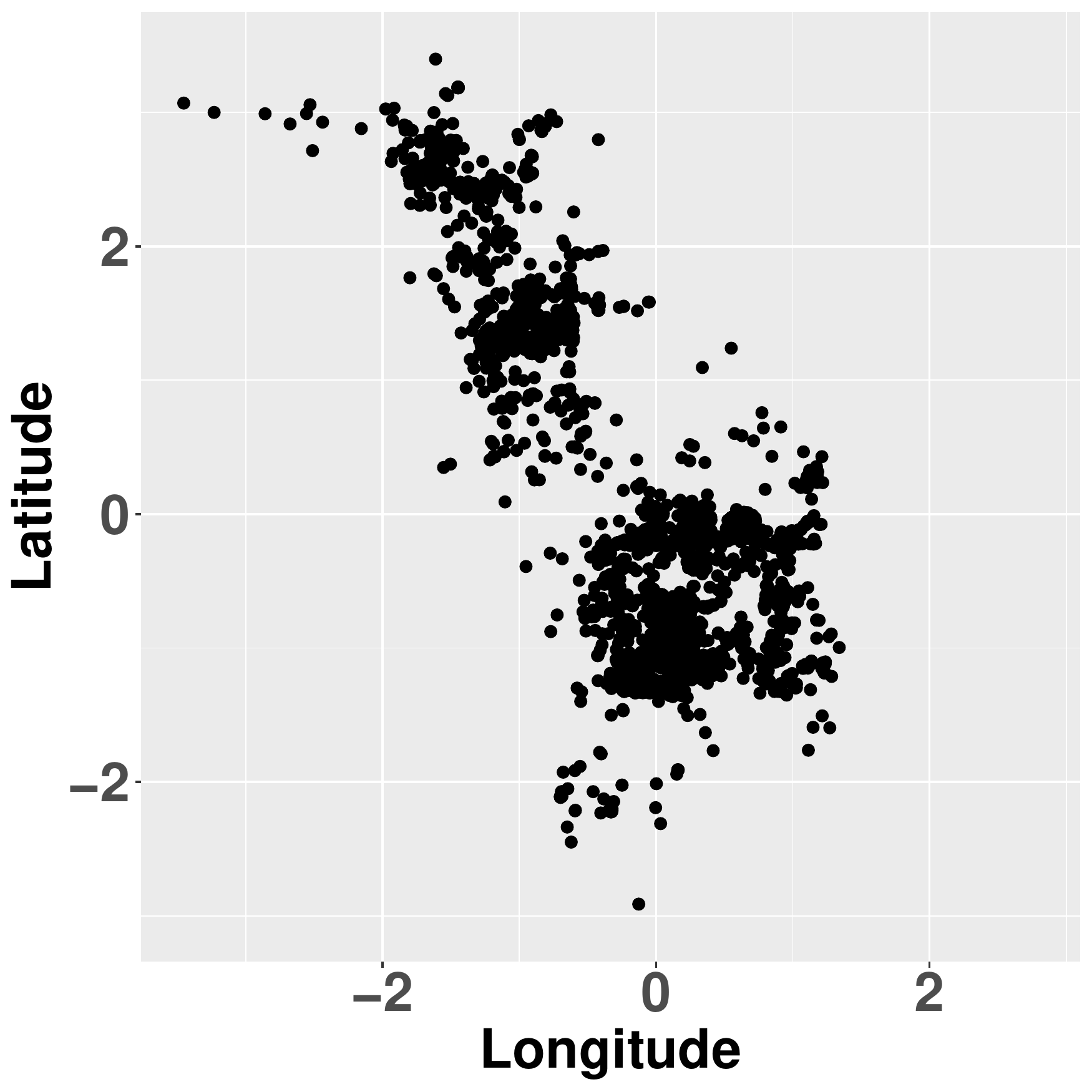}}}
  {\subfloat[Lucy]{\includegraphics[scale=0.22]{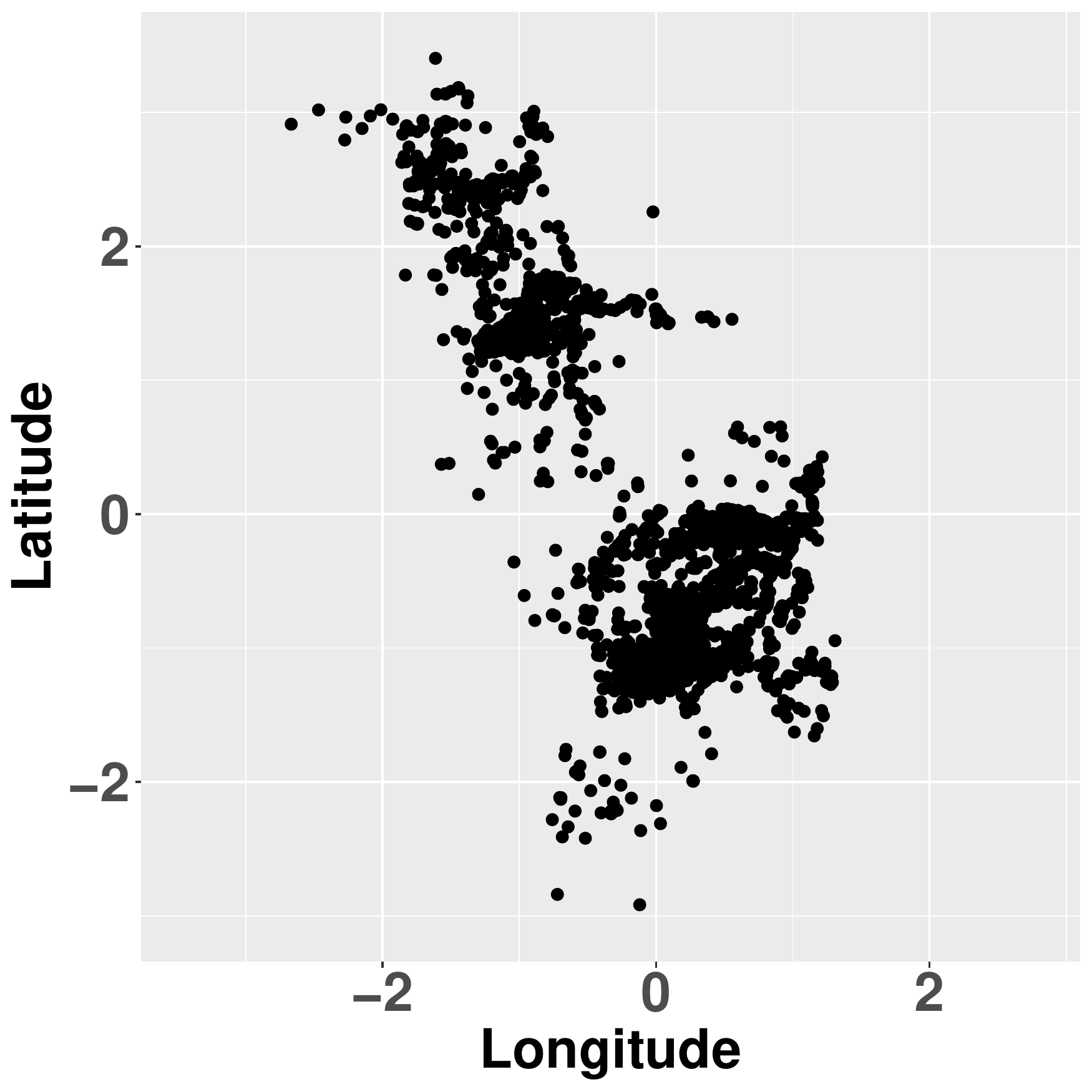}}}
  \caption{The observed spatial locations of the six dogs.}\label{newfig:sp_tot}
\end{figure}

We have the recorded coordinates of 6 dogs, taken  every 30 minutes at the Heatherlie  property in Australia, between   2012-11-10 15:30 and   2012-08-02 15:30. The data\footnote{The dataset is freely available from  the movebank repository \url{https://www.datarepository.movebank.org/handle/10255/move.395}} consist  of 4801 observations for each dog, with less than 1\% of missing points. To facilitate the specification of the prior  the   coordinates are centered  using the bivariate sample mean and scaled with a common standard deviation, computed using both  the X and Y coordinates, to  maintain the relative scale between the two coordinates; the recorded locations are shown in Figure \ref{newfig:sp_tot}.
The dogs    are called Woody, Sherlock, Alvin, Rosie, Bear, and Lucy. Rosie and Lucy are female,  while the other four are male, and
 Woody, Sherlock, Bear and Lucy form a  cohesive  group, which is responsible for  livestock protection, while Rosie, due to her advanced age, is solitary, and Alvin suffers from   social exclusion,  which restricted his movement \citep{bommel2}.\\
\indent    Maremma Sheepdogs  originate from Europe,  and  have been used   for centuries   to protect livestock from   potential predators \citep{Gehring}. They are trained to live with  the livestock from birth and, as a result, they develop   a strong bond with them  and an instinct to protect them. They can be fence-trained, but   are  generally allowed to move freely.
The use of livestock guardian dogs is  relatively new outside Europe, especially in Australia, and, due to their effectiveness,  interest in their use is increasing \citep{van2010guardian,bommel2}. Since   the  extension  of properties in Australia can be  as much as several thousand hectares, it is hard for the owner  to  supervise the dogs \citep{Bommel3} and  to have information about their behavior \citep{van2010guardian}.\\

\subsection{Comparison of the model and implementation details}
\begin{table}
\caption{Information criteria  for the  proposed model (M1), sHDP-HHMMs with  a common $G_0$ (M2), sHDP-HHMMs with animal-specific $G_{0,j}$ (M3). The model selected  by each index is indicated in bold. \label{newtab:cv} }
\centering
\fbox{
\begin{tabular}{c|ccc}
& M1& M2 & M3 \\
\hline
ICL &  \textbf{209593} & 	201880	& 192145\\
DIC5 &  \textbf{-457502}&-441264&	-407823\\
DIC7 & \textbf{-417544}&	-400786	& -376037\\
\hline \hline
\end{tabular}
}
\end{table}

We  compare the predictive performances between our proposal (M1) and two competitive models (M2 and M3), using  the  integrated completed likelihood (ICL) \citep{Biernacki:2000} and the deviance information criteria (DIC) $\text{DIC}_5$ and $\text{DIC}_7$ \citep{celeux2006deviance}.
In the first competitive model (M2), we assume that only the entire vector of STAP parameters can be shared between animals, which means that each time-series is an sHDP-HMM with a share based distribution.
 In terms of model formalization,
we  assume
\begin{align}
\{ \beta_{k}\}_{k \in \mathbb{N}}|\gamma &\sim \text{Gem}(\gamma),\\
\boldsymbol{\theta}_k|H_{\boldsymbol{\mu}} , H_{\boldsymbol{\eta}},  H_{\boldsymbol{\Sigma}},  H_{\tau} , H_{\rho} &\sim H_{\boldsymbol{\mu}}  \times H_{\boldsymbol{\eta}} \times  H_{\boldsymbol{\Sigma}} \times H_{\tau} \times H_{\rho},
\end{align}
 for the set of atoms and weights of $G_0$ (equation \eqref{neweq:g000}).
In the second competitor (M3), each animal follows the model of \cite{mastrantonio2020modeling},  which means they  are completely independent, and there is no sharing of parameters.
%. We then estimate the model of \cite{mastrantonio2020modeling} to each animal independently. IN M3,  i
Hence, we substitute $G_0$ with the animal-specific distribution
\begin{equation}
G_{0,j} = \sum_{k \in \mathbb{N}} \beta_{j,k} \delta_{\tilde{\boldsymbol{\theta}}_{j,k}},
\end{equation}
where
\begin{align}
\{ \beta_{j,k}\}_{k \in \mathbb{N}}|\gamma_j &\sim \text{Gem}(\gamma_j),\\
\tilde{\boldsymbol{\theta}}_{j,k}|H_{\boldsymbol{\mu}} , H_{\boldsymbol{\eta}},  H_{\boldsymbol{\Sigma}},  H_{\tau} , H_{\rho} &\sim H_{\boldsymbol{\mu}}  \times H_{\boldsymbol{\eta}} \times  H_{\boldsymbol{\Sigma}} \times H_{\tau} \times H_{\rho},\\
\tilde{\boldsymbol{\theta}}_{j,k}& = (\boldsymbol{\mu}_{j,k},\boldsymbol{\eta}_{j,k},\boldsymbol{\Sigma}_{j,k}, \tau_{j,k}, \rho_{j,k}),
\end{align}
and then
$$
\mathbf{s}_{j,t_{j,i+1}}|\mathbf{s}_{j,t_{j,i}},\mathbf{s}_{j,t_{j,i-1}},z_{j,t_{j,i}},\tilde{\boldsymbol{\theta}}_{j,z_{j,t_{j,i}}} \sim \text{STAP}(\tilde{\boldsymbol{\theta}}_{j, z_{j,t_{j,i}}} ).
$$
By changing the way  distribution $G_0$ is defined, we aim  to show that the main feature of our proposal, i.e., the sharing of sets and subsets of parameters, improves the model fitting and   leads to a better description of  the data.\\
\indent The models are implemented assuming  $H_{\boldsymbol{\mu}} \equiv N(\mathbf{0}, 20\mathbf{I})$, $H_{\boldsymbol{\eta}} \equiv N(\mathbf{0}, 20\mathbf{I})$, $H_{ {\tau}} \equiv U(0,1)$,  and $H_{\boldsymbol{\Sigma}} \equiv IW(3,\mathbf{I})$.
The distribution $H_{\rho} $ is assumed to be  a mixture of a $U(0,1)$ and two bulks of probability  on $0$ and $1$, with the 3 mixture weights  equal to 1/3. This   allows
$\rho_{k}$ (in M1 and M2) and $\rho_{j,k}$ (in M3)
to be, a posteriori, equal to 0 or 1 with a  greater probability than 0,  which allows us to detect  a pure CRW or BRW behavior.
We assume  $\alpha+\nu,\gamma_{\boldsymbol{\mu}},\gamma_{\boldsymbol{\eta}},\gamma_{\boldsymbol{\Sigma}},\gamma_{\rho},\gamma_{\tau},\gamma, \gamma_1, \dots , \gamma_m \sim G(0.01,0.01)$ and $\nu/(\alpha+\nu) \sim B(1,1)$,
which allows  us to easily  sample  from their full conditionals, see \cite{fox2011} and  Section \ref{I-sec:imp} of the  Appendix.  All the distributions are chosen to be weakly informative.
The  domain $\mathcal{D}$ is a square $[-5,5]\times [-5,5]$ and the parameter $\epsilon$ of the Geometric distribution is equal to 0.00001, which means that the distribution over the initial state, $z_{j,t_{j,0}}$, is approximately uniform over the positive integers.
Posterior estimates are obtained with 15000 iterations, burnin  7500, thin 3, and  thus 2500 samples are available for posterior inference.
Convergence has been checked by means of a visual inspection of the posterior chains and using the $\hat{R}$ statistics \citep{Gelman2013}.
Details on the MCMC algorithm,  implemented in  Julia 1.3 \citep{bezanson2017julia}, can be found in the Appendix, Section \ref{I-sec:imp}, and
the codes used to replicate the results, tables, and figures are available at \url{https://github.com/GianlucaMastrantonio/multiple_animals_movement_model}.\\
\indent In Table \ref{newtab:cv},   we can see that the three indices indicate that our model is the one with the best fit\footnote{It should be noted that a higher ICL and a lower DIC indicate a better fit.}, model M2 is the second, while M3 is always the last. Therefore, the joint modeling of the six dogs improves the  performances of the model (since M2 is always preferable to M3), but  the sharing of a subset of parameters also leads to a better description of the data (since M1 is better than M2).  We provide a  description of the results obtained with M2 and M3 in the Appendix, Section \ref{I-sec:commd}.

\subsection{Description and interpretation of the output}

\begin{figure}[t!]
  \centering
  {\subfloat[Woody - $\text{B}_{11}$]{\includegraphics[trim = 0 20 0 20,scale=0.27]{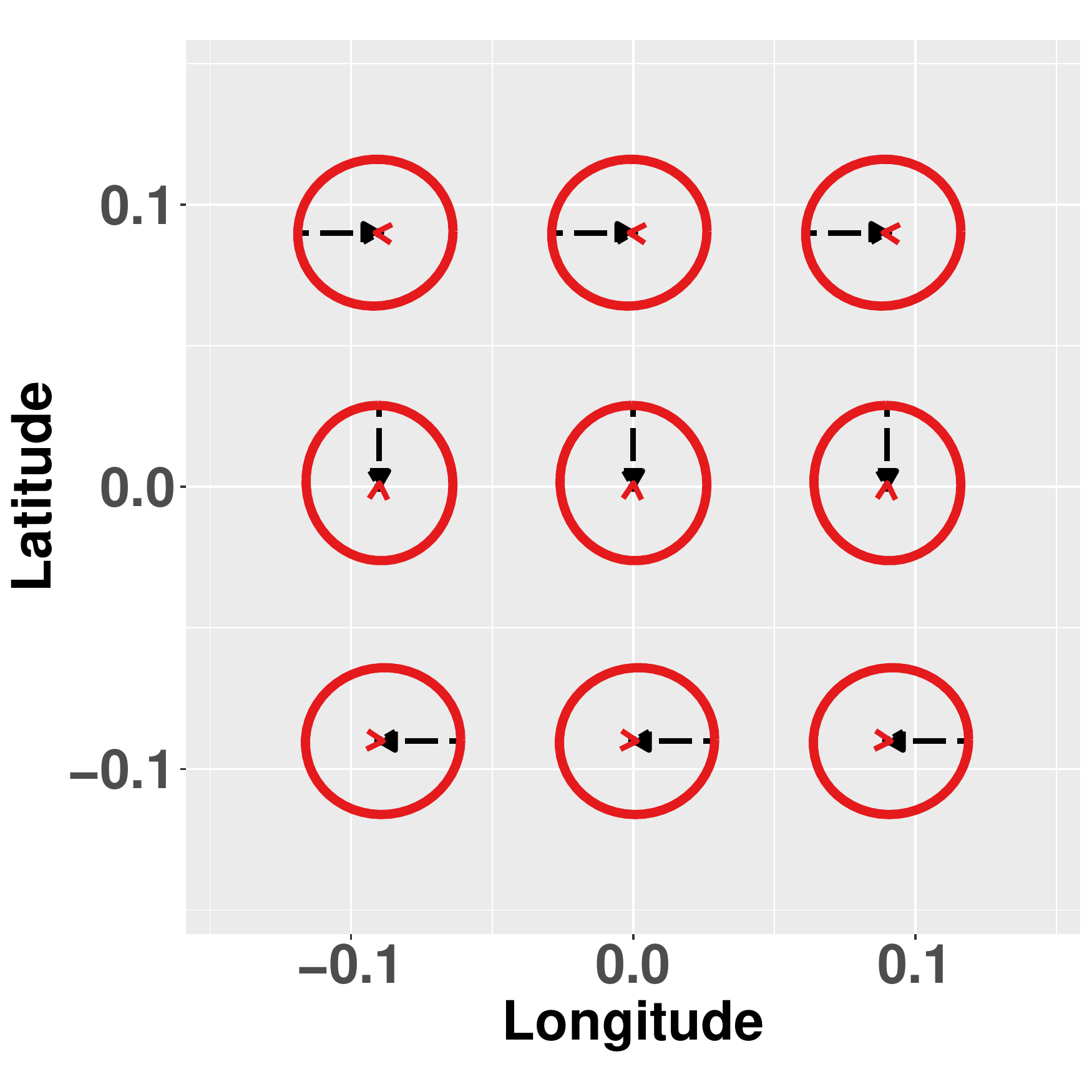}}}
  {\subfloat[Woody - $\text{B}_{12}$]{\includegraphics[trim = 0 20 0 20,scale=0.27]{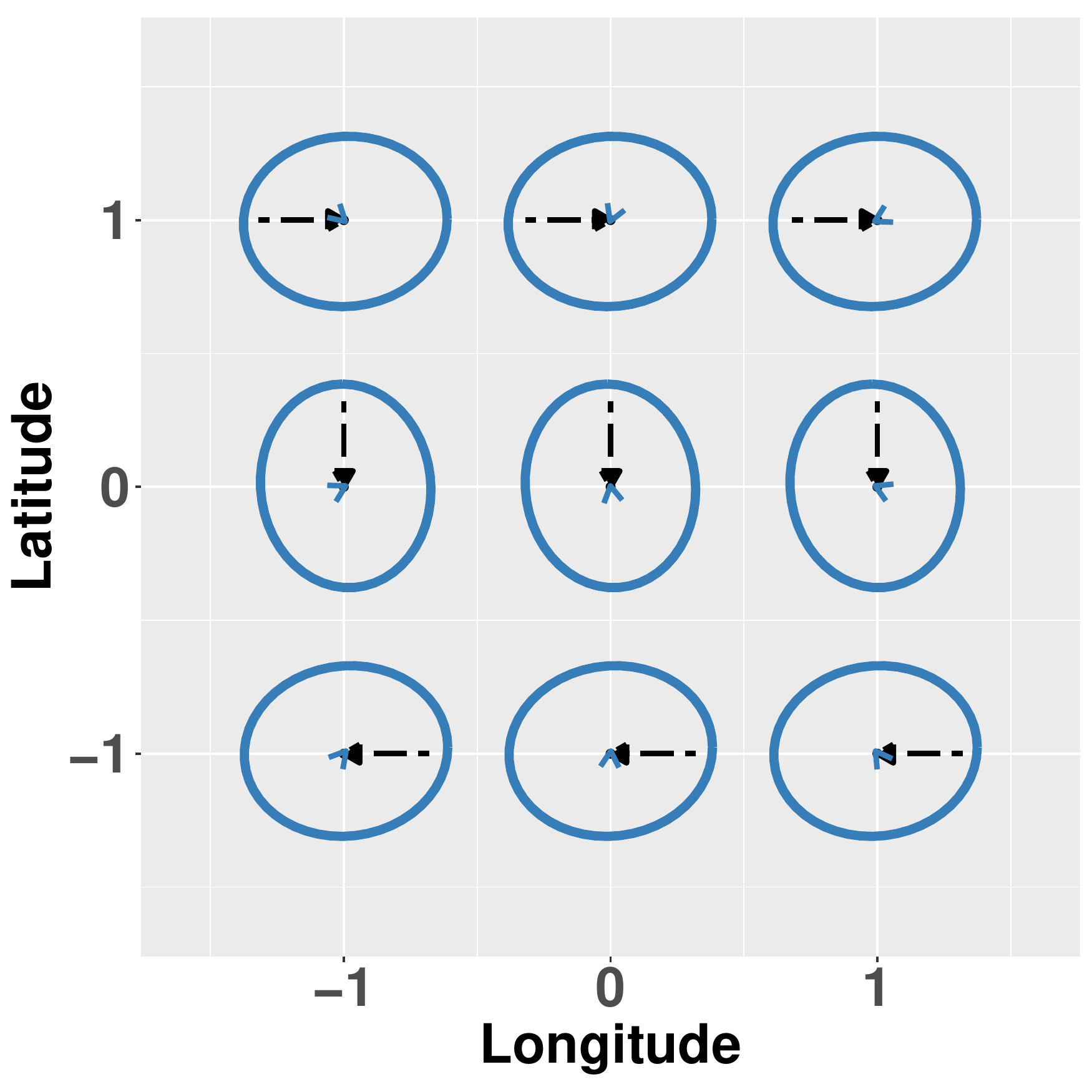}}}
  {\subfloat[Woody - $\text{B}_{13}$]{\includegraphics[trim = 0 20 0 20,scale=0.27]{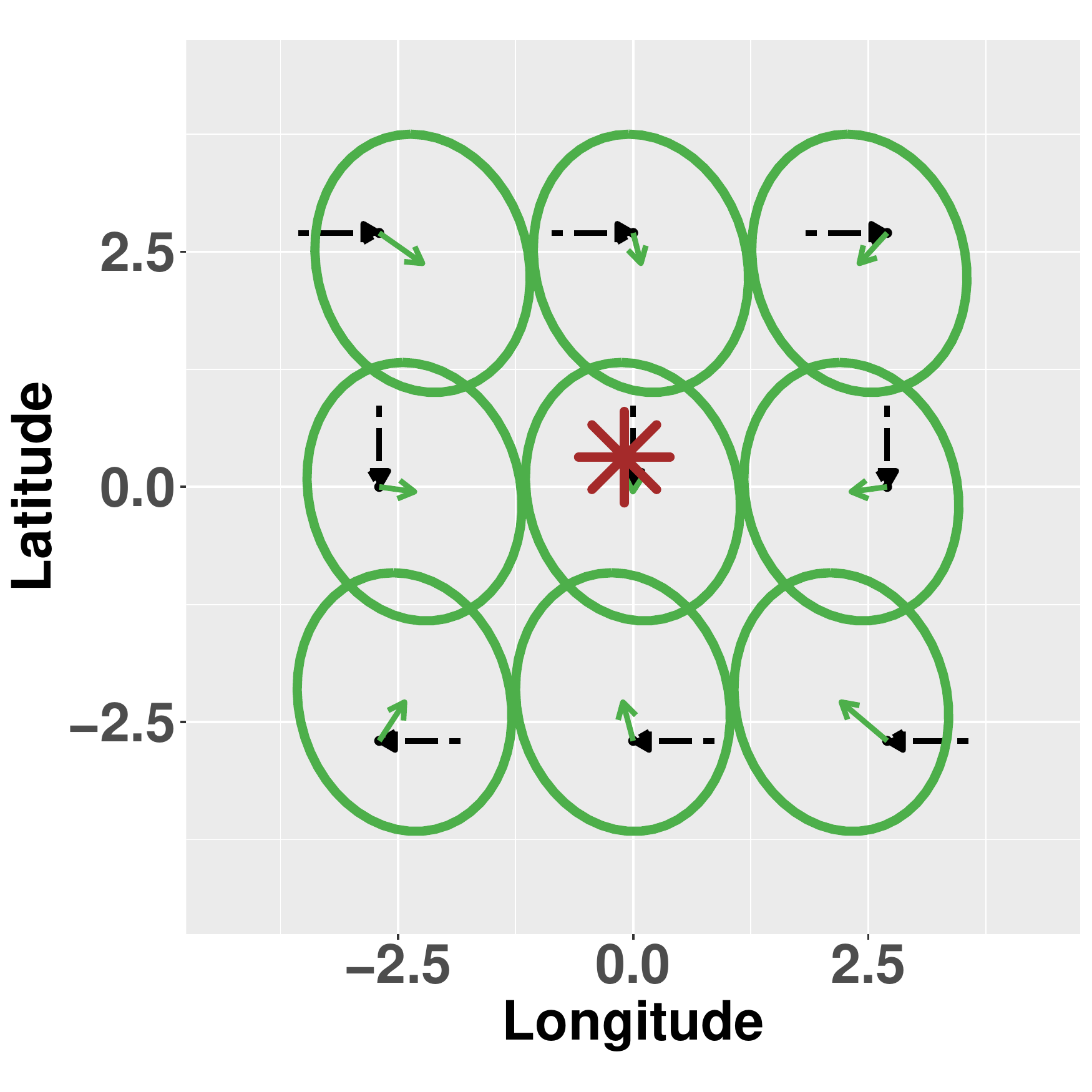}}}\\
  {\subfloat[Sherlock - $\text{B}_{21}$]{\includegraphics[trim = 0 20 0 20,scale=0.27]{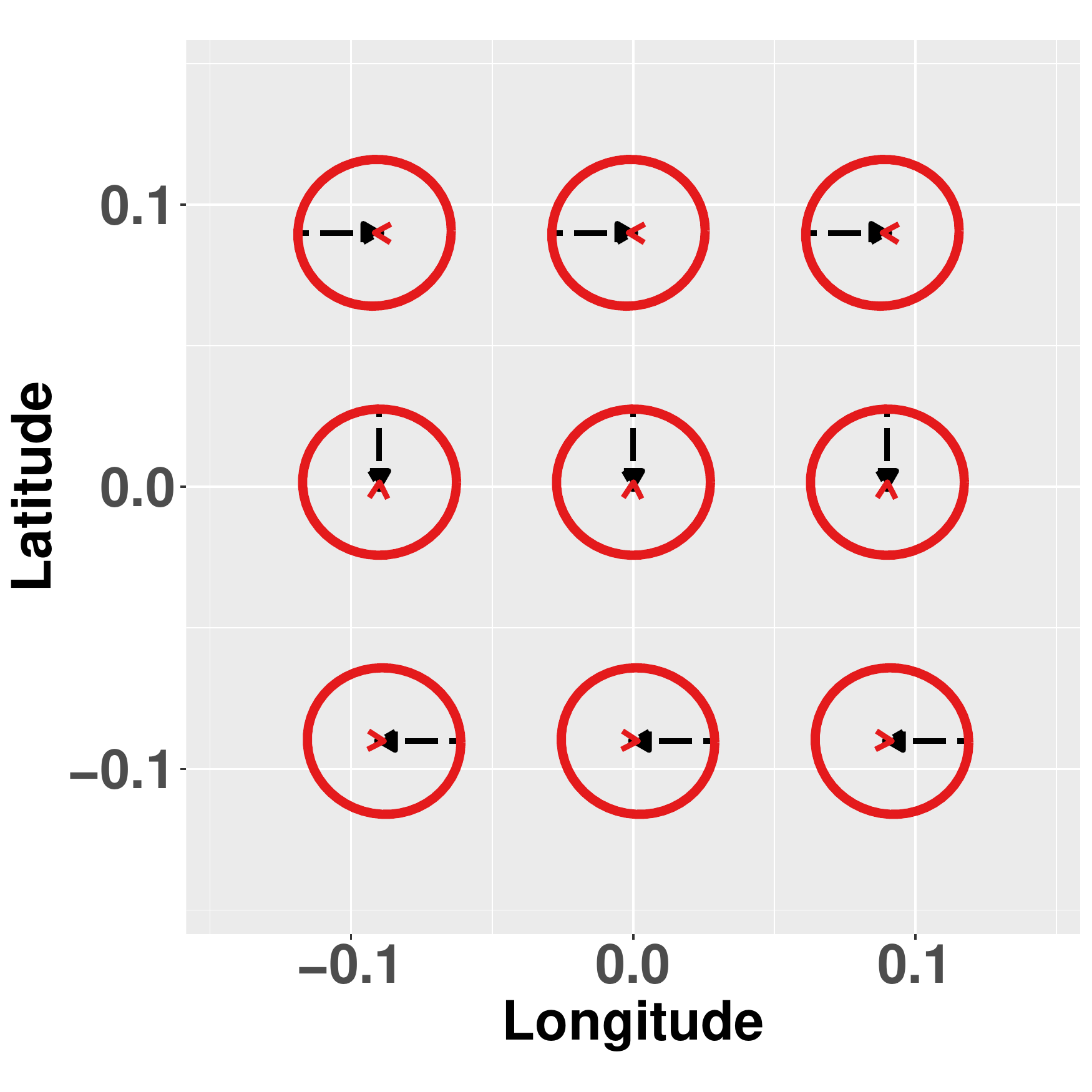}}}
  {\subfloat[Sherlock - $\text{B}_{22}$]{\includegraphics[trim = 0 20 0 20,scale=0.27]{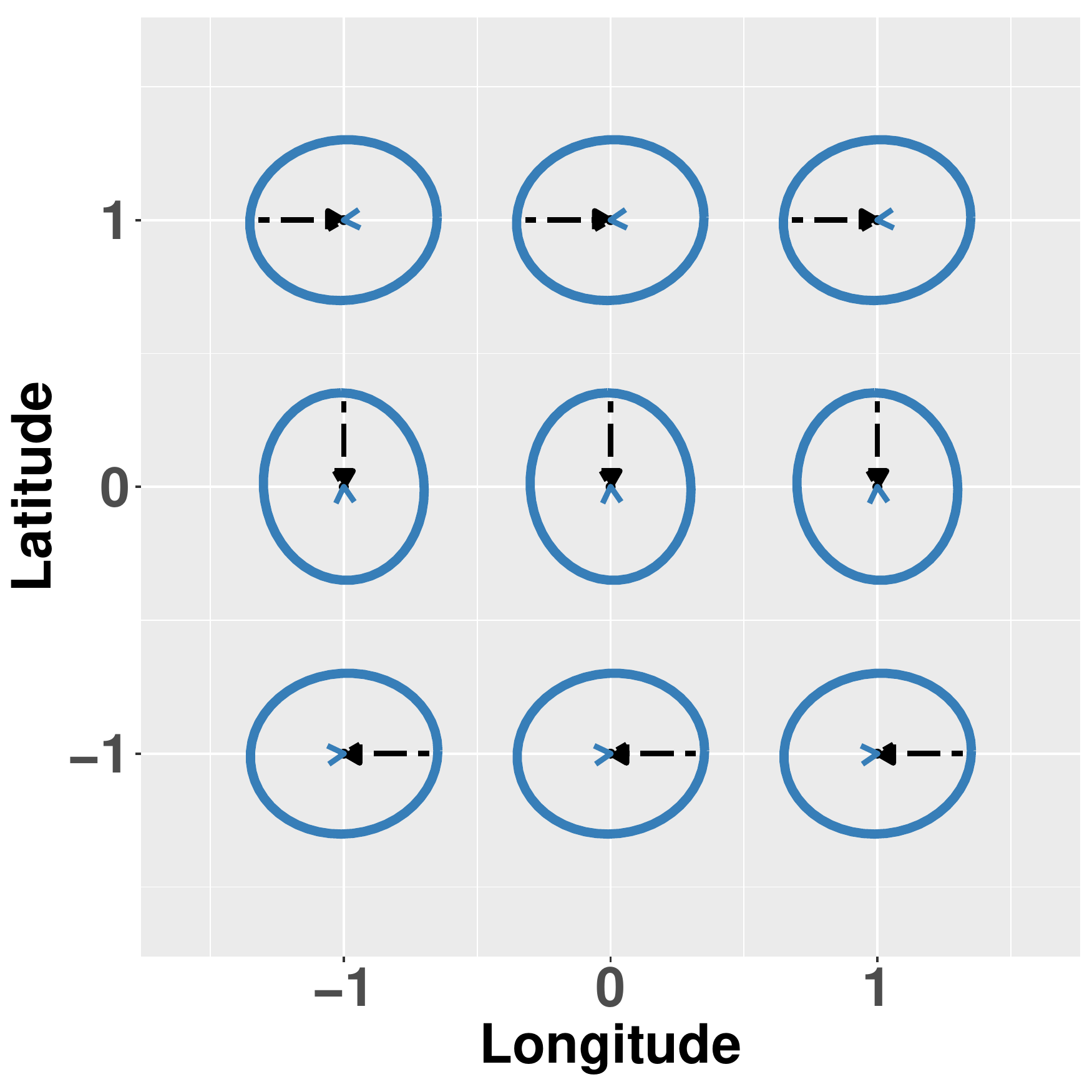}}}
  {\subfloat[Sherlock - $\text{B}_{23}$]{\includegraphics[trim = 0 20 0 20,scale=0.27]{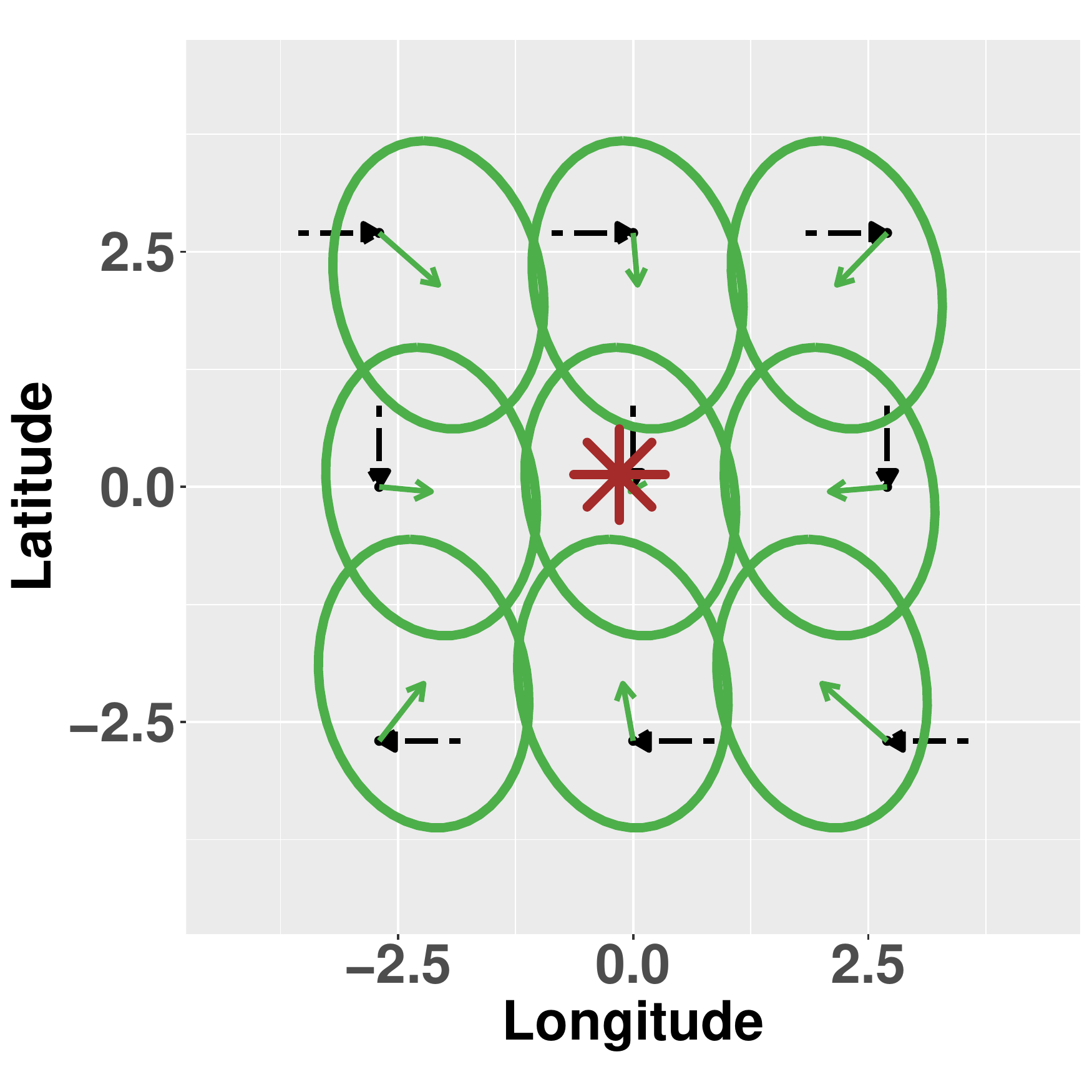}}}\\
	{\subfloat[Alvin - $\text{B}_{31}$]{\includegraphics[trim = 0 20 0 20,scale=0.27]{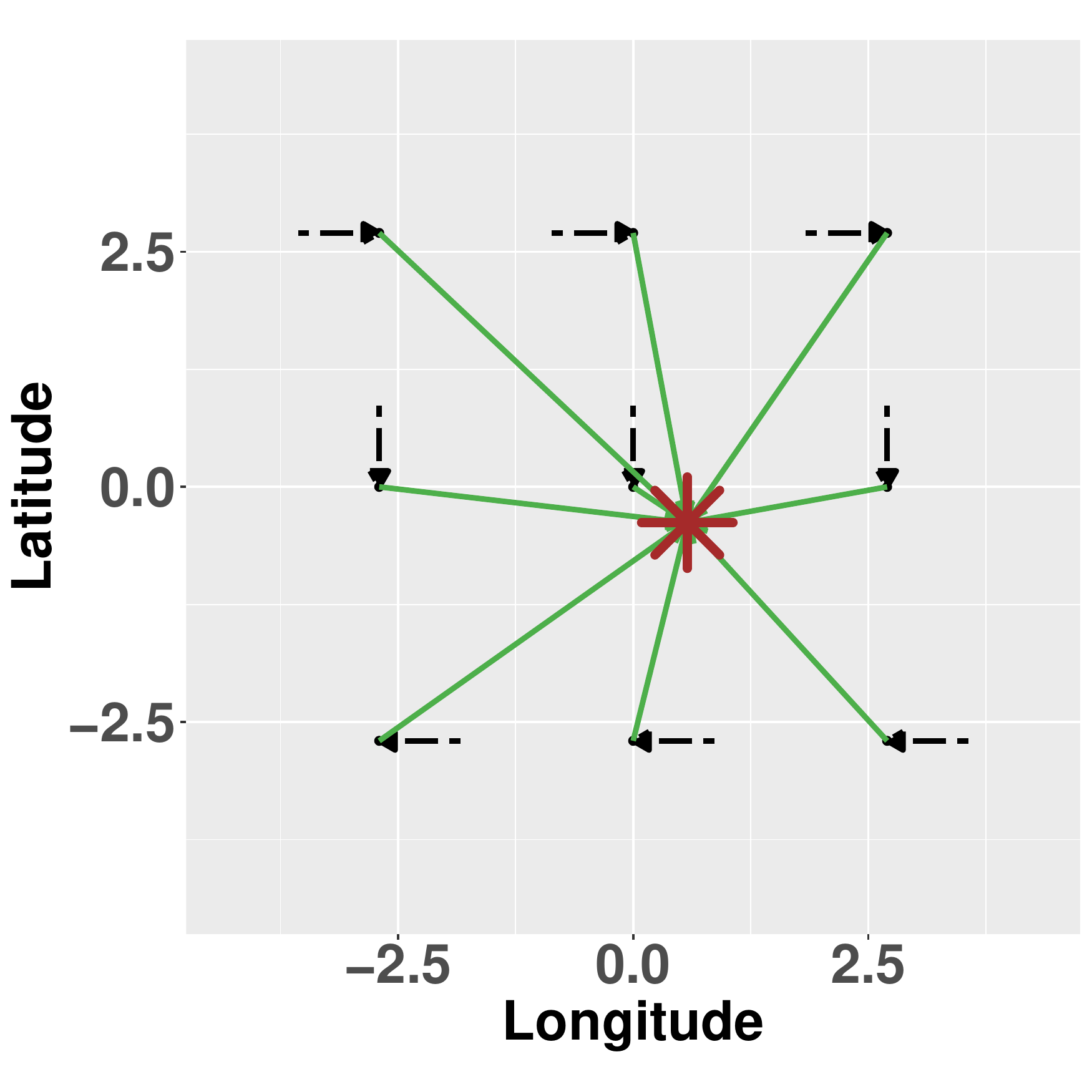}}}
	{\subfloat[Alvin - $\text{B}_{32}$]{\includegraphics[trim = 0 20 0 20,scale=0.27]{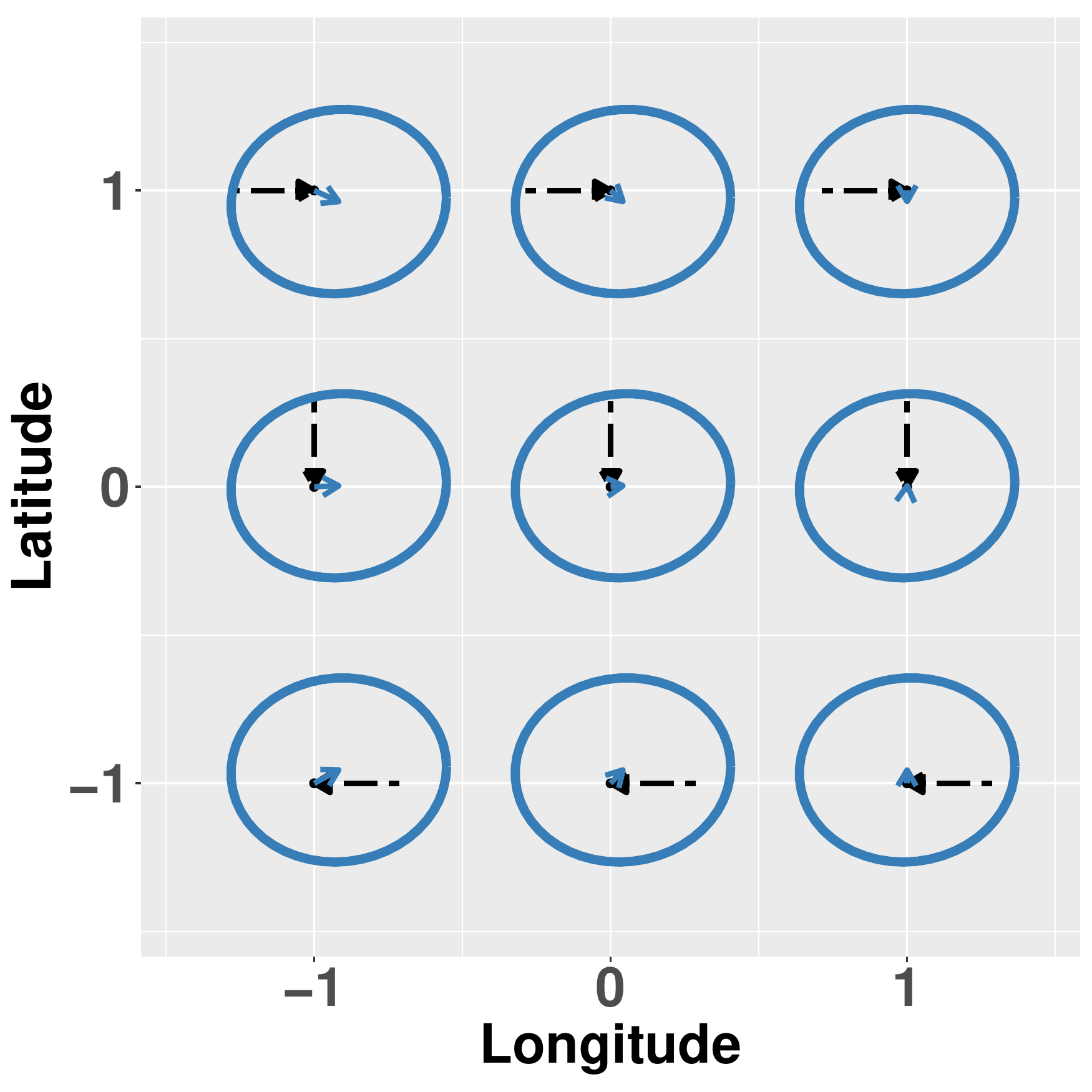}}}
	{\subfloat[Alvin - $\text{B}_{33}$]{\includegraphics[trim = 0 20 0 20,scale=0.27]{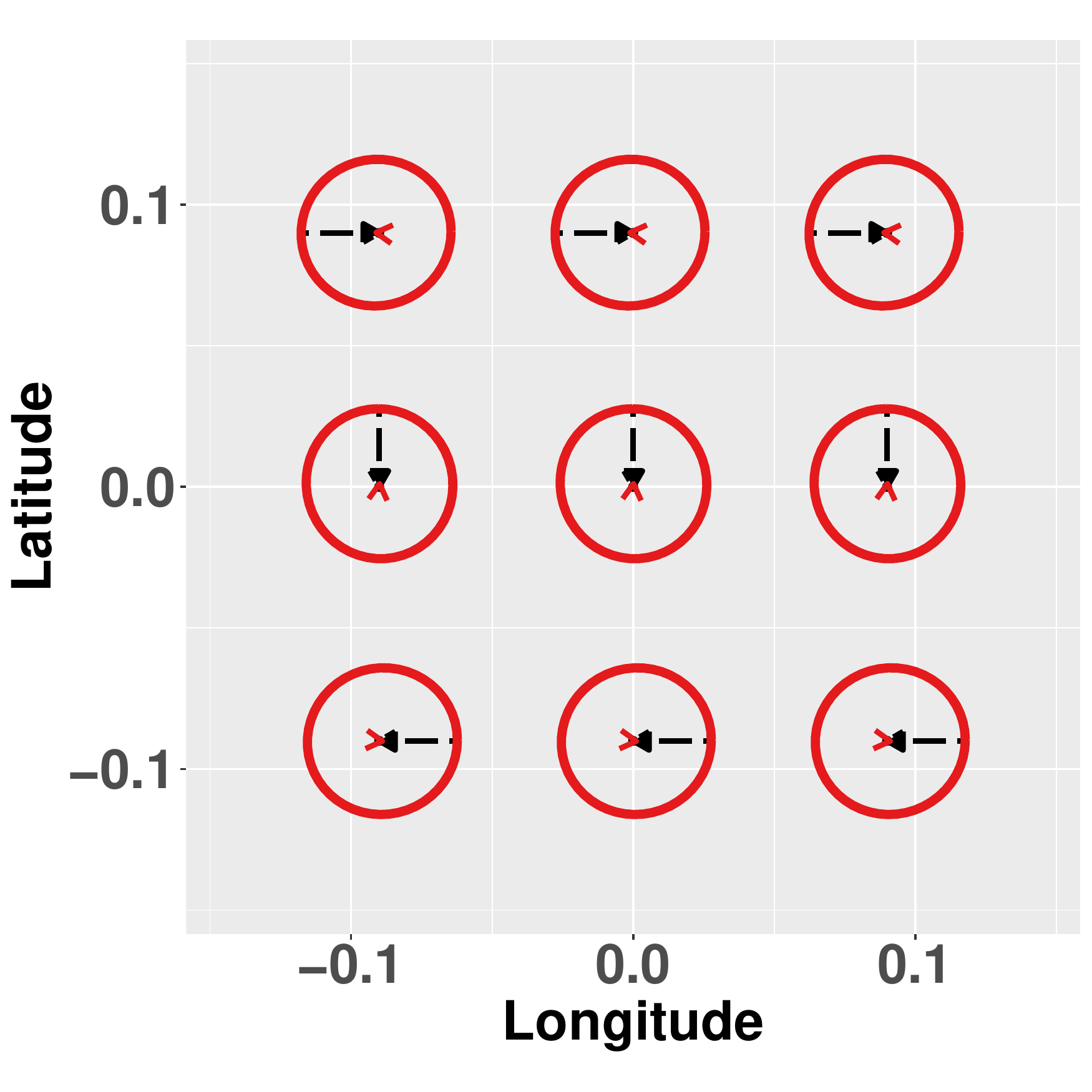}}}
	%{\subfloat[Alvin - $\text{B}_{34}$]{\includegraphics[trim = 0 20 0 20,scale=0.22]{3_4Mov}}}
  \caption{Graphical representation of the conditional distribution of $\mathbf{s}_{j,t_{j,i+1}}$ for the first three dogs,  for different possible values of $\mathbf{s}_{j,t_{j,i}}$ and  previous directions.
	The images has been obtained by using the posterior values that maximize the data likelihood of each animal, given the representative clusterization $\hat{z}_{j, t_{j,i}}$.
	The dashed arrow represents the movement between  $\mathbf{s}_{j,t_{j,i-1}}$ and $\mathbf{s}_{j,t_{j,i}}$. The solid arrow is $\protect\vec{\mathbf{F}}_{j,t_{j,i}}$,  while the ellipse is an area containing  95\% of the probability mass of  the  conditional  distribution of $\mathbf{s}_{j,t_{j,i+1}}$. The asterisk represents the  attractor,  and it is only shown for behaviors that have posterior values of
	${\rho}_{j,k}<0.9$ and ${\tau}_{j,k}>0.1$. %Figures with the same color are on the same spatial scales.
	}\label{newfig:a2_1}
\end{figure}

\begin{figure}[t!]
  \centering
  {\subfloat[Rosie - $\text{B}_{41}$]{\includegraphics[trim = 0 20 0 20,scale=0.27]{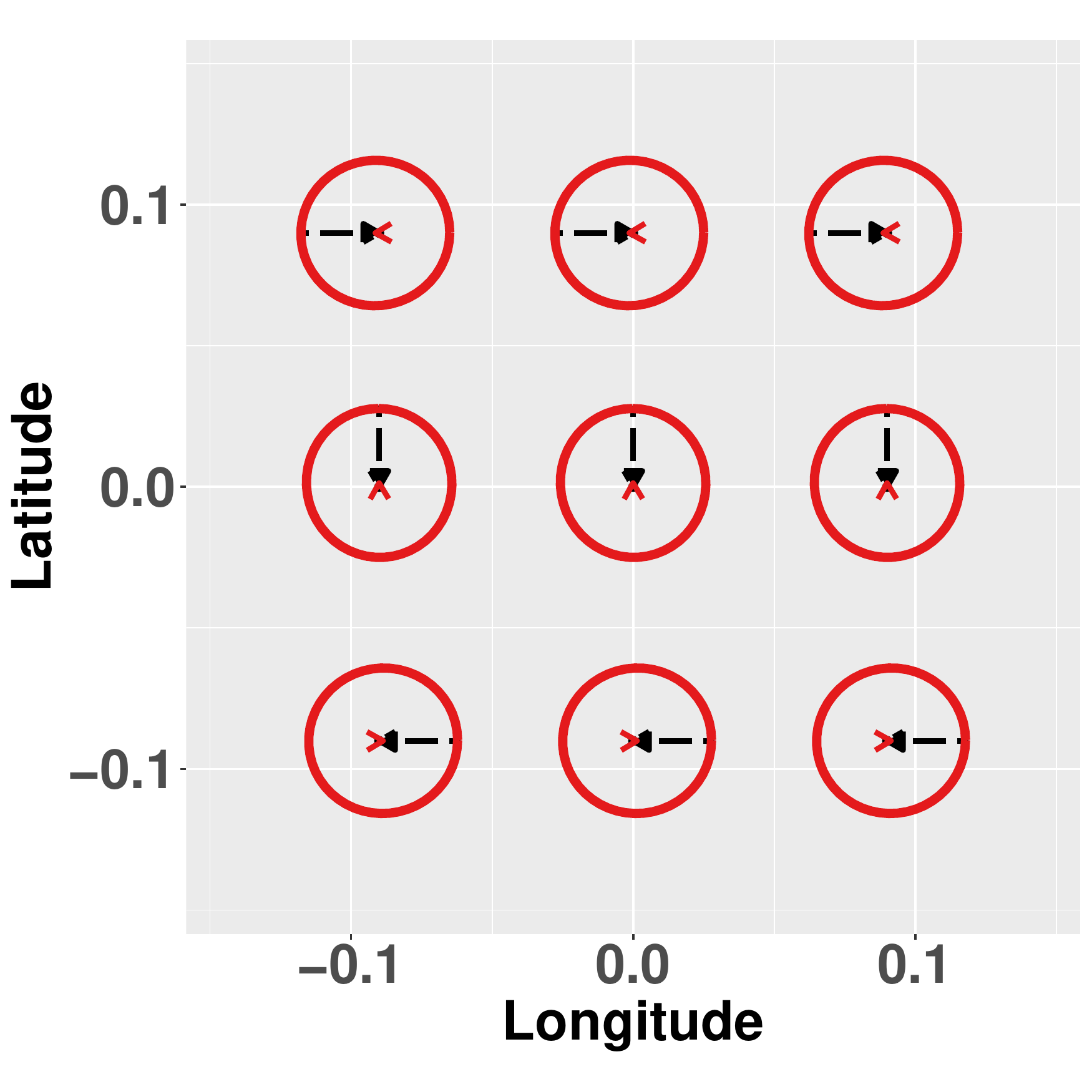}}}
  {\subfloat[Rosie - $\text{B}_{42}$]{\includegraphics[trim = 0 20 0 20,scale=0.27]{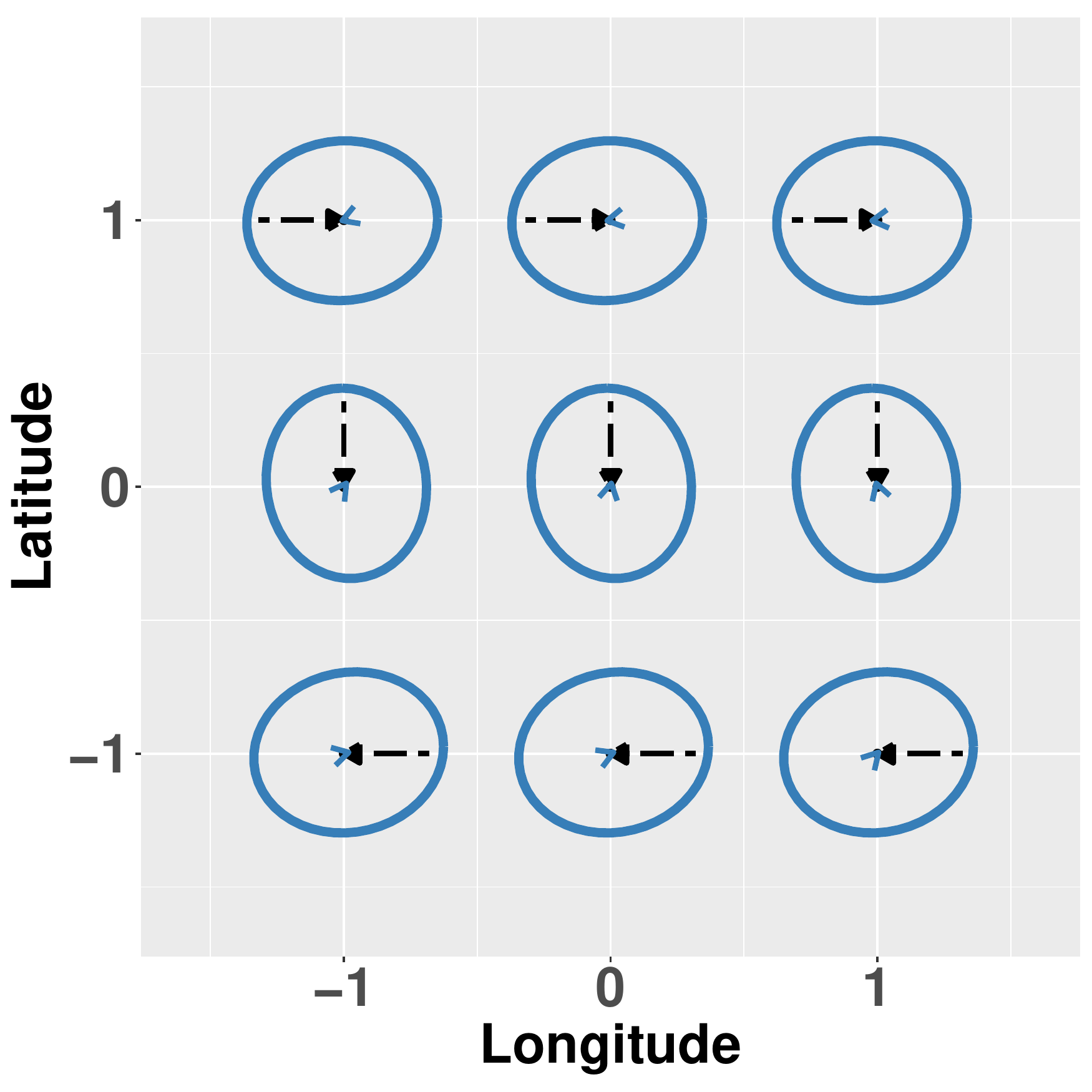}}}\\
  %{\subfloat[Rosie - $\text{B}_{43}$]{\includegraphics[trim = 0 20 0 20,scale=0.22]{4_3Mov}}}
	{\subfloat[Bear - $\text{B}_{51}$]{\includegraphics[trim = 0 20 0 20,scale=0.27]{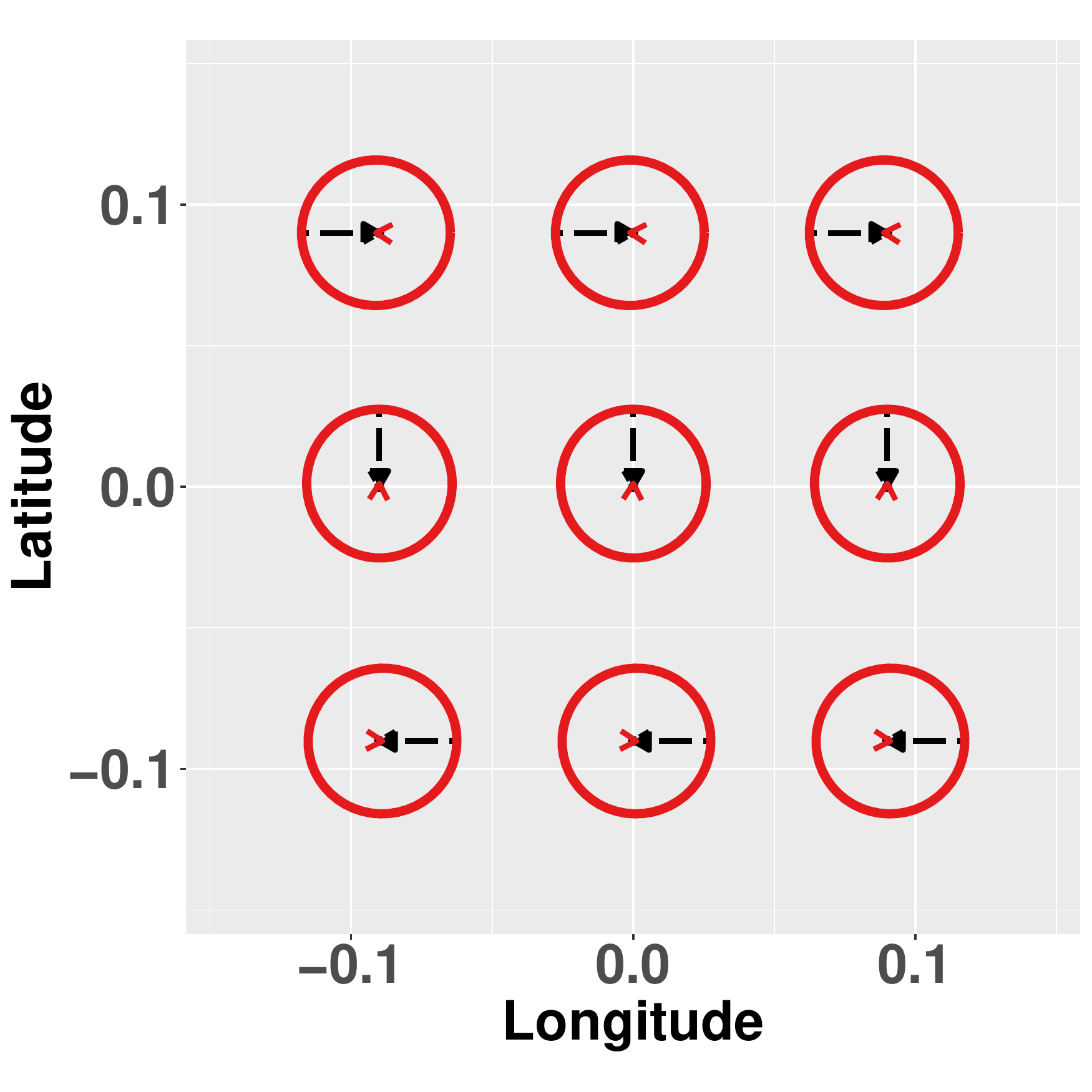}}}
	{\subfloat[Bear - $\text{B}_{52}$]{\includegraphics[trim = 0 20 0 20,scale=0.27]{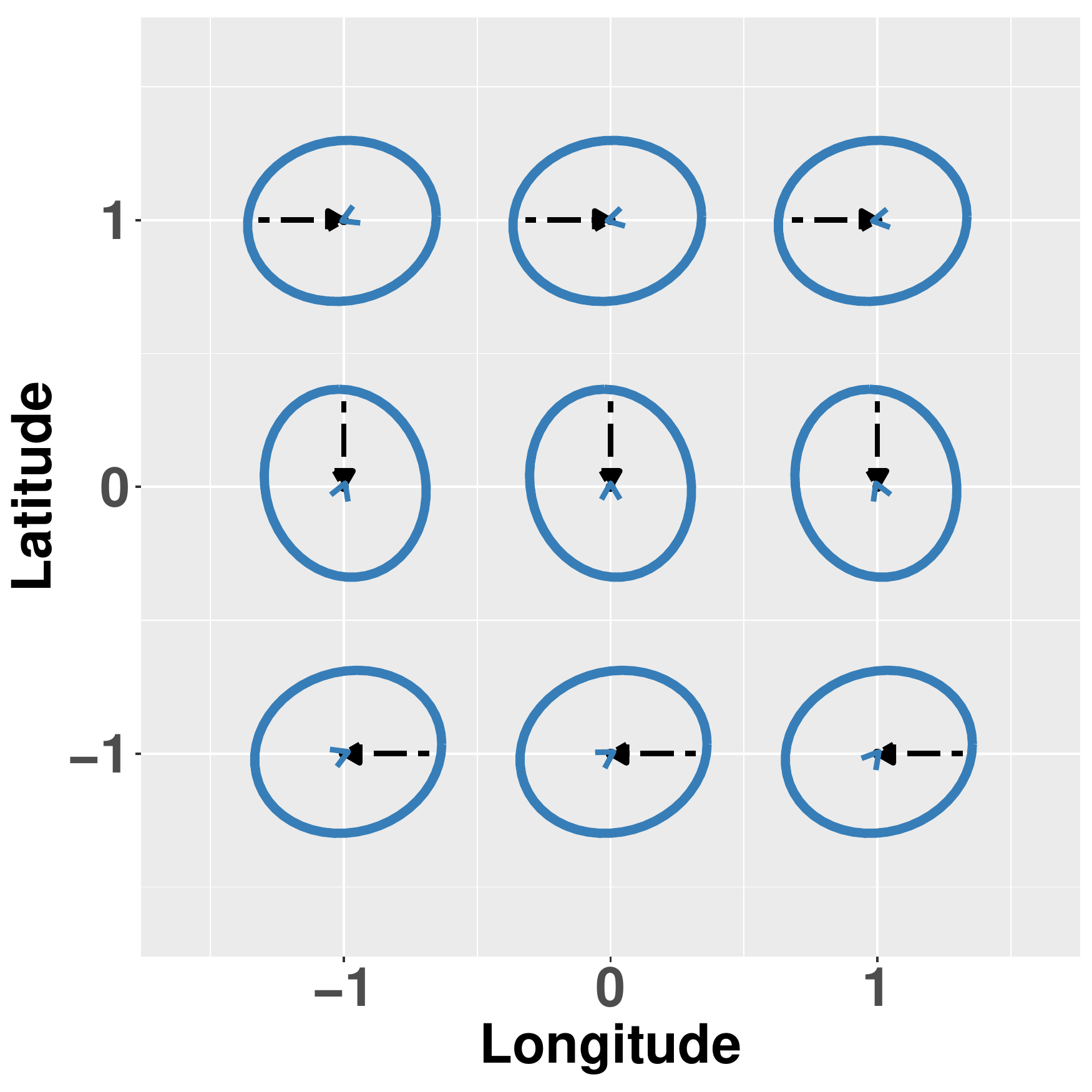}}}
	{\subfloat[Bear - $\text{B}_{53}$]{\includegraphics[trim = 0 20 0 20,scale=0.27]{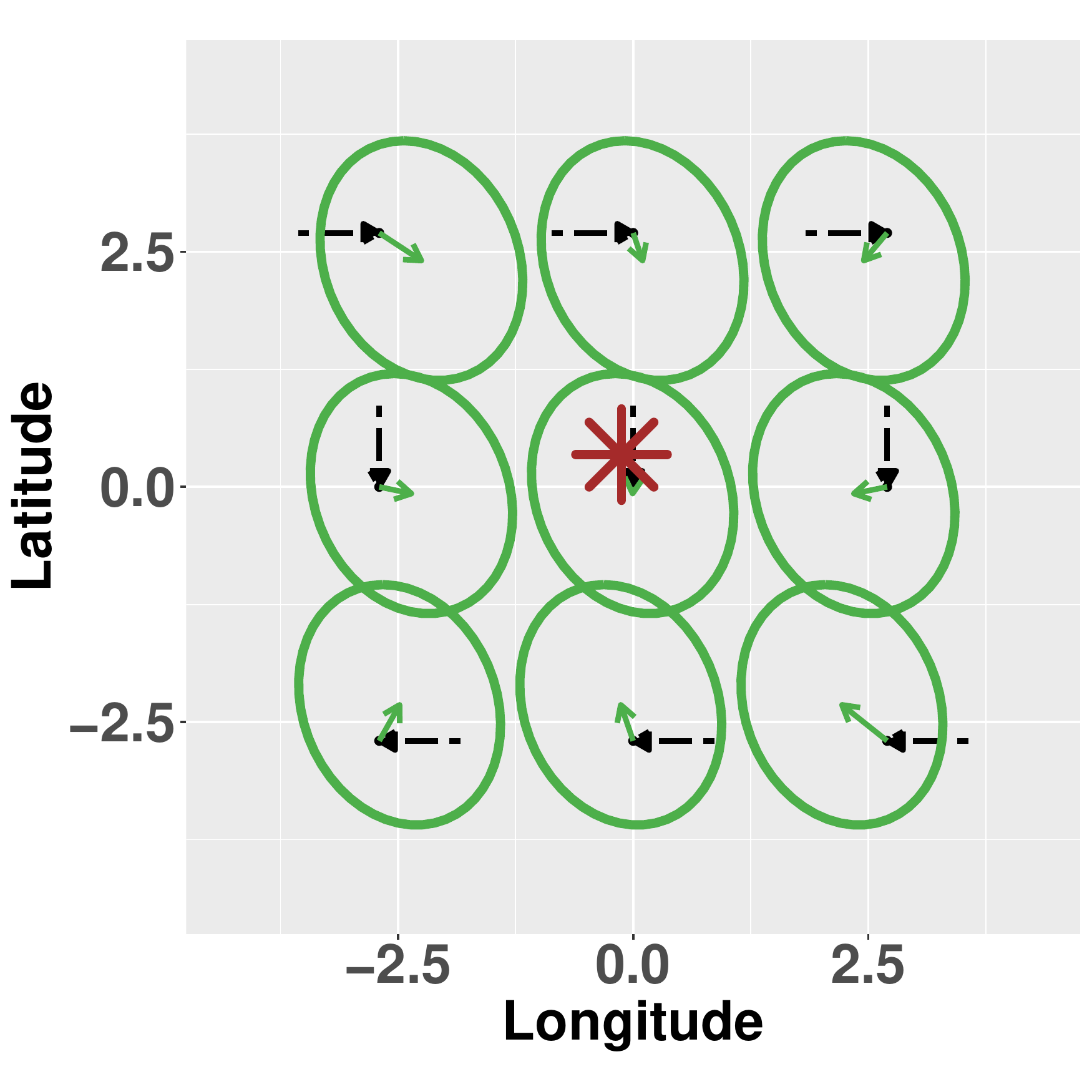}}}\\
	{\subfloat[Lucy - $\text{B}_{61}$]{\includegraphics[trim = 0 20 0 20,scale=0.27]{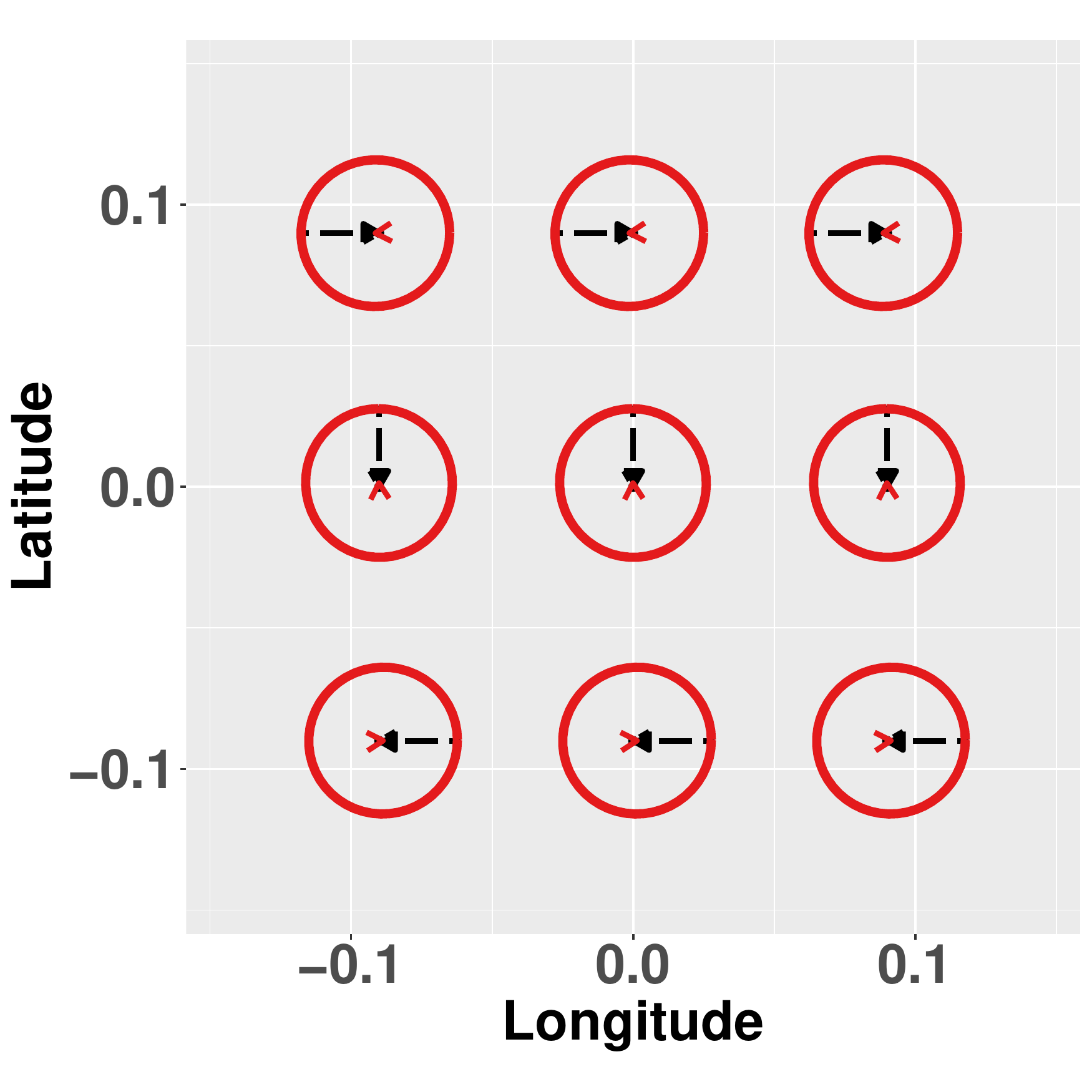}}}
	{\subfloat[Lucy - $\text{B}_{62}$]{\includegraphics[trim = 0 20 0 20,scale=0.27]{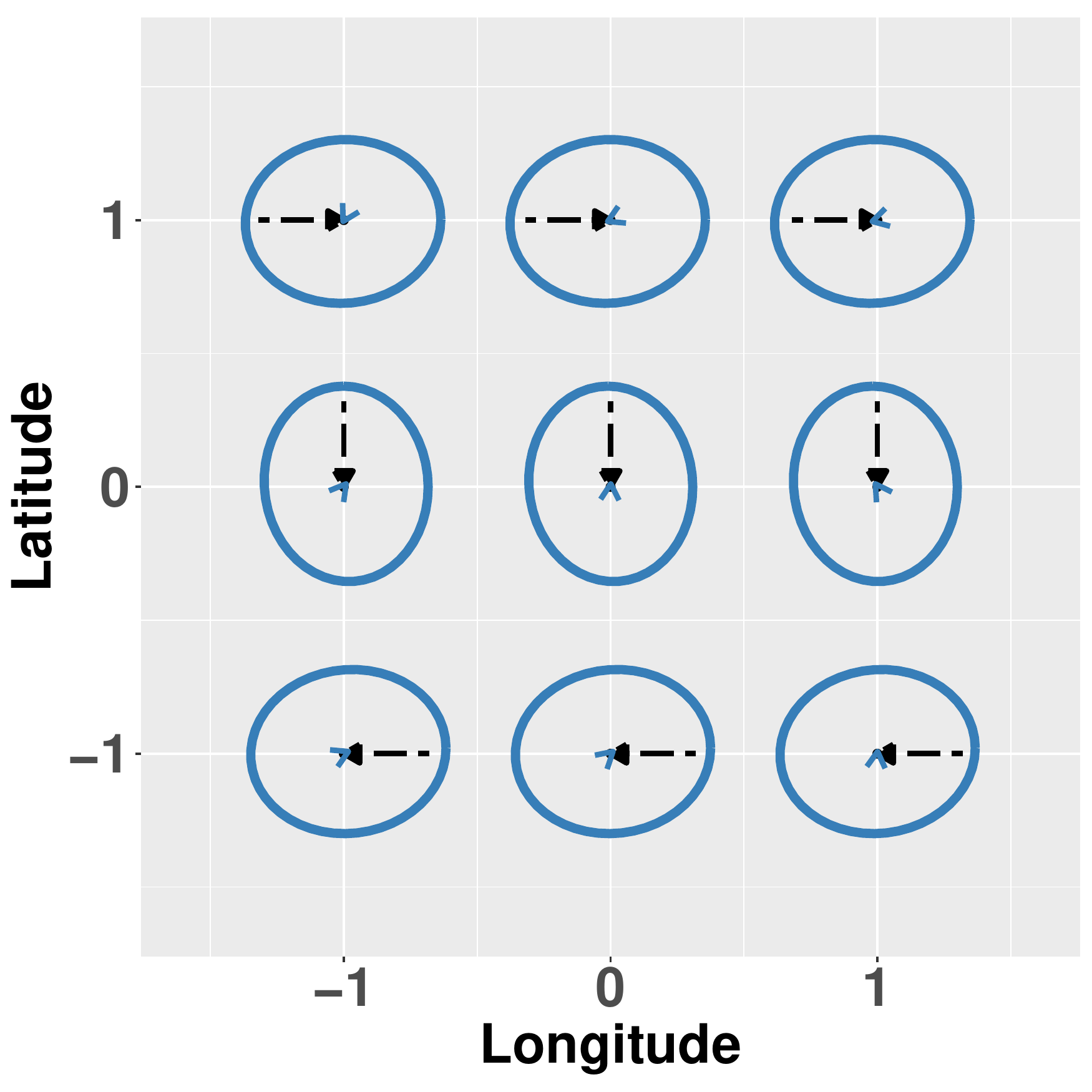}}}
	{\subfloat[Lucy - $\text{B}_{63}$]{\includegraphics[trim = 0 20 0 20,scale=0.27]{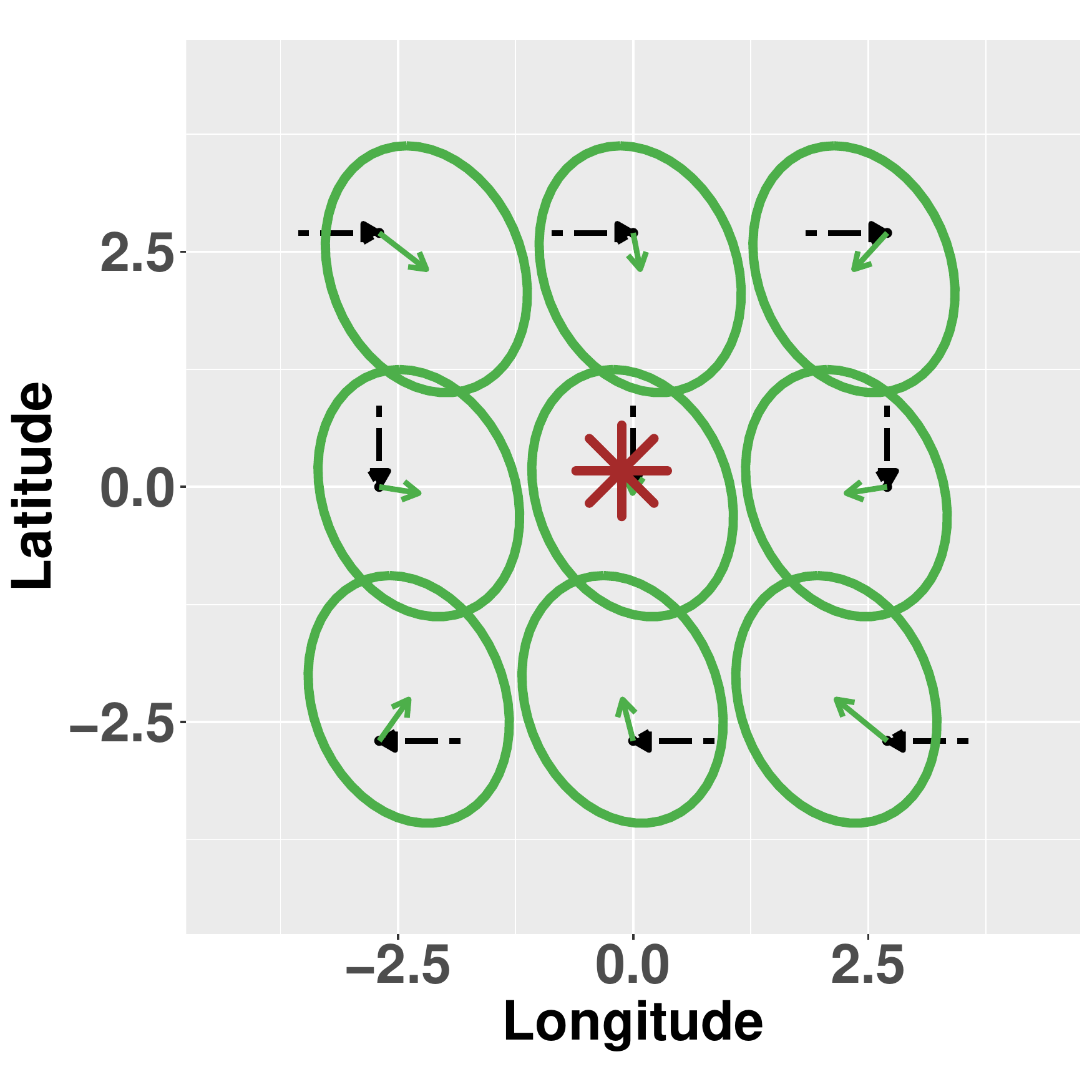}}}
  \caption{Graphical representation of the conditional distribution of $\mathbf{s}_{j,t_{j,i+1}}$ for the last three dogs,  for different possible values of $\mathbf{s}_{j,t_{j,i}}$ and  previous directions.
	The images has been obtained by using the posterior values that maximize the data likelihood of each animal, given the representative clusterization $\hat{z}_{j, t_{j,i}}$.
	The dashed arrow represents the movement between  $\mathbf{s}_{j,t_{j,i-1}}$ and $\mathbf{s}_{j,t_{j,i}}$. The solid arrow is $\protect\vec{\mathbf{F}}_{j,t_{j,i}}$,  while the ellipse is an area containing  95\% of the probability mass of  the  conditional  distribution of $\mathbf{s}_{j,t_{j,i+1}}$. The asterisk represents the  attractor,  and it is only shown for behaviors that have posterior values of
	${\rho}_{j,k}<0.9$ and ${\tau}_{j,k}>0.1$.
	 }\label{newfig:a2_new_2}
\end{figure}

\begin{figure}[t!]
  \centering
  {\subfloat[$\delta (\boldsymbol{\mu}_{j,k},\boldsymbol{\mu}_{j',k'})$]{\includegraphics[trim = 0 60 0 25,scale=0.38]{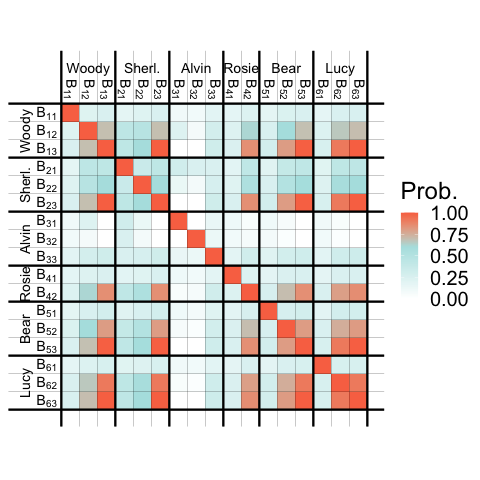}}}
  {\subfloat[$\delta (\boldsymbol{\eta}_{j,k},\boldsymbol{\eta}_{j',k'})$]{\includegraphics[trim = 0 60 0 25,scale=0.38]{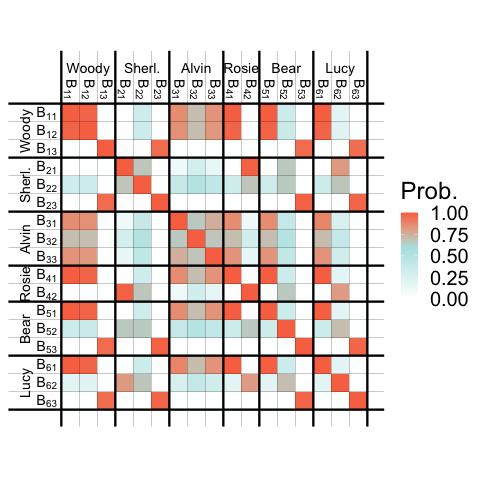}}}\\
  {\subfloat[$\delta (\tau_{j,k},\tau_{j',k'})$ ]{\includegraphics[trim = 0 60 0 25,scale=0.38]{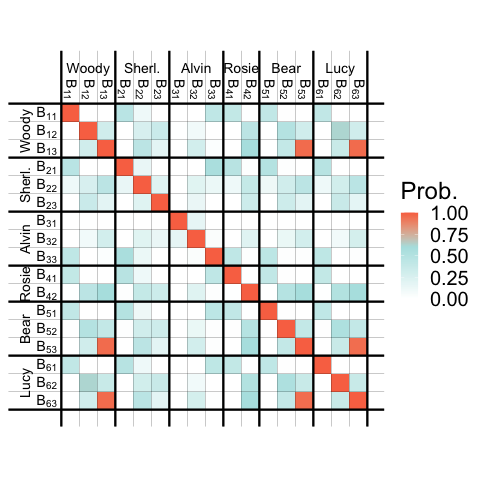}}}
  {\subfloat[$\delta (\boldsymbol{\Sigma}_{j,k},\boldsymbol{\Sigma}_{j',k'})$ ]{\includegraphics[trim = 0 60 0  25,scale=0.38]{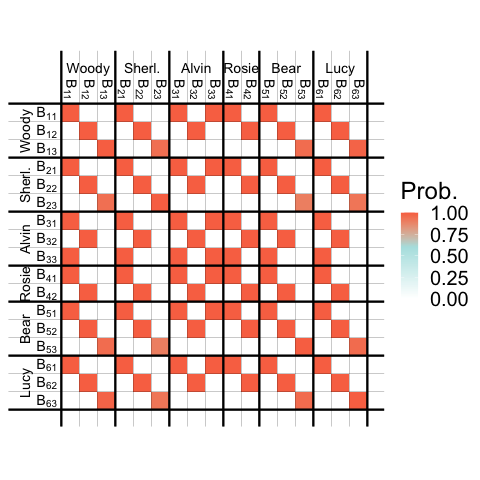}}}\\
  {\subfloat[$\delta (\rho_{j,k},\rho_{j',k'})$ ]{\includegraphics[trim = 0 60 0 25,scale=0.38]{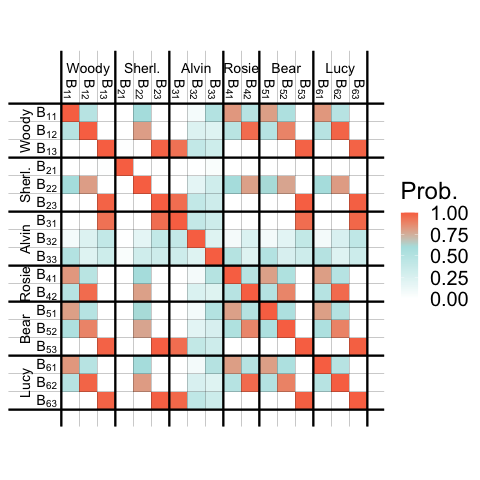}}}
  %{\subfloat[Parameter $z$]{\includegraphics[trim = 0 60 0 20,scale=0.38]{Sim_Zeta}}}
  \caption{Graphical representation of  the posterior mean of $\delta (\cdot,\cdot)$, which represents  the  probability of the  parameters in its argument having the same value for two behaviors.   }\label{newfig:a2_44}
\end{figure}

%
% From the MCMC output we have that for dog 1, 4, 5, and 6 the posterior distribution of $K_j$ has mode at 3 (behaviors) with probability mass $> 0.98$, while the mode of dog 2 is 4 with   probability mass $>0.99$, while dog 3 has mode 5 with probability mass 0.718 and the remaining mass is on the value $K_3=6$.

Using the algorithm proposed by \cite{wade2018}, we  find a  representative behavior $\hat{z}_{j,t_{j,i}}$   associated with each  animal and time, that we indicate as   \textit{MAP behavior}. We indicate the $k$-th behavior of the $j$-th dog based on $\hat{z}_{j,t_{j,i}}$ as  $\text{B}_{jk}$, and let  $n_{j,k}$ be  the number of times we have
$\hat{z}_{j,t_{j,i}}=k$, without any loss of generality, we assume
$n_{j,1}>n_{j,2}> \dots$, i.e., the behaviors are ordered with respect to the number of times they are observed. It should be noted that $\text{B}_{jk}$ is not the same as $\text{B}_{j'k}$, if $j \neq j'$, and therefore, to avoid confusion, we indicate  the vector of  STAP parameters for $\text{B}_{jk}$ with
$\boldsymbol{\theta}_{j,k}= (\boldsymbol{\mu}_{j,k},\boldsymbol{\eta}_{j,k},\boldsymbol{\Sigma}_{j,k},{\tau}_{j,k}, {\rho}_{j,k})$.  For easiness of interpretation, we  only discuss  behaviors that have been observed, on average,  at least once a day ($n_{j,k}>100$),  thus obtaining then 3 behaviors for each dogs, with the exception of  Rosie (dog 4) that has  2 behaviors; see  Tables   \ref{I-tab:anim1}-\ref{I-tab:anim6} in the Appendix, where   the posterior means
( $\hat{}$ ) and credible intervals (CIs)  for the STAP parameters,  $n_{j,k}$, and the transition probabilities for all the dogs and behaviors are shown. Using similar pictures to the ones  used in Figure \ref{newfig:EsModel},  we show a graphical  description of the  behaviors found by the model in Figures \ref{newfig:a2_1}  and \ref{newfig:a2_new_2};   the behaviors are represented on different spatial scales.
From the model output, we  computed  the posterior mean of the variable $\delta(\chi_{j,k},\chi_{j',k'})$, which  is the posterior probability that a STAP parameter assumes the same value in $\text{B}_{jk}$ and $\text{B}_{j'k'}$. These probabilities are depicted in  Figure \ref{newfig:a2_44} for all possible combinations of behaviors and animals. To take into account that identifiability  for $\boldsymbol{\mu}_{j,k}$ and
$\boldsymbol{\eta}_{j,k}$ is only granted   if $\rho_{j,k}<1$ and $\rho_{j,k}>0$, respectively (see equation \eqref{neweq:scondnew}),  we assume $\delta(\boldsymbol{\mu}_{j,k},\boldsymbol{\mu}_{j',k'})=0$,   if $\rho_{j,k}=1$ or $\rho_{j',k'}=1$,
and $\delta(\boldsymbol{\eta}_{j,k},\boldsymbol{\eta}_{j',k'})=0$ if $\rho_{j,k}=0$ or $\rho_{j',k'}=0$,   for $(i,j) \neq (i',j')$.

\paragraph{\textbf{Similarities between the MAP behaviors.}}

\begin{table}
\caption{The Adjusted Rand Index for all the pairs of dogs. \label{tab:rand} }
\centering
\fbox{
\begin{tabular}{c|ccccccc}
 & Woody & Sherlock & Alvin & Rosie & Bear & Lucy   \\
\hline
Woody  &1.000&	0.094&	0.006&	0.030	&0.181&	0.414\\
Sherlock&0.094&	1.000&	-0.004&	0.015&	0.125	&0.081\\
Alvin &0.006&	-0.004&	1.000&	0.047&	-0.002&	0.003\\
Rosie &0.030&	0.015&	0.047&	1.000	&0.021&	0.021\\
 Bear &0.181&	0.125&	-0.002&	0.021&	1.000&	0.159\\
 Lucy &0.414&	0.081&	0.003&	0.021	&0.159&	1.000\\
\hline \hline
\end{tabular}
}
\end{table}
One of the assumptions of our proposal is that the temporal evolution of the behaviors are  independent. To have a posterior confirmation that this hypothesis is true,  we computed the Adjusted Rand Index  \citep{Gates2017} for the MAP behaviors of each pair of animals. The Adjusted Rand Index,  which is a  measure of similarity (or agreement) between two clusterizations,
has  a value close to one, if there is a strong agreement, while  its value is close to zero (even negative) if the clusterization is very dissimilar. It should be noted that we are able to compute the index because the animals are observed in the same temporal-window.
The results in Table \ref{tab:rand}  show that the values of the index are very low,  with the exception of dogs 1 and 6, where the index is  0.414. We can  conclude that our hypothesis is reasonable.
\paragraph{\textbf{The dogs in the cohesive group.}}  We can clearly see, from Figures \ref{newfig:a2_1} and \ref{newfig:a2_new_2}, that
the four dogs that form a cohesive group (dog 1 Woody, dog 2 Sherlock, dog 5 Bear and dog 6 Lucy)  have similar behaviors.
In behavior $\text{B}_{j1}$,  the length of $\protect\vec{\mathbf{F}}_{j,t_{j,i}}$ is $\approx 0$, which means that the distribution of $\mathbf{s}_{j, t_{j,i+1}}$ is centered on the previous location ($\mathbf{s}_{j, t_{j,i}}$), there is no a preferable direction, since the ellipses are very close to a circle,  the speed is very low (see the size of the ellipses), and there is no attractor. Hence, the movement is only determined by the covariance matrix $\boldsymbol{\Sigma}_{j,1}$,  which is the same for all  the first behaviors ($\text{B}_{j1}$),
see Figure \ref{newfig:a2_44} (d).    The first behavior of the cohesive group  can  easily be interpreted as boundary-patrolling- or scent-marking-behavior, which  is a common behavior in this dog breed, and it has already  been observed   and with similar movement characteristics  \citep{MCgrew,mastrantonio2020modeling}.\\
\indent As in $\text{B}_{j1}$,  the length of $\protect\vec{\mathbf{F}}_{j,t_{j,i}}$  is $\approx 0$ in $\text{B}_{j2}$, there is no attractor but, since the  ellipses    rotate according to the previous directions, there is directional persistence. The  major and minor axes of the ellipses  have different lengths, with the major one in the same direction as $\phi_{j ,t_{j,i-1}}$, which means  that  we  can expect to observe  more movements on a straight line, i.e.,  in the same direction as the previous bearing angle $\phi_{j, t_{j,i-1}}$, or in direction $\phi_{j, t_{j,i-1}}-\pi$. The strength of directional persistence, measured by  parameter
$\rho_{j,2}$, is very similar for all the
$\text{B}_{j2}$, as we can see from  Figure  \ref{newfig:a2_44} (e), where the probability values are close to 1.   The movement speed in $\text{B}_{j2}$ increases compared to $\text{B}_{j1}$, and  $\text{B}_{j2}$ is fully characterized  by  $\boldsymbol{\Sigma}_{j,2}$ and $\rho_{j,2}$.
The strong directional persistence, the movement along a straight line,  and the higher speed  can lead to  $\text{B}_{j2}$ being  interpreted    as a defending-behavior, where   the dog  defends the territory and livestock  from    predators, that is, mostly wild dogs and foxes that are present in the area  \citep[see][]{Brook2012,Walton2017},   or an explore-behavior \citep{van2010guardian,mastrantonio2020modeling}. \\
\indent The third behavior is a BRW, since the  CIs of $\rho_{j,3}$ are very close to 0 and   the  ellipses  are independent of the previous direction (see Figures \ref{newfig:a2_1} and \ref{newfig:a2_new_2}). The CIs of $\tau_{j,3}$, which are  in $[0.1,0.28]$, indicate a moderate attraction to $\boldsymbol{\mu}_{j,3}$,
 see  Tables   \ref{I-tab:anim1}, \ref{I-tab:anim2}, \ref{I-tab:anim5} and \ref{I-tab:anim6} in the Appendix.
The four dogs have the same spatial attractor, $\boldsymbol{\mu}_{j,3}$, the same covariance matrix, $\boldsymbol{\Sigma}_{j,3}$, the same parameter $\rho_{j,k}$ and, with the exception of the second dog, the same $\tau_{j,k}$, as we can see from Figures \ref{newfig:a2_1}, \ref{newfig:a2_new_2},  and \ref{newfig:a2_44}.
The spatial attractor, due to the large variance  of the movement (the ellipses size), can be considered as a tendency of these dogs to move to the central patch of the space  and,  since we can see from Figure c2 of \cite{Bommel3} that the attractor is close to where the livestock is, it is  easy to interpret this behavior as the dog  attending livestock.\\
\indent  It is interesting to note that  the parameters that define  the three behaviors, i.e.,  $\boldsymbol{\Sigma}_{j,k}$ in $\text{B}_{j1}$, ($ \boldsymbol{\Sigma}_{j,k},\rho_{j,k}$ ) in $\text{B}_{j2}$, ($\boldsymbol{\mu}_{j,k}, \tau_{j,k}, \boldsymbol{\Sigma}_{j,k}, \rho_{j,k}$)  in $\text{B}_{j3}$,
have a high probability of being the same between dogs, see Figure \ref{newfig:a2_44},
which means that they behave in a similar manner.  In  transition matrix terms, we can see,  from  Tables   \ref{I-tab:anim1}, \ref{I-tab:anim2}, \ref{I-tab:anim5} and \ref{I-tab:anim6} in the Appendix,
that
$\hat{\pi}_{j,1,2}>\hat{\pi}_{j,1,3}$
and
$\hat{\pi}_{j,3,2}>\hat{\pi}_{j,3,1}$. Then, after patrolling  ($\text{B}_{j1}$), it  is more probable  that the dog  begins to  explore the space, or to defend the property from predators spotted during patrolling ($\text{B}_{j2}$), than to guard livestock ($\text{B}_{j3}$).   After attending livestock,  it is more probable that the dog switches to $\text{B}_{j2}$.
\paragraph{\textbf{The socially excluded dog and the old one.}}
The socially excluded dog (j=3) has 3 behaviors, one of which, $\text{B}_{31}$,   is different from all the other dogs' behaviors, while the other
two, $\text{B}_{32}$ and $\text{B}_{33}$,  are  similar to the ones of the cohesive group, e.g.,  $\text{B}_{32}$ is similar to $\text{B}_{j2}$ and $\text{B}_{33}$ is similar to $\text{B}_{j1}$, with $j=1,2,5,6$, see Figures \ref{newfig:a2_1} and \ref{newfig:a2_new_2}.
$\text{B}_{33}$  is only characterized  by the covariance matrix $\boldsymbol{\Sigma}_{j,3}$,  since like $\text{B}_{j1}$ in the cohesive group,
$\protect\vec{\mathbf{F}}_{3,t_{3,i}}$ is $\approx 0$ and there is no attractor or directional persistence.  Therefore, due to the  high probability of $\boldsymbol{\Sigma}_{3,3}$ having the same value as $\boldsymbol{\Sigma}_{j,3}$, with $j=1,2,5,6$, see Figure \ref{newfig:a2_44} (d), $\text{B}_{33}$  may be interpreted in the same way as the first behaviors of the   cohesive group, i.e., a boundary-patrolling behavior.\\
\indent $\text{B}_{32}$ is similar to the second behavior of the cohesive group  in terms of covariance matrix, see  Figure \ref{newfig:a2_44} (d), but there is a lack of directional persistence, and a slight bias toward an attractor located in the central area, see the direction of $\vec{\mathbf{F}}_{3,t_{3,i}}$.
With this behavior, the dog is exploring but, at the same time, staying in the proximity of the sheep paddock, see Figure \ref{newfig:sp_tot} (c).\\
\indent The CI of $\tau_{3,1}$  is very close to 1 and   $\text{B}_{31}$ therefore has a  strong attraction to the  coordinates $\hat{\boldsymbol{\mu}}_{3,1}=(0.575,-381)'$, as we can see from Table  \ref{I-tab:anim3} of the Appendix.  The  CIs  of  $\mu_{3,1,1}$ and $\mu_{3,1,2}$ are very small and  equal to  (0.575,0.576)  and (-0.381,0.380),  respectively, which means that the attractor is well localized in space. We can see from  Figure c2 of \cite{Bommel3},   that $\hat{\boldsymbol{\mu}}_{3,1}$
is close to the owner's homestead. This behavior cannot be interpreted as the dog attending livestock since, once it reaches the spatial attractor, it does not move very much, i.e., the sizes of the ellipses are very small.  Hence, this is probably  a behavior in which the dog stays close to the owner's house,  and rests.\\
\indent For the last dog, that is the old one, the  model has found only two behaviors. $\text{B}_{41}$ has the same characteristics as the first behavior of the cohesive group,  with the same values of the covariance matrix, while  $\text{B}_{42}$ is similar to the second behavior of the cohesive group, with the same covariance matrix and similar directional persistence, see Figures   \ref{newfig:a2_1},  \ref{newfig:a2_new_2}, and \ref{newfig:a2_44}.  Hence, they can be interpreted as the first and second behaviors of the cohesive group. It is interesting to note that, since this dog is very old, she is no longer able to attend livestock, which is the activity that requires the most energy    (higher speed), but she is still working, that is,  checking the boundaries ($\text{B}_{41}$), exploring the space,  and  defending the territory ($\text{B}_{42}$).

\section{Final remarks} \label{newsec:rem}

In this work, we have proposed a new HMM which can be used to model trajectory-tracking data of multiple animals and which, according to   the classification given by \cite{Scharf2020}, is part of the indirect approach. Our model allows  subsets of parameters to be shared between  animals and behaviors, and  the number of latent behaviors to be selected during the model fitting.  The emission distribution is the STAP,  but other distributions can be used  by changing the model formalization accordingly. \\
\indent The model was used to help understand the behavior of  6 Maremma Sheepdogs, observed  in a property in Australia. The results show that there are many common features between the animals, such as the  attractive point,  and most of them share the same number of behaviors as well as the same parameter values.
The obtained results are easily interpretable, and  the rich output offers an insight into the similarities between animals.\\
\indent  %In our proposal,  we cannot  evaluate  if two animals change behavior at the same time-point.
%This is a possible extension   that we are currently working  on.
As a possible extension, we are currently exploring the  use of covariates to model the probability that behaviors share parameters, and we are working on a different formalization that makes the model able to detect whether different animals  tend to  follow the same   behaviors at the same time-points.

%On the other hands, our proposal has a richer output and is able to give an isight on the similarities between animal which make its use more interesting in applied context and then preferable \citep{Pohle2017}.

\section*{Acknowledgments}
The author would like to thank the Editor-in-Chief, the Associate Editor and the two anonymous
reviewers for their comments that have greatly improved the manuscript.
The work of the  author has partially been developed under the MIUR grant Dipartimenti di Eccellenza 2018 - 2022 (E11G18000350001), conferred to the Dipartimento di Scienze Matematiche - DISMA, Politecnico di Torino.

%We acknowledge the support of the Italian Ministry of Education, University, and Research (MIUR), grant  \textit{Dipartimenti di Eccellenza}, CUP: E11G18000350001, conferred to Dipartimento di Scienze Matematiche - DISMA, Politecnico di Torino.
%The work of the first three  authors  is partially developed under the PRIN2015 supported-project Environmental processes and human activities: capturing their interactions via statistical methods (EPHAStat) funded by MIUR (Italian Ministry of Education, University and Scientific Research) (20154X8K23-SH3).

%
% \begin{appendices}
% \counterwithin{figure}{section}
% \counterwithin{table}{section}
%
%
% \section{MCMC algorithm}
%
%
% \end{appendices}

\begin{appendices}

\section{Markov chain Monte Carlo algorithm}\label{I-sec:imp}

In the next two sections, we show how to sample the posterior values  of the HMM parameters, Section \ref{I-HMM}, and STAP parameters, Section \ref{I-STAP}.
The algorithm is based on  the works  of \cite{mastrantonio2020modeling} and  \cite{fox2011}.

\subsection{HMM} \label{I-HMM}
In order to implement the sample of the HMM parameters, we use the
\textit{degree L weak limit approximation}  \citep{Ishwaran2002} of a DP, which approximates the distribution of  the infinite-dimensional probability vector GEM distributed, with a finite one of dimension L. We thus obtain
\begin{align}
\boldsymbol{\beta}_{\boldsymbol{\mu}}^*|\gamma_{\boldsymbol{\mu}}  &\sim \text{Dir}\left(\frac{\gamma_{\boldsymbol{\mu}}}{L}, \dots , \frac{\gamma_{\boldsymbol{\mu}}}{L}\right),\\
\boldsymbol{\beta}_{\boldsymbol{\eta}}^*|\gamma_{\boldsymbol{\eta}}  &\sim \text{Dir}\left(\frac{\gamma_{\boldsymbol{\eta}}}{L}, \dots , \frac{\gamma_{\boldsymbol{\eta}}}{L}\right),\\
\boldsymbol{\beta}_{\boldsymbol{\Sigma}}^*|\gamma_{\boldsymbol{\Sigma}}  &\sim \text{Dir}\left(\frac{\gamma_{\boldsymbol{\Sigma}}}{L}, \dots , \frac{\gamma_{\boldsymbol{\Sigma}}}{L}\right),\label{I-eq:betadit}\\
\boldsymbol{\beta}_{\tau}^*|\gamma_{\tau}  &\sim \text{Dir}\left(\frac{\gamma_{\tau}}{L}, \dots , \frac{\gamma_{\tau}}{L}\right),\\
\boldsymbol{\beta}_{\rho}^*|\gamma_{\rho}  &\sim \text{Dir}\left(\frac{\gamma_{\rho}}{L}, \dots , \frac{\gamma_{\rho}}{L}\right),
\end{align}
and
\begin{equation} \label{I-eq:pi}
\boldsymbol{\pi}_{j,l}|\alpha, \nu, \boldsymbol{\beta} \sim \text{Dir}(\alpha{\beta}_1+ \nu\delta(l,1),\alpha{\beta}_2+ \nu\delta(l,2), \dots ,\alpha{\beta}_{L^5}+ \nu\delta(l,L^5)).
\end{equation}
Hence, $\boldsymbol{\beta}$ and $\boldsymbol{\pi}_{j,l}$ are both  finite-dimensional (see  \cite{fox2011}) with $L$ and $L^5$ elements.
As $L \rightarrow \infty$, the  Dirichlet distributions in \eqref{I-eq:betadit} converges to the associated GEMs, and equation \eqref{I-eq:pi} converges to a $\text{DP}\left(\alpha+\nu,   \frac{\alpha\boldsymbol{\beta}+ \nu\delta_{l}}{\alpha+\nu}\right)$. This approximation greatly simplifies the implementation, and it is accurate as long as the number of unique values of each  STAP parameter, estimated by means of the model, is   $<<L$. For our implementation, we assume $L=100$.

\paragraph{Sampling $\boldsymbol{\pi}_{j,l}$}  Let $n_{j,l,k}$ be the number of transitions of the  $j$-th animal from behavior $l$ to behavior $k$:
\begin{equation}
n_{j,l,k} = \sum_{i=1}^{T-1}\delta( z_{j, t_{j,i-1}},l) \delta( z_{j, t_{j,i}},k).
\end{equation}
 Since \eqref{I-eq:pi} has a  finite dimension, $\boldsymbol{\pi}_{j,l}$ can be  sampled as in a standard HMM,  and its full conditional is
\begin{equation} \label{I-eq:pi_p}
\boldsymbol{\pi}_{j,l} \sim \text{Dir}(\alpha\boldsymbol{\beta}_1+ \nu\delta(l,1)+n_{j,l,1},\alpha\boldsymbol{\beta}_2+ \nu\delta(l,2)+n_{j,l,2}, \dots , \alpha\boldsymbol{\beta}_{L^5}+ \nu\delta(l,L^5)+n_{j,l,L^5} ).
\end{equation}
It should be noted that, even though the dimension of $\boldsymbol{\pi}_{j,l}$ is very large, we do not have to sample all of its components but only the ones with $n_{j,l,k}>0$. This can  easily be done using the stick-breaking sample scheme  \citep{sethuraman:stick}.
% which allow also to sample components with $n_{j,l,k}=0$ when needed.
%\textcolor{red}{DIre che possa simulare inponento decresacenza e questo aiuta nel beam sample... forse posso metter eprima z sample}

\paragraph{Sampling $\boldsymbol{\beta}_{\boldsymbol{\mu}}^*$, $\boldsymbol{\beta}_{\boldsymbol{\eta}}^*$, $\boldsymbol{\beta}_{\boldsymbol{\Sigma}}^*$, $\boldsymbol{\beta}_{\tau}^*$, $\boldsymbol{\beta}_{\rho}^*$}

In order to be able to sample these parameters, we use the results of \cite{fox2011}, which  show  that, by introducing the latent variables
\begin{align}
x_{j,l,k,h}& \sim  \text{Ber}\left( \frac{\alpha \beta_k +\nu\delta(l,k)}{ h-1+  \alpha \beta_k +\nu\delta(l,k)}  \right), h = 1,\ \dots,  n_{j,l,k} , \label{I-exq}\\
w_{j,l} &\sim  \text{Binomial}\left(b_{j,l,l}, \frac{\nu}{\nu+\beta_{j}\alpha} \right), \label{I-eq:wwww}
\end{align}
where the variables $x_{j,l,k,h}$ in  \eqref{I-exq}  only need to be sampled if $n_{j,l,k}>0$,
 the vector
 $\boldsymbol{\beta}$
 enters into the density of $\{\boldsymbol{\pi}_{j,l}\}_{l=1}^{L^5}$ as
\begin{equation} \label{I-eq:ddddd}
\prod_{l=1}^{L^5}\prod_{k=1}^{L^5}\beta_{k}^{\bar{b}_{j,l,k}}= \prod_{k=1}^{L^5}\beta_{k}^{\bar{b}_{j,.,k}} =  \prod_{k=1}^{L^5}\left( \beta_{\boldsymbol{\mu},{\lambda_{\boldsymbol{\mu},k}}}^* \beta_{\boldsymbol{\eta},{\lambda_{\boldsymbol{\eta},k}}}^* \beta_{\boldsymbol{\Sigma},{\lambda_{\boldsymbol{\Sigma},k}}}^*
\beta_{\tau,{\lambda_{\tau,k}}}^* \beta_{\rho,{\lambda_{\rho,k}}}^*  \right)^{\bar{b}_{j,\cdot,k}},
\end{equation}
where
\begin{align}
\bar{b}_{j,\cdot,k}& = \sum_{l=1}^{L^5}\bar{b}_{j,l,k},\\
\bar{b}_{j,l,k} & =
\begin{cases}
{b}_{j,l,k} & \text{if } l\neq k,\\
{b}_{j,l,k}-w_{j,l} & \text{if } l = k,
\end{cases}\\
b_{j,l,k} & =
\begin{cases}
0 & \text{if } n_{j,l,k}=0,\\
\sum_{h=1}^{n_{j,l,k}}x_{j,l,k,h}& \text{otherwise}.
\end{cases}
\end{align}
Hence, the full conditionals are
\begin{align}
\boldsymbol{\beta}_{\boldsymbol{\mu}}^*& \sim \text{Dir}\left(\frac{\gamma_{\boldsymbol{\mu}}}{L}+\bar{b}_{\boldsymbol{\mu},1},\frac{\gamma_{\boldsymbol{\mu}}}{L}+\bar{b}_{\boldsymbol{\mu},2}, \dots ,
 \frac{\gamma_{\boldsymbol{\mu}}}{L}+\bar{b}_{\boldsymbol{\mu},L}\right),\\
\boldsymbol{\beta}_{\boldsymbol{\eta}}^*& \sim \text{Dir}\left(\frac{\gamma_{\boldsymbol{\eta}}}{L}+\bar{b}_{\boldsymbol{\eta},1},\frac{\gamma_{\boldsymbol{\eta}}}{L}+\bar{b}_{\boldsymbol{\eta},2}, \dots ,
\frac{\gamma_{\boldsymbol{\eta}}}{L}+\bar{b}_{\boldsymbol{\eta},L}\right),\\
\boldsymbol{\beta}_{\boldsymbol{\Sigma}}^*& \sim \text{Dir}\left(\frac{\gamma_{\boldsymbol{\Sigma}}}{L}+\bar{b}_{\boldsymbol{\Sigma},1},\frac{\gamma_{\boldsymbol{\Sigma}}}{L}+\bar{b}_{\boldsymbol{\Sigma},2}, \dots ,
 \frac{\gamma_{\boldsymbol{\Sigma}}}{L}+\bar{b}_{\boldsymbol{\Sigma},L}\right),\\
\boldsymbol{\beta}_{\tau}^*& \sim \text{Dir}\left(\frac{\gamma_{\tau}}{L}+\bar{b}_{\tau,1},\frac{\gamma_{\tau}}{L}+\bar{b}_{\tau,2}, \dots , \frac{\gamma_{\tau}}{L}+\bar{b}_{\tau,L}
\right),\\%
\boldsymbol{\beta}_{\rho}^*& \sim \text{Dir}\left(\frac{\gamma_{\rho}}{L}+\bar{b}_{\rho,1},\frac{\gamma_{\rho}}{L}+\bar{b}_{\rho,2}, \dots , \frac{\gamma_{\rho}}{L}+
\bar{b}_{\rho,L}
\right),
\end{align}
where
\begin{align}
\bar{b}_{\boldsymbol{\mu},p}= &\sum_{j=1}^m \sum_{k=1}^{L^5}\bar{b}_{j,\cdot,k}\delta( \lambda_{\boldsymbol{\mu},k} ,p  ),\\
\bar{b}_{\boldsymbol{\eta},p}= &\sum_{j=1}^m \sum_{k=1}^{L^5}\bar{b}_{j,\cdot,k}\delta( \lambda_{\boldsymbol{\eta},k} ,p  ),\\
\bar{b}_{\boldsymbol{\Sigma},p}= &\sum_{j=1}^m \sum_{k=1}^{L^5}\bar{b}_{j,\cdot,k}\delta( \lambda_{\boldsymbol{\Sigma},k} ,p  ),\\
\bar{b}_{\tau,p}= &\sum_{j=1}^m \sum_{k=1}^{L^5}\bar{b}_{j,\cdot,k}\delta( \lambda_{\tau,k} ,p  ),\\
\bar{b}_{\rho,p}= &\sum_{j=1}^m \sum_{k=1}^{L^5}\bar{b}_{j,\cdot,k}\delta( \lambda_{\rho,k} ,p  ).
\end{align}
% Then, the full conditionals are
% \begin{align}
% \boldsymbol{\beta}_{\boldsymbol{\mu}}^*& \sim \text{Dir}\left(\frac{\gamma_{\boldsymbol{\mu}}}{L}+\bar{b}_{\boldsymbol{\mu},1},\frac{\gamma_{\boldsymbol{\mu}}}{L}+\bar{b}_{\boldsymbol{\mu},2}, \dots ,
%  \frac{\gamma_{\boldsymbol{\mu}}}{L}+\bar{b}_{\boldsymbol{\mu},L}\right),\\
% \boldsymbol{\beta}_{\boldsymbol{\eta}}^*& \sim \text{Dir}\left(\frac{\gamma_{\boldsymbol{\eta}}}{L}+\bar{b}_{\boldsymbol{\eta},1},\frac{\gamma_{\boldsymbol{\eta}}}{L}+\bar{b}_{\boldsymbol{\eta},2}, \dots ,
% \frac{\gamma_{\boldsymbol{\eta}}}{L}+\bar{b}_{\boldsymbol{\eta},L}\right),\\
% \boldsymbol{\beta}_{\boldsymbol{\Sigma}}^*& \sim \text{Dir}\left(\frac{\gamma_{\boldsymbol{\Sigma}}}{L}+\bar{b}_{\boldsymbol{\Sigma},1},\frac{\gamma_{\boldsymbol{\Sigma}}}{L}+\bar{b}_{\boldsymbol{\Sigma},2}, \dots ,
%  \frac{\gamma_{\boldsymbol{\Sigma}}}{L}+\bar{b}_{\boldsymbol{\Sigma},L}\right),\\
% \boldsymbol{\beta}_{\tau}^*& \sim \text{Dir}\left(\frac{\gamma_{\tau}}{L}+\bar{b}_{\tau,1},\frac{\gamma_{\tau}}{L}+\bar{b}_{\tau,2}, \dots , \frac{\gamma_{\tau}}{L}+\bar{b}_{\tau,L}
% \right),\\%
% %
% \boldsymbol{\beta}_{\rho}^*& \sim \text{Dir}\left(\frac{\gamma_{\rho}}{L}+\bar{b}_{\rho,1},\frac{\gamma_{\rho}}{L}+\bar{b}_{\rho,2}, \dots , \frac{\gamma_{\rho}}{L}+
% \bar{b}_{\rho,L}
% \right).\\
% \end{align}

\paragraph{ Sampling $\gamma_{\boldsymbol{\mu}}$, $\gamma_{\boldsymbol{\eta}}$, $\gamma_{\boldsymbol{\Sigma}}$, $\gamma_{\tau}$, $\gamma_{\rho}$}

Since each $\gamma_{\chi}$ is the hyperparameter of a DP, we can use   the approach proposed by \cite{Escobar1995}, which was later extended to the sHDP-HMM framework by \cite{fox2011}. If the prior is    $G(a_{\gamma},b_{\gamma})$ and
if we introduce  the  latent variables
\begin{align}
\xi_{\boldsymbol{\mu}}& \sim  \text{B}(\gamma_{\boldsymbol{\mu}}+1, \sum_{p=1}^L\bar{b}_{\boldsymbol{\mu},p}),\\
\xi_{\boldsymbol{\eta}} &\sim  \text{B}(\gamma_{\boldsymbol{\eta}}+1, \sum_{p=1}^L\bar{b}_{\boldsymbol{\eta},p}),\\
\xi_{\boldsymbol{\Sigma}} &\sim  \text{B}(\gamma_{\boldsymbol{\Sigma}}+1, \sum_{p=1}^L\bar{b}_{\boldsymbol{\Sigma},p}),\\
\xi_{\tau}& \sim  \text{B}(\gamma_{\tau}+1, \sum_{p=1}^L\bar{b}_{\tau,p}),\\
\xi_{\rho} &\sim  \text{B}(\gamma_{\rho}+1, \sum_{p=1}^L\bar{b}_{\rho,p}),
\end{align}
and let
\begin{align}
K_{\boldsymbol{\mu}} &= \sum_{p=1}^L \mathbb{I}(\bar{b}_{\boldsymbol{\mu},p}>0),\\
K_{\boldsymbol{\eta}}& = \sum_{p=1}^L \mathbb{I}(\bar{b}_{\boldsymbol{\eta},p}>0),\\
K_{\boldsymbol{\Sigma}}& = \sum_{p=1}^L \mathbb{I}(\bar{b}_{\boldsymbol{\Sigma},p}>0),\\
K_{\tau} &= \sum_{p=1}^L \mathbb{I}(\bar{b}_{\tau,p}>0),\\
K_{\rho} &= \sum_{p=1}^L \mathbb{I}(\bar{b}_{\rho,p}>0).
\end{align}
where $\mathbb{I}(\cdot)$ is equal to 1 if the argument is true, and zero otherwise,
the full conditionals  become
\begin{align}
\gamma_{\boldsymbol{\mu}} &\sim \pi_{\boldsymbol{\mu}}\text{G}(a_{\gamma}+K_{\boldsymbol{\mu}},b_{\gamma}-\log(\xi_{\boldsymbol{\mu}} ))+(1-\pi_{\boldsymbol{\mu}}) \text{G}(a_{\gamma}+K_{\boldsymbol{\mu}}-1,b_{\gamma}-\log(\xi_{\boldsymbol{\mu}} )),\\
\gamma_{\boldsymbol{\eta}} &\sim \pi_{\boldsymbol{\eta}}\text{G}(a_{\gamma}+K_{\boldsymbol{\eta}},b_{\gamma}-\log(\xi_{\boldsymbol{\eta}} ))+(1-\pi_{\boldsymbol{\eta}}) \text{G}(a_{\gamma}+K_{\boldsymbol{\eta}}-1,b_{\gamma}-\log(\xi_{\boldsymbol{\mu}} )),\\
\gamma_{\boldsymbol{\Sigma}} &\sim \pi_{\boldsymbol{\Sigma}}\text{G}(a_{\gamma}+K_{\boldsymbol{\Sigma}},b_{\gamma}-\log(\xi_{\boldsymbol{\Sigma}} ))+(1-\pi_{\boldsymbol{\Sigma}}) \text{G}(a_{\gamma}+K_{\boldsymbol{\Sigma}}-1,b_{\gamma}-\log(\xi_{\boldsymbol{\Sigma}} )),\\
\gamma_{\tau} &\sim \pi_{\tau}\text{G}(a_{\gamma}+K_{\tau},b_{\gamma}-\log(\xi_{\tau} ))+(1-\pi_{\tau}) \text{G}(a_{\gamma}+K_{\tau}-1,b_{\gamma}-\log(\xi_{\tau} )),\\
\gamma_{\rho} &\sim \pi_{\rho}\text{G}(a_{\gamma}+K_{\rho},b_{\gamma}-\log(\xi_{\rho} ))+(1-\pi_{\rho}) \text{G}(a_{\gamma}+K_{\rho}-1,b_{\gamma}-\log(\xi_{\rho} )),
\end{align}
where
\begin{align}
\pi_{\boldsymbol{\mu}} &= \frac{a_{\gamma}+K_{\boldsymbol{\mu}}-1}{(b_{\gamma}-\log(\xi_{\boldsymbol{\mu}})) \sum_{p=1}^L\bar{b}_{\boldsymbol{\mu},p}  },\\
\pi_{\boldsymbol{\eta}} &= \frac{a_{\gamma}+K_{\boldsymbol{\eta}}-1}{(b_{\gamma}-\log(\xi_{\boldsymbol{\eta}})) \sum_{p=1}^L\bar{b}_{\boldsymbol{\eta},p}  },\\
\pi_{\boldsymbol{\Sigma}} &= \frac{a_{\gamma}+K_{\boldsymbol{\Sigma}}-1}{(b_{\gamma}-\log(\xi_{\boldsymbol{\Sigma}})) \sum_{p=1}^L\bar{b}_{\boldsymbol{\Sigma},p}  },\\
\pi_{\tau} &= \frac{a_{\gamma}+K_{\tau}-1}{(b_{\gamma}-\log(\xi_{\tau})) \sum_{p=1}^L\bar{b}_{\tau,p}  },\\
\pi_{\rho} &= \frac{a_{\gamma}+K_{\rho}-1}{(b_{\gamma}-\log(\xi_{\rho})) \sum_{p=1}^L\bar{b}_{\rho,p}  }.
\end{align}

\paragraph{Sampling $\nu/(\alpha+\nu)$}

The sampling of this parameter follows directly from \cite{fox2011}, if we assume a $B(a_{\nu}, b_{\nu})$  prior. This parameter enters  into the distribution of the transition matrix as in the model of \cite{fox2011},  with the only difference being that we have to take into consideration that  $m$ different HMMs  are used.
%We have that  $\sum_{l=1}^{K_j} w_{j,l}$ is the number of success in $\sum_{l=1}^l\sum_{k=1}^l b_{j,l,k}$ independent $\text{Ber}(\nu/(\alpha+\nu))$ trials, see equation \eqref{eq:wwww}.
Hence, the full conditional of $\nu/(\alpha+\nu)$ is
$$
\text{B}\left(\sum_{j=1}^m\sum_{l=1}^{K_j} w_{j,l}+a_{\nu},\sum_{l=1}^L\sum_{k=1}^L b_{j,l,k} -\sum_{j=1}^m\sum_{l=1}^{K_j} w_{j,l}  +b_{\nu} \right).
$$
\paragraph{Sampling $\alpha+\nu$}  \cite{fox2011}  showed that the contribution of the  j-th transition matrix to the full conditional of $\alpha+\nu$ is
\begin{equation} \label{I-eq:kr}
(\alpha+\nu)^{\sum_{l=1}^L\sum_{k=1}^L b_{j,l,k}}\prod_{l=1}^{K_j} \frac{\Gamma(\alpha+\nu )}{\Gamma(\alpha+\nu+ \sum_{k=1}^L n_{j,l,k} ) }.
\end{equation}
Since conditionally to $\boldsymbol{\beta}$ the animals are independent,
we can  use the results of  \cite{fox2011}. Hence, if we  introduce the following latent variables
\begin{align}
r_{1,j,l} & \sim B(\alpha+\nu+1, \sum_{k=1}^L n_{j,l,k}),\\
r_{2,j,l} & \sim \text{Ber}\left(   \frac{\sum_{k=1}^L n_{j,l,k}}{\sum_{k=1}^L n_{j,l,k}+\alpha+\nu} \right),
\end{align}
we can  express the full conditional of $\alpha+\nu$ as
$$
G\left(a_{\alpha+\nu}+\sum_{j=1}^m\sum_{l=1}^{K_j} w_{j,l}- \sum_{j=1}^m\sum_{l=1}^L  r_{2,j,l}, b_{\alpha+\nu} -  \sum_{j=1}^m\sum_{l=1}^L  \log(r_{1,j,l})\right).
$$

\paragraph{Sampling $z_{j, t_{j,i}}$}

The full conditional of $ z_{j, t_{j,i}}$ is multinomial, with one trial and the  $k$-th probability  proportional to
\begin{equation} \label{I-eq:sampz}
\begin{cases}
\pi_{j,z_{j, t_{j,i-1}},k}\pi_{j,k,z_{j,  t_{j,i+1}}}f(s_{j,t_{j,i+1}}|s_{j,t_{j,i}},s_{j,t_{j,i-1}}, \boldsymbol{\theta}_{k}) \,  \text{ if  }  i \geq 1,\\
(1-\epsilon)^{k-1}\epsilon   \pi_{j,k,z_{j, k, t_{j,1}}}\,  \text{ otherwise,  }
\end{cases}
\end{equation}
where $f(s_{j,t_{j,i+1}}|s_{j,t_{j,i}},s_{j,t_{j,i-1}}, \boldsymbol{\theta}_{k})$ is the STAP density and $((1-\epsilon)^{k-1}\epsilon)$ is the Geometric probability mass over the value $k$.
In order to be able to sample from the full conditional,   we have to compute \eqref{I-eq:sampz} for all the  possible $L^5$ behaviors, which is unfeasible. For this reason, we decided to exploit   the beam sample scheme of \cite{VanGael2008} which, as a result of the introduction of latent variables, is able to  drastically reduce the number of elements that need to be computed. The idea is to introduce the variables
$$
u_{j,t_{j,i}} \sim U(0,\pi_{j,z_{j, t_{j,i-1}},z_{j, t_{j,i}}} ),
$$
and then, conditionally to all $u_{j,t_{j,i}}$,   the variable $ z_{j, t_{j,i}}$ is only allowed to assume  values that satisfy the constraints $\pi_{j,z_{j, t_{j,i-1}},z_{j, t_{j,i}}}>u_{j,t_{j,i}} $ and  $\pi_{j,z_{j, t_{j,i}},z_{j, t_{j,i+1}}}>u_{j,t_{j,i+1}} $. Hence,  equation \eqref{I-eq:sampz} only needs to be evaluated  for a finite set of elements, which is   often very small.  The values of $\pi_{j,l,k}$
for empty behaviors can  easily be obtained using the stick breaking sample scheme.\\
\indent In order to increase the  convergence speed, we also  suggest using  another updating strategy for  $ z_{j, t_{j,i}}$.
Let us introduce the  new random variables $ z_{\boldsymbol{\mu},j, t_{j,i}}$,  $ z_{\boldsymbol{\eta},j, t_{j,i}}$,  $ z_{\boldsymbol{\Sigma},j, t_{j,i}}$,  $ z_{\tau,j, t_{j,i}}$, and  $ z_{\rho,j, t_{j,i}}$, whose values represent what elements of $\boldsymbol{\mu}^*$, $\boldsymbol{\eta}^*$, $\boldsymbol{\Sigma}^*$, $\boldsymbol{\tau}^*$,
and $\boldsymbol{\rho}^*$, respectively, are in $\boldsymbol{\theta}$, at time $t_{j,i}$ andfor the $j$-th animal. There is a one-to-one relationship between $z_{j, t_{j,i}}$ and ($ z_{\boldsymbol{\mu},j, t_{j,i}}, z_{\boldsymbol{\eta},j, t_{j,i}}, z_{\boldsymbol{\Sigma},j, t_{j,i}},  z_{\tau,j, t_{j,i}}, z_{\rho,j, t_{j,i}}$), i.e., if we know the value of
$z_{j, t_{j,i}}$, we also know ($ z_{\boldsymbol{\mu},j, t_{j,i}}, z_{\boldsymbol{\eta},j, t_{j,i}}, z_{\boldsymbol{\Sigma},j, t_{j,i}},  z_{\tau,j, t_{j,i}}, z_{\rho,j, t_{j,i}}$), and vice-versa,
and sampling the former or the latter is therefore equivalent, but if we use ($ z_{\boldsymbol{\mu},j, t_{j,i}}, z_{\boldsymbol{\eta},j, t_{j,i}}, z_{\boldsymbol{\Sigma},j, t_{j,i}},  z_{\tau,j, t_{j,i}}, z_{\rho,j, t_{j,i}})$, we have the possibility of updating  a parameter at the time which, according to our experience, increases the convergence speed.
% The advanted of using ($ z_{\boldsymbol{\mu},j, t_{j,i}}, z_{\boldsymbol{\eta},j, t_{j,i}}, z_{\boldsymbol{\Sigma},j, t_{j,i}},  z_{\tau,j, t_{j,i}}, z_{\rho,j, t_{j,i}}$) is that we can update a parameter at the time and, for example, if we want to update $z_{\boldsymbol{\mu},j, t_{j,i}}$ we have to compute  \eqref{eq:sampz} only for the $L$ behaviors that have unchanging values of  $ z_{\boldsymbol{\eta},j, t_{j,i}}$,  $ z_{\boldsymbol{\Sigma},j, t_{j,i}}$,  $ z_{\tau,j, t_{j,i}}$, and  $ z_{\rho,j, t_{j,i}}$.

\subsection{STAP parameters} \label{I-STAP}

Conditionally to the latent behaviors, the animals are independent and, therefore,  the sampling of the STAP parameters follows directly from \cite{mastrantonio2020modeling}, if we take  into account that the parameters are shared between animals and behaviors. For this reasons, let us define $\mathcal{I}_{\boldsymbol{\mu},j,p} $ as the set of indices $i$ of the $j$-th animal, so that $\boldsymbol{\mu}_{z_{j,t_{j,i}}} =  \boldsymbol{\mu}_p^*$; the sets $\mathcal{I}_{\boldsymbol{\eta},j,p} $, $\mathcal{I}_{\boldsymbol{\Sigma},j,p} $,
 $\mathcal{I}_{\tau,j,p} $, and $\mathcal{I}_{\rho,j,p} $ are defined in a similar way. If  the sets $\mathcal{I}_{\chi,j,p} $ are empty  for a given parameter $\chi_p^*$ and all animals, then the full conditional is equal to the distribution $H_{\chi}$ which, given the  chosen distributions,  is   easy to sample from. We do not need to sample all the $L$ parameters, but only the  ones that are observed in at least one of the non-empty behaviors, or the ones that are needed to sample the variables $z_{j, t_{j,i}}$.
%  \begin{align} \label{I-eq:likefull}
% &\prod_{i \in \mathcal{I}_j} \left| \boldsymbol{\Sigma}_j \right|^{-\frac{1}{2}}\exp\left(-\frac{ \left(\mathbf{s}_{t_{i+1}}-\mathbf{s}_{t_{i}}- \mathbf{M}_{t_i,j}   \right)' \mathbf{V}_{t_i,j}^{-1}  \left(\mathbf{s}_{t_{i+1}}-\mathbf{s}_{t_{i}}- \mathbf{M}_{t_i,j}   \right)  }{2} \right),\\
% &d   \\
% &d
% \end{align}

 \paragraph*{Sampling $\boldsymbol{\mu}_p^*$}
If we assume $\boldsymbol{\mu}_p^*\sim N\left(\mathbf{B}_{\boldsymbol{\mu}},\mathbf{W}_{\boldsymbol{\mu}}\right)$,  the full conditional is $ N\left(\mathbf{B}_{\boldsymbol{\mu},p},\mathbf{W}_{\boldsymbol{\mu},p}\right)$
with
\begin{align}
\mathbf{W}_{\boldsymbol{\mu},p}& =  \left( \sum_{j=1}^m\sum_{i \in \mathcal{I}_{\boldsymbol{\mu},j,p}}   (1-\rho_{z_{j,t_{j,i}}})^2\tau_{z_{j,t_{j,i}}}^2  \mathbf{V}_{\boldsymbol{\mu},j,t_{j,i}}^{-1}   +\mathbf{W}_{\boldsymbol{\mu}}^{-1}  \right)^{-1},\\
\mathbf{B}_{\boldsymbol{\mu},p}& =  \mathbf{W}_{\boldsymbol{\mu},p} \left( \sum_{j=1}^m\sum_{i \in \mathcal{I}_{\boldsymbol{\mu},j,p}}    (1-\rho_{z_{j,t_{j,i}}})\tau_{z_{j,t_{j,i}}}  \mathbf{V}_{\boldsymbol{\mu},j,t_{j,i}}^{-1}
 \mathbf{M}_{\boldsymbol{\mu},j,t_{j,i}}   +\mathbf{W}_{\boldsymbol{\mu}}^{-1}\mathbf{B}_{\boldsymbol{\mu}}  \right).
\end{align}
and
\begin{align}
\mathbf{V}_{\boldsymbol{\mu},j,t_{j,i}}& = \mathbf{R}(\rho_{z_{j,t_{j,i}}}
 \phi_{j,t_{j,i-1}})\boldsymbol{\Sigma}_{z_{j,t_{j,i}}}^{-1}\mathbf{R}'(\rho_{z_{j,t_{j,i}}}\phi_{j,t_{j,i-1}})\\
\mathbf{M}_{\boldsymbol{\mu},j,t_{j,i}} & =  \mathbf{s}_{j,t_{j,i+1}}- \mathbf{s}_{j,t_{j,i}} +(1-\rho_{z_{j,t_{j,i}}} )\tau_{z_{j,t_{j,i}}}  \mathbf{s}_{j,t_{j,i}} - \rho_{z_{j,t_{j,i}}} \mathbf{R}
\left(\phi_{j,t_{i-1}}\right) \boldsymbol{\eta}_{z_{j,t_{j,i}}} \\
\end{align}

\paragraph*{Sampling $\boldsymbol{\eta}_p^*$}
If we assume $\boldsymbol{\eta}_p^*\sim N\left(\mathbf{B}_{\boldsymbol{\eta}},\mathbf{W}_{\boldsymbol{\eta}}\right)$, the full conditional is $ N\left(\mathbf{B}_{\boldsymbol{\eta},p},\mathbf{W}_{\boldsymbol{\eta},p}\right)$
with
\begin{align}
\mathbf{W}_{\boldsymbol{\mu},p}& =  \left( \sum_{j=1}^m\sum_{i \in \mathcal{I}_{\boldsymbol{\eta},j,p}}   \rho_{z_{j,t_{j,i}}}^2  \mathbf{R}'(  \phi_{j,t_{j,i-1}}) \mathbf{V}_{\boldsymbol{\eta},j,t_{j,i}}^{-1}  \mathbf{R}(  \phi_{j,t_{j,i-1}})  +\mathbf{W}_{\boldsymbol{\eta}}^{-1}  \right)^{-1},\\
\mathbf{B}_{\boldsymbol{\eta},p}& =  \mathbf{W}_{\boldsymbol{\eta},p} \left( \sum_{j=1}^m\sum_{i \in \mathcal{I}_{\boldsymbol{\eta},j,p}}    \rho_{z_{j,t_{j,i}}} \mathbf{R}'(  \phi_{j,t_{j,i-1}})\mathbf{V}_{\boldsymbol{\eta},j,t_{j,i}}^{-1}
 \mathbf{M}_{\boldsymbol{\mu},j,t_{j,i}}   +\mathbf{W}_{\boldsymbol{\eta}}^{-1}\mathbf{B}_{\boldsymbol{\eta}}  \right).
\end{align}
and
\begin{align}
\mathbf{V}_{\boldsymbol{\mu},j,t_{j,i}}& = \mathbf{R}(\rho_{z_{j,t_{j,i}}}
 \phi_{j,t_{j,i-1}})\boldsymbol{\Sigma}_{z_{j,t_{j,i}}}^{-1}\mathbf{R}'(\rho_{z_{j,t_{j,i}}}\phi_{j,t_{j,i-1}})\\
\mathbf{M}_{\boldsymbol{\mu},j,t_{j,i}} & =  \mathbf{s}_{j,t_{j,i+1}}- \mathbf{s}_{j,t_{j,i}} -(1-\rho_{z_{j,t_{j,i}}} )\tau_{z_{j,t_{j,i}}} (\boldsymbol{\mu}_{z_{j,t_{j,i}}}-\mathbf{s}_{j,t_{j,i}})  \\
\end{align}

\paragraph*{Sampling $\tau_p^*$}

If we assume $\tau_p^* \sim U(0,1)$,   the full conditional is therefore a truncated normal   $N\left(\mathbf{B}_{\tau,p},\mathbf{W}_{\tau,p}\right)I(0,1)$ with
\begin{align}
\mathbf{W}_{\tau,p}& =  \left( \sum_{j=1}^m\sum_{i \in \mathcal{I}_{ {\tau},j,p}}  (1-\rho_{z_{j,t_{j,i}}}) ^2  (\boldsymbol{\mu}_{z_{j,t_{j,i}}}-\mathbf{s}_{j,t_{j,i}})'     \mathbf{V}_{\tau,j,t_{j,i}}^{-1} (\boldsymbol{\mu}_{z_{j,t_{j,i}}}-\mathbf{s}_{j,t_{j,i}})    \right)^{-1},\\
\mathbf{B}_{\tau,p}& =  \mathbf{W}_{\tau,p} \left( \sum_{j=1}^m\sum_{i \in \mathcal{I}_{ {\tau},j,p}}    (1-\rho_{z_{j,t_{j,i}}})   (\boldsymbol{\mu}_{z_{j,t_{j,i}}}-\mathbf{s}_{j,t_{j,i}})' \mathbf{V}_{\boldsymbol{\eta},j,t_{j,i}}^{-1}
 \mathbf{M}_{\tau,j,t_{j,i}}   \right).
\end{align}
and
\begin{align}
\mathbf{V}_{\tau,j,t_{j,i}}& = \mathbf{R}(\rho_{z_{j,t_{j,i}}}
 \phi_{j,t_{j,i-1}})\boldsymbol{\Sigma}_{z_{j,t_{j,i}}}^{-1}\mathbf{R}'(\rho_{z_{j,t_{j,i}}}\phi_{j,t_{j,i-1}})\\
\mathbf{M}_{\tau,j,t_{j,i}} & =  \mathbf{s}_{j,t_{j,i+1}}- \mathbf{s}_{j,t_{j,i}} -\rho_{z_{j,t_{j,i}}} \mathbf{R}
\left(\phi_{j,t_{i-1}}\right) \boldsymbol{\eta}_{z_{j,t_{j,i}}}  \\
\end{align}

\paragraph*{Sampling $\boldsymbol{\Sigma}_p^*$}
Since $\boldsymbol{\Sigma}_p^* \sim \text{IW}(a_{\boldsymbol{\Sigma}},\mathbf{C}_{\boldsymbol{\Sigma}})$, the full conditional is  $IW(a_{\boldsymbol{\Sigma},p},\mathbf{C}_{\boldsymbol{\Sigma},p})$ with
\begin{align}
a_{\boldsymbol{\Sigma},j}& = a_{\boldsymbol{\Sigma}}+\sum_{j=1}^m\sum_{i \in \mathcal{I}_{\boldsymbol{\Sigma},j,p}} 1 ,\\
\mathbf{C}_{\boldsymbol{\Sigma},j} & =\mathbf{C}_{\boldsymbol{\Sigma}}+ \sum_{j=1}^m\sum_{i \in \mathcal{I}_{\boldsymbol{\Sigma},j,p}} \left( \mathbf{R}'(\rho_{z_{j,t_{j,i}}})  \mathbf{M}_{\boldsymbol{\Sigma},j,t_{j,i}}  \right)\left( \mathbf{R}'(\rho_{z_{j,t_{j,i}}} ) \mathbf{M}_{\boldsymbol{\Sigma},j,t_{j,i}} \right)'.
\end{align}
and
\begin{align}
\mathbf{M}_{\boldsymbol{\Sigma},j,t_{j,i}} & =    \mathbf{s}_{j,t_{j,i+1}}- \mathbf{s}_{j,t_{j,i}} - (1-\rho_{z_{j,t_{j,i}}} )\tau_{z_{j,t_{j,i}}} (\boldsymbol{\mu}_{z_{j,t_{j,i}}}-\mathbf{s}_{j,t_{j,i}})-\rho_{z_{j,t_{j,i}}} \mathbf{R}
 \left(\phi_{j,t_{i-1}}\right) \boldsymbol{\eta}_{z_{j,t_{j,i}}}
\end{align}

\paragraph*{Sampling $\rho_p^*$}

It is not  possible to find a closed-form update for $\rho_p^*$, since it is part of the rotation matrix argument, and   a Metropolis algorithm should therefore be used.  Special care should be taken in defining the proposal in order to be able to propose a value in [0,1].

\paragraph*{Sampling the missing locations}

It is not possible to derive a closed-form full conditional for a missing $\mathbf{s}_{j,t_{j,i}}$, and a Metropolis algorithm should therefore be used.

\section{Real data application - Additional results} \label{I-sec:commd}

\begin{table}[t!]
\caption{\label{I-tab:anim1}   M1 - Woody: posterior means and CIs of the model parameters (j=1).}
%\scriptsize
  \centering
  \fbox{
\begin{tabular}{c|ccc}
& k=1& k=2 & k=3  \\
 \hline
$\mu_{j,k,1}$   &  0.088  &  0.082  &  -0.147  \\
(CI)  & (-2.536 1.164) & (-0.268 0.752) & (-0.282 -0.007)  \\
$\mu_{j,k,2}$   &  0.151  &  0.011  &  0.148  \\
(CI)  & (-0.381 3.043) & (-0.381 0.31) & (-0.013 0.33)  \\
$\eta_{j,k,1}$   &  -0.001  &  -0.001  &  13.688  \\
(CI)  & (-0.001 -0.001) & (-0.002 -0.001) & (4.141 18.192)  \\
$\eta_{j,k,2}$   &  0  &  0  &  1.444  \\
(CI)  & (0 0) & (0 0) & (-1.857 4.694)  \\
$\tau_{j,k}$   &  0.151  &  0.281  &  0.128  \\
(CI)  & (0 0.992) & (0.11 0.589) & (0.102 0.159)  \\
$\rho_{j,k}$   &  0.984  &  0.965  &  0.008  \\
(CI)  & (0.948 1) & (0.941 0.984) & (0.005 0.01)  \\
$\boldsymbol{\Sigma}_{j,k,1,1}$   &  0  &  0.022  &  0.201  \\
(CI)  & (0 0) & (0.02 0.025) & (0.178 0.23)  \\
$\boldsymbol{\Sigma}_{j,k,1,2}$   &  0  &  0.001  &  -0.037  \\
(CI)  & (0 0) & (0 0.001) & (-0.053 -0.025)  \\
$\boldsymbol{\Sigma}_{j,k,2,2}$   &  0  &  0.016  &  0.29  \\
(CI)  & (0 0) & (0.014 0.018) & (0.251 0.335)  \\
${\pi}_{j,1,k}$   &  0.772  &  0.197  &  0.03  \\
(CI)  & (0.755 0.789) & (0.18 0.215) & (0.02 0.042)  \\
${\pi}_{j,2,k}$   &  0.354  &  0.49  &  0.156  \\
(CI)  & (0.325 0.384) & (0.442 0.539) & (0.123 0.19)  \\
${\pi}_{j,3,k}$   &  0.149  &  0.388  &  0.462  \\
(CI)  & (0.116 0.187) & (0.318 0.467) & (0.381 0.534)  \\
$n_{j,k}$   &  2745  &  1553  &  502  \\
\end{tabular}
}
\end{table}

\begin{table}[t!]
   \caption{\label{I-tab:anim2}   M1 - Sherlock: posterior means and CIs of the model parameters (j=2).}
%\scriptsize
  \centering
  \fbox{
\begin{tabular}{c|ccccc}
 & k=1& k=2 & k=3  \\
 \hline
 $\mu_{j,k,1}$   &  0.34  &  0.056  &  -0.147  \\
 (CI)  & (-0.265 1.207) & (-0.289 0.852) & (-0.282 -0.007)  \\
 $\mu_{j,k,2}$   &  -0.03  &  0.069  &  0.147  \\
 (CI)  & (-0.381 0.368) & (-0.381 0.344) & (-0.014 0.33)  \\
 $\eta_{j,k,1}$   &  -0.012  &  -0.008  &  14.055  \\
 (CI)  & (-0.018 -0.007) & (-0.016 -0.001) & (10.181 18.192)  \\
 $\eta_{j,k,2}$   &  0.002  &  0.001  &  1.551  \\
 (CI)  & (-0.001 0.004) & (-0.001 0.004) & (-1.773 4.827)  \\
 $\tau_{j,k}$   &  0  &  0.169  &  0.204  \\
 (CI)  & (0 0) & (0 0.983) & (0.129 0.277)  \\
 $\rho_{j,k}$   &  0.139  &  0.972  &  0.007  \\
 (CI)  & (0.086 0.233) & (0.942 1) & (0.005 0.01)  \\
 $\boldsymbol{\Sigma}_{j,k,1,1}$   &  0  &  0.022  &  0.204  \\
 (CI)  & (0 0) & (0.02 0.025) & (0.18 0.24)  \\
 $\boldsymbol{\Sigma}_{j,k,1,2}$   &  0  &  0.001  &  -0.038  \\
 (CI)  & (0 0) & (0 0.001) & (-0.056 -0.024)  \\
 $\boldsymbol{\Sigma}_{j,k,2,2}$   &  0  &  0.016  &  0.296  \\
 (CI)  & (0 0) & (0.014 0.018) & (0.256 0.381)  \\
 ${\pi}_{j,1,k}$   &  0.781  &  0.179  &  0.039  \\
 (CI)  & (0.766 0.796) & (0.162 0.196) & (0.028 0.052)  \\
 ${\pi}_{j,2,k}$   &  0.374  &  0.521  &  0.104  \\
 (CI)  & (0.346 0.402) & (0.49 0.553) & (0.083 0.128)  \\
 ${\pi}_{j,3,k}$   &  0.213  &  0.385  &  0.402  \\
 (CI)  & (0.169 0.261) & (0.295 0.492) & (0.301 0.485)  \\
 $n_{j,k}$   &  2927  &  1473  &  400  \\
\end{tabular}
}
\end{table}

\begin{table}[t!]
  \caption{\label{I-tab:anim3}   M1 - Alvin: posterior means and CIs of the model parameters (j=3).}
%\scriptsize
  \centering
  \fbox{
\begin{tabular}{c|ccccccc}
 & k=1& k=2 & k=3& k=4& k=5  \\
 \hline
 $\mu_{j,k,1}$   &  0.575  &  0.921  &  0.168  &  0.625  &  -2.344  \\
 (CI)  & (0.575 0.576) & (0.577 2.02) & (-0.329 1.078) & (0.557 0.835) & (-6.044 -0.793)  \\
 $\mu_{j,k,2}$   &  -0.381  &  0.094  &  0.061  &  -0.225  &  3.651  \\
 (CI)  & (-0.381 -0.38) & (-0.186 0.987) & (-0.381 1.014) & (-0.381 0.122) & (1.556 8.708)  \\
 $\eta_{j,k,1}$   &  0.144  &  0.014  &  0.012  &  0.866  &  1.133  \\
 (CI)  & (-0.013 -0.001) & (-0.015 -0.001) & (-0.015 -0.001) & (-0.016 14.445) & (-0.016 14.947)  \\
 $\eta_{j,k,2}$   &  0.025  &  0.003  &  0.001  &  0.128  &  0.097  \\
 (CI)  & (0 0.003) & (0 0.003) & (0 0.003) & (-0.001 2.61) & (-0.002 2.252)  \\
 $\tau_{j,k}$   &  0.991  &  0.258  &  0.125  &  0.435  &  0.663  \\
 (CI)  & (0.984 0.997) & (0.024 0.994) & (0 0.991) & (0.224 0.585) & (0.222 0.995)  \\
 $\rho_{j,k}$   &  0.007  &  0.454  &  0.617  &  0.022  &  0.029  \\
 (CI)  & (0 0.01) & (0.005 0.97) & (0.005 1) & (0.004 0.142) & (0 0.233)  \\
 $\boldsymbol{\Sigma}_{j,k,1,1}$   &  0  &  0.022  &  0  &  0.214  &  0.023  \\
 (CI)  & (0 0) & (0.02 0.025) & (0 0) & (0.181 0.326) & (0.02 0.025)  \\
 $\boldsymbol{\Sigma}_{j,k,1,2}$   &  0  &  0.001  &  0  &  -0.046  &  0  \\
 (CI)  & (0 0) & (0 0.001) & (0 0) & (-0.142 -0.025) & (0 0.001)  \\
 $\boldsymbol{\Sigma}_{j,k,2,2}$   &  0  &  0.016  &  0  &  0.32  &  0.017  \\
 (CI)  & (0 0) & (0.014 0.018) & (0 0) & (0.258 0.568) & (0.014 0.018)  \\
 ${\pi}_{j,1,k}$   &  0.89  &  0.096  &  0.009  &  0.004  &  0.001  \\
 (CI)  & (0.878 0.901) & (0.085 0.107) & (0.004 0.015) & (0.002 0.007) & (0 0.002)  \\
 ${\pi}_{j,2,k}$   &  0.318  &  0.421  &  0.21  &  0.05  &  0  \\
 (CI)  & (0.282 0.354) & (0.377 0.463) & (0.18 0.243) & (0.027 0.079) & (0 0)  \\
 ${\pi}_{j,3,k}$   &  0.112  &  0.209  &  0.659  &  0.019  &  0  \\
 (CI)  & (0.086 0.14) & (0.176 0.244) & (0.619 0.699) & (0.007 0.036) & (0 0)  \\
 ${\pi}_{j,4,k}$   &  0.3  &  0.18  &  0.121  &  0.396  &  0  \\
 (CI)  & (0.202 0.408) & (0.083 0.294) & (0.057 0.204) & (0.295 0.511) & (0 0)  \\
 ${\pi}_{j,5,k}$   &  0  &  0  &  0.651  &  0  &  0.305  \\
 (CI)  & (0 0) & (0 0) & (0.223 0.996) & (0 0) & (0 0.753)  \\
 $n_{j,k}$   &  3304  &  792  &  608  &  92  &  4  \\
\end{tabular}
}
\end{table}

\begin{table}[t!]
  \caption{\label{I-tab:anim4}  M1 -  Rosie: posterior means and CIs of the model parameters (j=4).}
%\scriptsize
  \centering
  \fbox{
\begin{tabular}{c|ccc}
 & k=1& k=2 & k=3 \\
 \hline
 $\mu_{j,k,1}$   &  0.123  &  -0.07  &  0.103  \\
 (CI)  & (-1.216 1.095) & (-0.314 0.771) & (-0.981 1.122)  \\
 $\mu_{j,k,2}$   &  0.117  &  0.145  &  0.136  \\
 (CI)  & (-0.381 1.747) & (-0.381 0.383) & (-0.381 1.678)  \\
 $\eta_{j,k,1}$   &  -0.001  &  -0.012  &  0.671  \\
 (CI)  & (-0.001 -0.001) & (-0.018 -0.007) & (-0.016 14.411)  \\
 $\eta_{j,k,2}$   &  0  &  0.001  &  0.088  \\
 (CI)  & (0 0) & (-0.001 0.004) & (-0.001 1.771)  \\
 $\tau_{j,k}$   &  0.159  &  0.189  &  0.219  \\
 (CI)  & (0 0.992) & (0.049 0.499) & (0 0.993)  \\
 $\rho_{j,k}$   &  0.984  &  0.965  &  0.737  \\
 (CI)  & (0.948 1) & (0.941 0.986) & (0.005 1)  \\
 $\boldsymbol{\Sigma}_{j,k,1,1}$   &  0  &  0.022  &  0.211  \\
 (CI)  & (0 0) & (0.02 0.025) & (0.179 0.31)  \\
 $\boldsymbol{\Sigma}_{j,k,1,2}$   &  0  &  0.001  &  -0.04  \\
 (CI)  & (0 0) & (0 0.001) & (-0.083 -0.024)  \\
 $\boldsymbol{\Sigma}_{j,k,2,2}$   &  0  &  0.016  &  0.302  \\
 (CI)  & (0 0) & (0.014 0.018) & (0.253 0.457)  \\
 ${\pi}_{j,1,k}$   &  0.835  &  0.165  &  0.001  \\
 (CI)  & (0.821 0.847) & (0.152 0.178) & (0 0.002)  \\
 ${\pi}_{j,2,k}$   &  0.411  &  0.582  &  0.007  \\
 (CI)  & (0.384 0.439) & (0.553 0.61) & (0.002 0.016)  \\
 ${\pi}_{j,3,k}$   &  0.05  &  0.455  &  0.482  \\
 (CI)  & (0 0.281) & (0.162 0.755) & (0.221 0.727)  \\
 $n_{j,k}$   &  3421  &  1362  &  17  \\
\end{tabular}
}
\end{table}

\begin{table}[t!]
  \caption{ \label{I-tab:anim5}   M1 - Bear: posterior means and CIs of the model parameters (j=5).}
%\scriptsize
\centering
\fbox{
\begin{tabular}{c|ccc}
 & k=1& k=2 & k=3  \\
 \hline
 $\mu_{j,k,1}$   &  0.108  &  -0.049  &  -0.147  \\
 (CI)  & (-1.009 1.091) & (-0.287 0.726) & (-0.283 -0.006)  \\
 $\mu_{j,k,2}$   &  0.095  &  0.114  &  0.148  \\
 (CI)  & (-0.381 1.709) & (-0.381 0.344) & (-0.013 0.332)  \\
 $\eta_{j,k,1}$   &  -0.001  &  -0.008  &  14.031  \\
 (CI)  & (-0.001 -0.001) & (-0.016 -0.001) & (10.04 18.192)  \\
 $\eta_{j,k,2}$   &  0  &  0.001  &  1.529  \\
 (CI)  & (0 0) & (0 0.004) & (-1.812 4.744)  \\
 $\tau_{j,k}$   &  0.151  &  0.236  &  0.128  \\
 (CI)  & (0 0.992) & (0 0.549) & (0.101 0.161)  \\
 $\rho_{j,k}$   &  0.979  &  0.969  &  0.007  \\
 (CI)  & (0.947 1) & (0.941 1) & (0.005 0.01)  \\
 $\boldsymbol{\Sigma}_{j,k,1,1}$   &  0  &  0.022  &  0.199  \\
 (CI)  & (0 0) & (0.02 0.025) & (0.162 0.228)  \\
 $\boldsymbol{\Sigma}_{j,k,1,2}$   &  0  &  0.001  &  -0.037  \\
 (CI)  & (0 0) & (0 0.001) & (-0.054 -0.024)  \\
 $\boldsymbol{\Sigma}_{j,k,2,2}$   &  0  &  0.016  &  0.287  \\
 (CI)  & (0 0) & (0.014 0.018) & (0.244 0.333)  \\
 ${\pi}_{j,1,k}$   &  0.773  &  0.21  &  0.017  \\
 (CI)  & (0.756 0.789) & (0.194 0.227) & (0.009 0.026)  \\
 ${\pi}_{j,2,k}$   &  0.386  &  0.528  &  0.085  \\
 (CI)  & (0.359 0.415) & (0.492 0.565) & (0.065 0.107)  \\
 ${\pi}_{j,3,k}$   &  0.069  &  0.426  &  0.504  \\
 (CI)  & (0.035 0.109) & (0.331 0.53) & (0.408 0.591)  \\
 $n_{j,k}$   &  2852  &  1624  &  324  \\
\end{tabular}
}
\end{table}

\begin{table}[t!]
  \caption{\label{I-tab:anim6}    M1 -   Lucy: posterior means and CIs of the model parameters (j=6).}
%\scriptsize
  \centering
\fbox{
\begin{tabular}{c|cccc}
 & k=1& k=2 & k=3  \\
 \hline
 $\mu_{j,k,1}$   &  0.134  &  -0.092  &  -0.147  \\
 (CI)  & (-1.006 1.107) & (-0.284 0.576) & (-0.282 -0.006)  \\
 $\mu_{j,k,2}$   &  0.088  &  0.118  &  0.148  \\
 (CI)  & (-0.381 1.747) & (-0.381 0.33) & (-0.013 0.33)  \\
 $\eta_{j,k,1}$   &  -0.001  &  -0.01  &  13.918  \\
 (CI)  & (-0.001 -0.001) & (-0.017 -0.001) & (9.441 18.192)  \\
 $\eta_{j,k,2}$   &  0  &  0.001  &  1.522  \\
 (CI)  & (0 0) & (-0.001 0.004) & (-1.773 4.744)  \\
 $\tau_{j,k}$   &  0.152  &  0.259  &  0.128  \\
 (CI)  & (0 0.991) & (0.107 0.548) & (0.101 0.159)  \\
 $\rho_{j,k}$   &  0.984  &  0.965  &  0.007  \\
 (CI)  & (0.948 1) & (0.941 0.984) & (0.005 0.01)  \\
 $\boldsymbol{\Sigma}_{j,k,1,1}$   &  0  &  0.022  &  0.201  \\
 (CI)  & (0 0) & (0.02 0.025) & (0.177 0.229)  \\
 $\boldsymbol{\Sigma}_{j,k,1,2}$   &  0  &  0.001  &  -0.037  \\
 (CI)  & (0 0) & (0 0.001) & (-0.053 -0.024)  \\
 $\boldsymbol{\Sigma}_{j,k,2,2}$   &  0  &  0.016  &  0.289  \\
 (CI)  & (0 0) & (0.014 0.018) & (0.248 0.333)  \\
 ${\pi}_{j,1,k}$   &  0.729  &  0.24  &  0.031  \\
 (CI)  & (0.71 0.747) & (0.221 0.259) & (0.02 0.043)  \\
 ${\pi}_{j,2,k}$   &  0.34  &  0.552  &  0.108  \\
 (CI)  & (0.315 0.366) & (0.516 0.59) & (0.085 0.133)  \\
 ${\pi}_{j,3,k}$   &  0.157  &  0.375  &  0.467  \\
 (CI)  & (0.12 0.195) & (0.301 0.46) & (0.381 0.542)  \\
 $n_{j,k}$   &  2537  &  1825  &  438  \\
\end{tabular}
}
\end{table}

\begin{table}[t!]
\caption{\label{I-tab:anim1single}  M2 -   Woody: posterior means and CIs of the model parameters (j=1).}
%\scriptsize
  \centering
  \fbox{
\begin{tabular}{c|ccc}
	& k=1& k=2 & k=3  \\
	\hline
$\mu_{j,k,1}$   &  -0.236  &  -0.287  &  -0.164  \\
(CI)  & (-8.941 8.273) & (-0.831 0.221) & (-0.334 0.002)  \\
$\mu_{j,k,2}$   &  -0.161  &  0.02  &  0.222  \\
(CI)  & (-8.892 8.281) & (-0.497 0.507) & (0.009 0.436)  \\
$\eta_{j,k,1}$   &  -0.002  &  -0.008  &  17.502  \\
(CI)  & (-0.002 -0.001) & (-0.012 -0.004) & (12.141 23.362)  \\
$\eta_{j,k,2}$   &  0  &  0.004  &  1.508  \\
(CI)  & (0 0) & (0.001 0.007) & (-2.707 5.842)  \\
$\tau_{j,k}$   &  0.503  &  0.543  &  0.144  \\
(CI)  & (0.021 0.973) & (0.192 0.957) & (0.118 0.175)  \\
$\rho_{j,k}$   &  1  &  0.985  &  0.006  \\
(CI)  & (1 1) & (0.968 0.993) & (0.004 0.008)  \\
$\boldsymbol{\Sigma}_{j,k,1,1}$   &  0  &  0.023  &  0.197  \\
(CI)  & (0 0) & (0.02 0.026) & (0.172 0.227)  \\
$\boldsymbol{\Sigma}_{j,k,1,2}$   &  0  &  0  &  -0.037  \\
(CI)  & (0 0) & (0 0.001) & (-0.054 -0.022)  \\
$\boldsymbol{\Sigma}_{j,k,2,2}$   &  0  &  0.017  &  0.291  \\
(CI)  & (0 0) & (0.014 0.019) & (0.253 0.34)  \\
${\pi}_{j,1,k}$   &  0.77  &  0.2  &  0.03  \\
(CI)  & (0.754 0.787) & (0.183 0.218) & (0.02 0.041)  \\
${\pi}_{j,2,k}$   &  0.354  &  0.493  &  0.153  \\
(CI)  & (0.326 0.383) & (0.444 0.542) & (0.119 0.189)  \\
${\pi}_{j,3,k}$   &  0.15  &  0.388  &  0.462  \\
(CI)  & (0.115 0.187) & (0.32 0.467) & (0.384 0.529)  \\
$n_{j,k}$   &  2742  &  1565  &  493  \\
\end{tabular}
}
\end{table}

\begin{table}[t!]
   \caption{\label{I-tab:anim2single}  M2 -   Sherlock: posterior means and CIs of the model parameters (j=2).}
%\scriptsize
  \centering
  \fbox{
\begin{tabular}{c|ccccc}
 & k=1& k=2 & k=3 & k=4& k=5 \\
 \hline
 $\mu_{j,k,1}$   &  -0.236  &  -0.287  &  -0.164  &  0.165  &  0.593  \\
 (CI)  & (-8.941 8.273) & (-0.831 0.221) & (-0.334 0.002) & (-8.559 8.967) & (0.587 0.6)  \\
 $\mu_{j,k,2}$   &  -0.161  &  0.02  &  0.222  &  -0.189  &  -0.369  \\
 (CI)  & (-8.892 8.281) & (-0.497 0.507) & (0.009 0.436) & (-8.188 8.476) & (-0.374 -0.364)  \\
 $\eta_{j,k,1}$   &  -0.002  &  -0.008  &  17.502  &  -0.051  &  -0.08  \\
 (CI)  & (-0.002 -0.001) & (-0.012 -0.004) & (12.141 23.362) & (-8.867 8.43) & (-8.868 8.416)  \\
 $\eta_{j,k,2}$   &  0  &  0.004  &  1.508  &  0.055  &  0.013  \\
 (CI)  & (0 0) & (0.001 0.007) & (-2.707 5.842) & (-8.746 8.624) & (-8.35 8.601)  \\
 $\tau_{j,k}$   &  0.503  &  0.543  &  0.144  &  0.487  &  0.953  \\
 (CI)  & (0.021 0.973) & (0.192 0.957) & (0.118 0.175) & (0.017 0.975) & (0.93 0.975)  \\
 $\rho_{j,k}$   &  1  &  0.985  &  0.006  &  0.418  &  0  \\
 (CI)  & (1 1) & (0.968 0.993) & (0.004 0.008) & (0 1) & (0 0)  \\
 $\boldsymbol{\Sigma}_{j,k,1,1}$   &  0  &  0.023  &  0.197  &  4.858  &  0.005  \\
 (CI)  & (0 0) & (0.02 0.026) & (0.172 0.227) & (0.146 18.378) & (0.004 0.007)  \\
 $\boldsymbol{\Sigma}_{j,k,1,2}$   &  0  &  0  &  -0.037  &  -1.274  &  0.001  \\
 (CI)  & (0 0) & (0 0.001) & (-0.054 -0.022) & (-5.08 4.609) & (0 0.001)  \\
 $\boldsymbol{\Sigma}_{j,k,2,2}$   &  0  &  0.017  &  0.291  &  3.102  &  0.003  \\
 (CI)  & (0 0) & (0.014 0.019) & (0.253 0.34) & (0.14 13.914) & (0.003 0.004)  \\
 ${\pi}_{j,1,k}$   &  0.781  &  0.179  &  0.04  &  0  &  0  \\
 (CI)  & (0.765 0.797) & (0.162 0.197) & (0.027 0.052) & (0 0.001) & (0 0)  \\
 ${\pi}_{j,2,k}$   &  0.375  &  0.521  &  0.103  &  0  &  0  \\
 (CI)  & (0.346 0.403) & (0.489 0.555) & (0.081 0.126) & (0 0.003) & (0 0.001)  \\
 ${\pi}_{j,3,k}$   &  0.213  &  0.383  &  0.403  &  0  &  0  \\
 (CI)  & (0.168 0.261) & (0.295 0.491) & (0.301 0.486) & (0 0) & (0 0.005)  \\
 ${\pi}_{j,4,k}$   &  0.287  &  0.197  &  0.029  &  0.453  &  0.012  \\
 (CI)  & (0 0.791) & (0 0.725) & (0 0.245) & (0 0.978) & (0 0.15)  \\
 ${\pi}_{j,5,k}$   &  0.342  &  0.183  &  0.027  &  0.001  &  0.428  \\
 (CI)  & (0 0.852) & (0 0.641) & (0 0.231) & (0 0) & (0 0.978)  \\
 $n_{j,k}$   &  2925  &  1475  &  398  &  1  &  1  \\
\end{tabular}
}
\end{table}

\begin{table}[t!]
  \caption{\label{I-tab:anim3single}  M2 -  Alvin: posterior means and CIs of the model parameters (j=3).}
%\scriptsize
  \centering
  \fbox{
\begin{tabular}{c|cccccc}
 & k=1& k=2 & k=3& k=4 & k=5 \\
 \hline
 $\mu_{j,k,1}$   &  -0.236  &  0.593  &  -0.287  &  0.797  &  -1.992  \\
 (CI)  & (-8.941 8.273) & (0.587 0.6) & (-0.831 0.221) & (0.59 1.043) & (-5.598 -0.642)  \\
 $\mu_{j,k,2}$   &  -0.161  &  -0.369  &  0.02  &  -0.189  &  3.423  \\
 (CI)  & (-8.892 8.281) & (-0.374 -0.364) & (-0.497 0.507) & (-0.459 0.057) & (1.675 8.501)  \\
 $\eta_{j,k,1}$   &  -0.002  &  -0.08  &  -0.008  &  0.02  &  -0.034  \\
 (CI)  & (-0.002 -0.001) & (-8.868 8.416) & (-0.012 -0.004) & (-8.51 8.206) & (-8.861 8.627)  \\
 $\eta_{j,k,2}$   &  0  &  0.013  &  0.004  &  -0.09  &  -0.004  \\
 (CI)  & (0 0) & (-8.35 8.601) & (0.001 0.007) & (-8.92 8.665) & (-8.818 9.066)  \\
 $\tau_{j,k}$   &  0.503  &  0.953  &  0.543  &  0.399  &  0.64  \\
 (CI)  & (0.021 0.973) & (0.93 0.975) & (0.192 0.957) & (0.287 0.513) & (0.232 0.978)  \\
 $\rho_{j,k}$   &  1  &  0  &  0.985  &  0.001  &  0.002  \\
 (CI)  & (1 1) & (0 0) & (0.968 0.993) & (0 0.004) & (0 0.026)  \\
 $\boldsymbol{\Sigma}_{j,k,1,1}$   &  0  &  0.005  &  0.023  &  0.171  &  0.236  \\
 (CI)  & (0 0) & (0.004 0.007) & (0.02 0.026) & (0.124 0.238) & (0.074 0.692)  \\
 $\boldsymbol{\Sigma}_{j,k,1,2}$   &  0  &  0.001  &  0  &  0.028  &  -0.077  \\
 (CI)  & (0 0) & (0 0.001) & (0 0.001) & (-0.013 0.076) & (-0.408 0.085)  \\
 $\boldsymbol{\Sigma}_{j,k,2,2}$   &  0  &  0.003  &  0.017  &  0.257  &  0.252  \\
 (CI)  & (0 0) & (0.003 0.004) & (0.014 0.019) & (0.178 0.37) & (0.083 0.802)  \\
 ${\pi}_{j,1,k}$   &  0.861  &  0.081  &  0.047  &  0.01  &  0.001  \\
 (CI)  & (0.849 0.874) & (0.068 0.096) & (0.032 0.06) & (0.006 0.016) & (0 0.002)  \\
 ${\pi}_{j,2,k}$   &  0.509  &  0.418  &  0.057  &  0.015  &  0  \\
 (CI)  & (0.46 0.557) & (0.369 0.468) & (0.03 0.086) & (0 0.036) & (0 0.004)  \\
 ${\pi}_{j,3,k}$   &  0.159  &  0.341  &  0.415  &  0.076  &  0.008  \\
 (CI)  & (0.113 0.212) & (0.274 0.409) & (0.343 0.484) & (0.034 0.128) & (0 0.02)  \\
 ${\pi}_{j,4,k}$   &  0.11  &  0.386  &  0.15  &  0.352  &  0  \\
 (CI)  & (0.046 0.186) & (0.282 0.491) & (0.059 0.258) & (0.254 0.46) & (0 0.003)  \\
 ${\pi}_{j,5,k}$   &  0.381  &  0.003  &  0.173  &  0.173  &  0.249  \\
 (CI)  & (0.114 0.706) & (0 0.038) & (0.005 0.483) & (0.023 0.45) & (0 0.628)  \\
 $n_{j,k}$   &  3537  &  818  &  342  &  96  &  7  \\
\end{tabular}
}
\end{table}

\begin{table}[t!]
  \caption{\label{I-tab:anim4single}  M2 -  Rosie: posterior means and CIs of the model parameters (j=4).}
%\scriptsize
  \centering
  \fbox{
\begin{tabular}{c|ccccc}
 & k=1& k=2 & k=3  \\
 \hline
 $\mu_{j,k,1}$   &  -0.236  &  -0.287  &  -0.164  \\
 (CI)  & (-8.941 8.273) & (-0.831 0.221) & (-0.334 0.002)  \\
 $\mu_{j,k,2}$   &  -0.161  &  0.02  &  0.222  \\
 (CI)  & (-8.892 8.281) & (-0.497 0.507) & (0.009 0.436)  \\
 $\eta_{j,k,1}$   &  -0.002  &  -0.008  &  17.502  \\
 (CI)  & (-0.002 -0.001) & (-0.012 -0.004) & (12.141 23.362)  \\
 $\eta_{j,k,2}$   &  0  &  0.004  &  1.508  \\
 (CI)  & (0 0) & (0.001 0.007) & (-2.707 5.842)  \\
 $\tau_{j,k}$   &  0.503  &  0.543  &  0.144  \\
 (CI)  & (0.021 0.973) & (0.192 0.957) & (0.118 0.175)  \\
 $\rho_{j,k}$   &  1  &  0.985  &  0.006  \\
 (CI)  & (1 1) & (0.968 0.993) & (0.004 0.008)  \\
 $\boldsymbol{\Sigma}_{j,k,1,1}$   &  0  &  0.023  &  0.197  \\
 (CI)  & (0 0) & (0.02 0.026) & (0.172 0.227)  \\
 $\boldsymbol{\Sigma}_{j,k,1,2}$   &  0  &  0  &  -0.037  \\
 (CI)  & (0 0) & (0 0.001) & (-0.054 -0.022)  \\
 $\boldsymbol{\Sigma}_{j,k,2,2}$   &  0  &  0.017  &  0.291  \\
 (CI)  & (0 0) & (0.014 0.019) & (0.253 0.34)  \\
 ${\pi}_{j,1,k}$   &  0.834  &  0.165  &  0.001  \\
 (CI)  & (0.82 0.848) & (0.152 0.179) & (0 0.003)  \\
 ${\pi}_{j,2,k}$   &  0.411  &  0.583  &  0.006  \\
 (CI)  & (0.383 0.439) & (0.554 0.61) & (0.002 0.014)  \\
 ${\pi}_{j,3,k}$   &  0.17  &  0.313  &  0.506  \\
 (CI)  & (0.018 0.374) & (0.111 0.563) & (0.254 0.734)  \\
 $n_{j,k}$   &  3419  &  1363  &  18  \\
\end{tabular}
}
\end{table}

\begin{table}[t!]
  \caption{ \label{I-tab:anim5single} M2 -  Bear: posterior means and CIs of the model parameters (j=5).}
%\scriptsize
\centering
\fbox{
\begin{tabular}{c|ccc}
 & k=1& k=2 & k=3  \\
 \hline
 $\mu_{j,k,1}$   &  -0.236  &  -0.287  &  -0.164  \\
 (CI)  & (-8.941 8.273) & (-0.831 0.221) & (-0.334 0.002)  \\
 $\mu_{j,k,2}$   &  -0.161  &  0.02  &  0.222  \\
 (CI)  & (-8.892 8.281) & (-0.497 0.507) & (0.009 0.436)  \\
 $\eta_{j,k,1}$   &  -0.002  &  -0.008  &  17.502  \\
 (CI)  & (-0.002 -0.001) & (-0.012 -0.004) & (12.141 23.362)  \\
 $\eta_{j,k,2}$   &  0  &  0.004  &  1.508  \\
 (CI)  & (0 0) & (0.001 0.007) & (-2.707 5.842)  \\
 $\tau_{j,k}$   &  0.503  &  0.543  &  0.144  \\
 (CI)  & (0.021 0.973) & (0.192 0.957) & (0.118 0.175)  \\
 $\rho_{j,k}$   &  1  &  0.985  &  0.006  \\
 (CI)  & (1 1) & (0.968 0.993) & (0.004 0.008)  \\
 $\boldsymbol{\Sigma}_{j,k,1,1}$   &  0  &  0.023  &  0.197  \\
 (CI)  & (0 0) & (0.02 0.026) & (0.172 0.227)  \\
 $\boldsymbol{\Sigma}_{j,k,1,2}$   &  0  &  0  &  -0.037  \\
 (CI)  & (0 0) & (0 0.001) & (-0.054 -0.022)  \\
 $\boldsymbol{\Sigma}_{j,k,2,2}$   &  0  &  0.017  &  0.291  \\
 (CI)  & (0 0) & (0.014 0.019) & (0.253 0.34)  \\
 ${\pi}_{j,1,k}$   &  0.773  &  0.211  &  0.016  \\
 (CI)  & (0.756 0.789) & (0.195 0.228) & (0.008 0.025)  \\
 ${\pi}_{j,2,k}$   &  0.384  &  0.532  &  0.084  \\
 (CI)  & (0.354 0.414) & (0.494 0.568) & (0.065 0.105)  \\
 ${\pi}_{j,3,k}$   &  0.071  &  0.429  &  0.499  \\
 (CI)  & (0.038 0.11) & (0.336 0.533) & (0.398 0.585)  \\
 $n_{j,k}$   &  2850  &  1639  &  311  \\
\end{tabular}
}
\end{table}

\begin{table}[t!]
  \caption{\label{I-tab:anim6single}  Lucy: posterior means and CIs of the model parameters (j=6).}
%\scriptsize
  \centering
\fbox{
\begin{tabular}{c|cccc}
 & k=1& k=2 & k=3& k=4  \\
 \hline
 $\mu_{j,k,1}$   &  -0.236  &  -0.287  &  -0.05  &  -1.992  \\
 (CI)  & (-8.941 8.273) & (-0.831 0.221) & (-0.414 0.318) & (-5.598 -0.642)  \\
 $\mu_{j,k,2}$   &  -0.161  &  0.02  &  0.244  &  3.423  \\
 (CI)  & (-8.892 8.281) & (-0.497 0.507) & (-0.147 0.679) & (1.675 8.501)  \\
 $\eta_{j,k,1}$   &  -0.002  &  -0.008  &  6.462  &  -0.034  \\
 (CI)  & (-0.002 -0.001) & (-0.012 -0.004) & (-5.942 15.635) & (-8.861 8.627)  \\
 $\eta_{j,k,2}$   &  0  &  0.004  &  1.203  &  -0.004  \\
 (CI)  & (0 0) & (0.001 0.007) & (-5.868 7.427) & (-8.818 9.066)  \\
 $\tau_{j,k}$   &  0.503  &  0.543  &  0.114  &  0.64  \\
 (CI)  & (0.021 0.973) & (0.192 0.957) & (0.077 0.162) & (0.232 0.978)  \\
 $\rho_{j,k}$   &  1  &  0.985  &  0.011  &  0.002  \\
 (CI)  & (1 1) & (0.968 0.993) & (0 0.067) & (0 0.026)  \\
 $\boldsymbol{\Sigma}_{j,k,1,1}$   &  0  &  0.023  &  0.196  &  0.236  \\
 (CI)  & (0 0) & (0.02 0.026) & (0.16 0.25) & (0.074 0.692)  \\
 $\boldsymbol{\Sigma}_{j,k,1,2}$   &  0  &  0  &  -0.031  &  -0.077  \\
 (CI)  & (0 0) & (0 0.001) & (-0.059 -0.008) & (-0.408 0.085)  \\
 $\boldsymbol{\Sigma}_{j,k,2,2}$   &  0  &  0.017  &  0.263  &  0.252  \\
 (CI)  & (0 0) & (0.014 0.019) & (0.209 0.34) & (0.083 0.802)  \\
 ${\pi}_{j,1,k}$   &  0.729  &  0.24  &  0.031  &  0  \\
 (CI)  & (0.711 0.747) & (0.22 0.26) & (0.019 0.044) & (0 0)  \\
 ${\pi}_{j,2,k}$   &  0.34  &  0.552  &  0.108  &  0  \\
 (CI)  & (0.314 0.368) & (0.511 0.592) & (0.083 0.134) & (0 0)  \\
 ${\pi}_{j,3,k}$   &  0.158  &  0.371  &  0.471  &  0  \\
 (CI)  & (0.122 0.198) & (0.29 0.461) & (0.381 0.551) & (0 0.001)  \\
 ${\pi}_{j,4,k}$   &  0.232  &  0.262  &  0.005  &  0.452  \\
 (CI)  & (0 0.728) & (0 0.776) & (0 0.058) & (0 0.978)  \\
 $n_{j,k}$   &  2535  &  1834  &  430  &  1  \\
\end{tabular}
}
\end{table}

\begin{figure}[t!]
  \centering
  {\subfloat[Woody - $\text{B}_{11}$]{\includegraphics[trim = 0 20 0 20,scale=0.27]{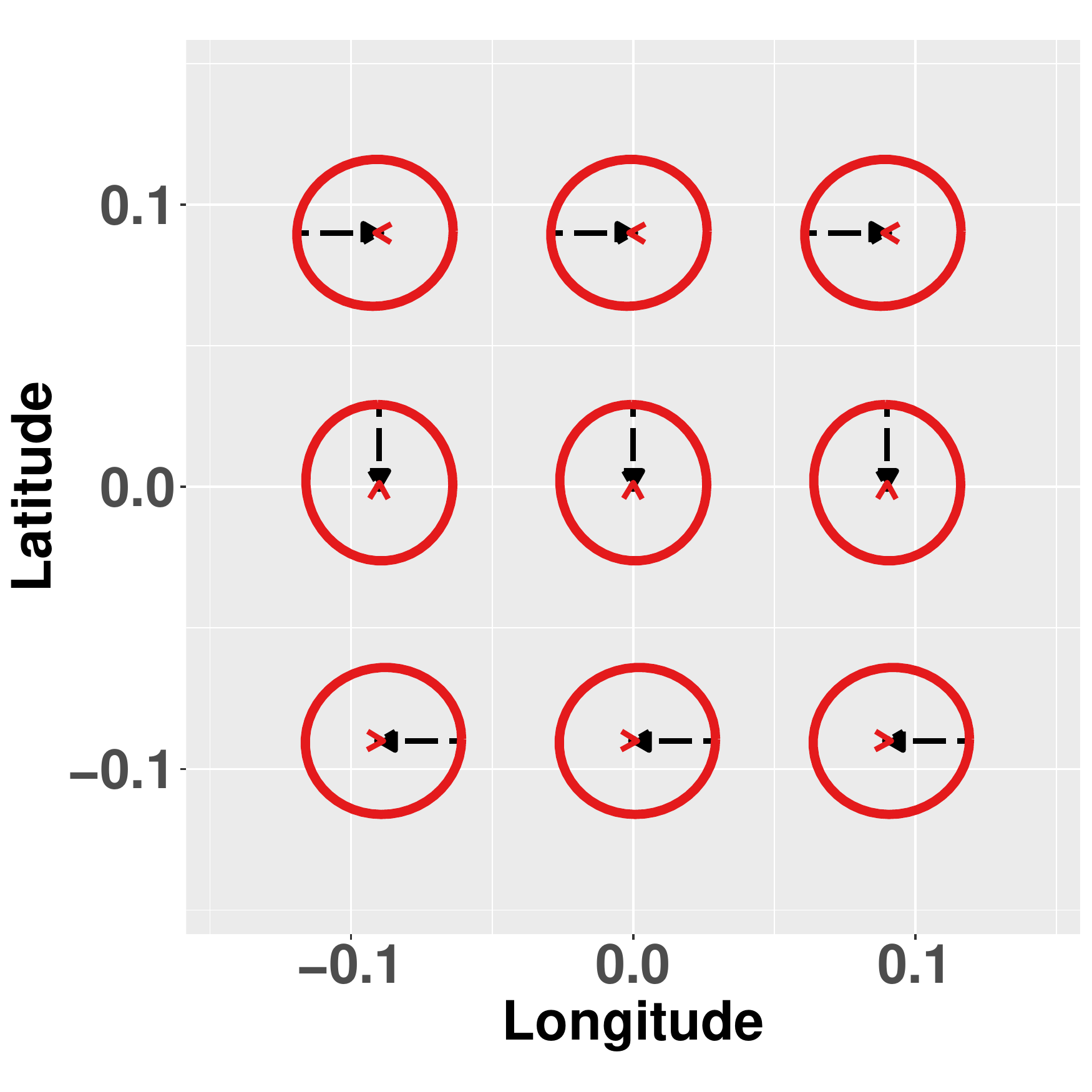}}}
  {\subfloat[Woody - $\text{B}_{12}$]{\includegraphics[trim = 0 20 0 20,scale=0.27]{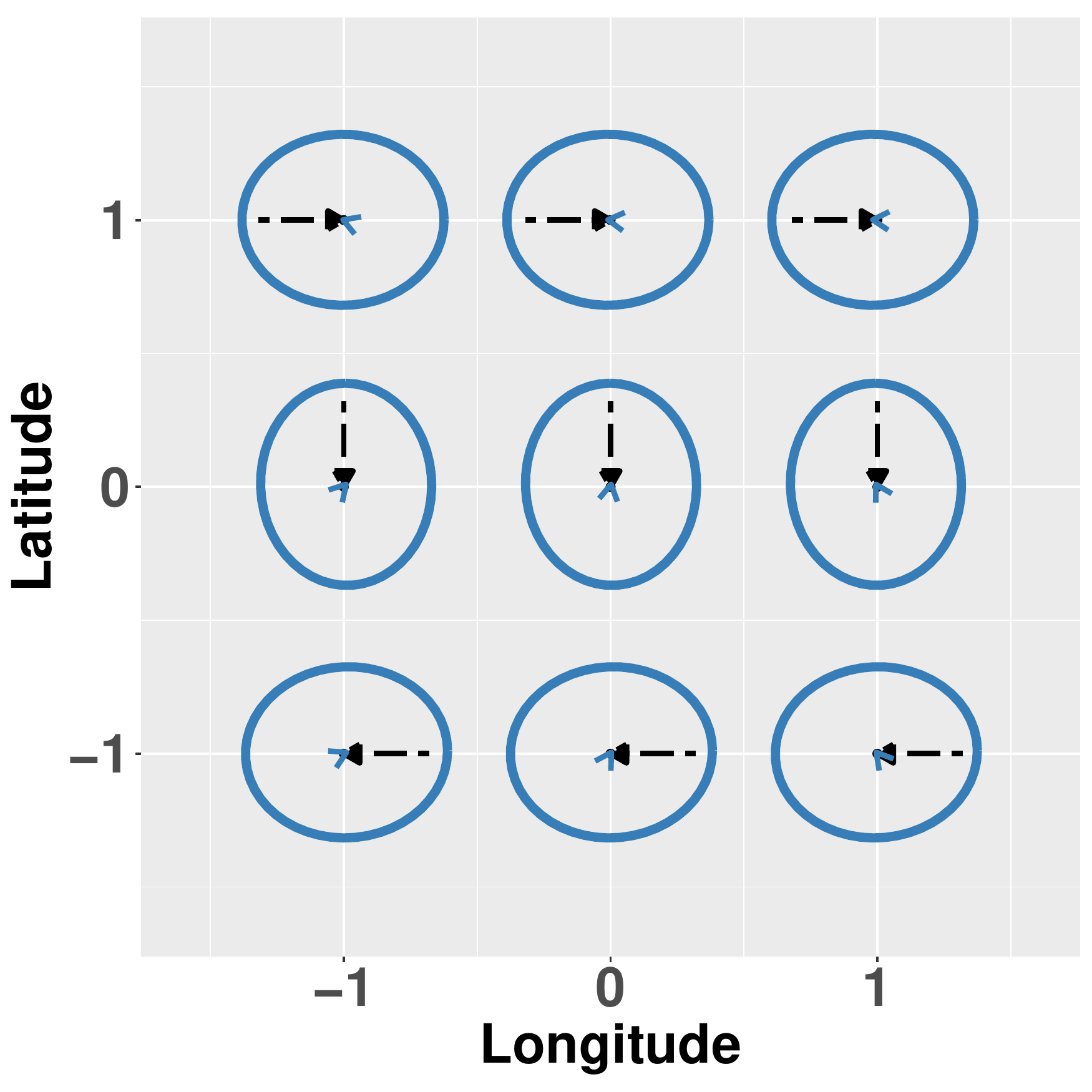}}}
  {\subfloat[Woody - $\text{B}_{13}$]{\includegraphics[trim = 0 20 0 20,scale=0.27]{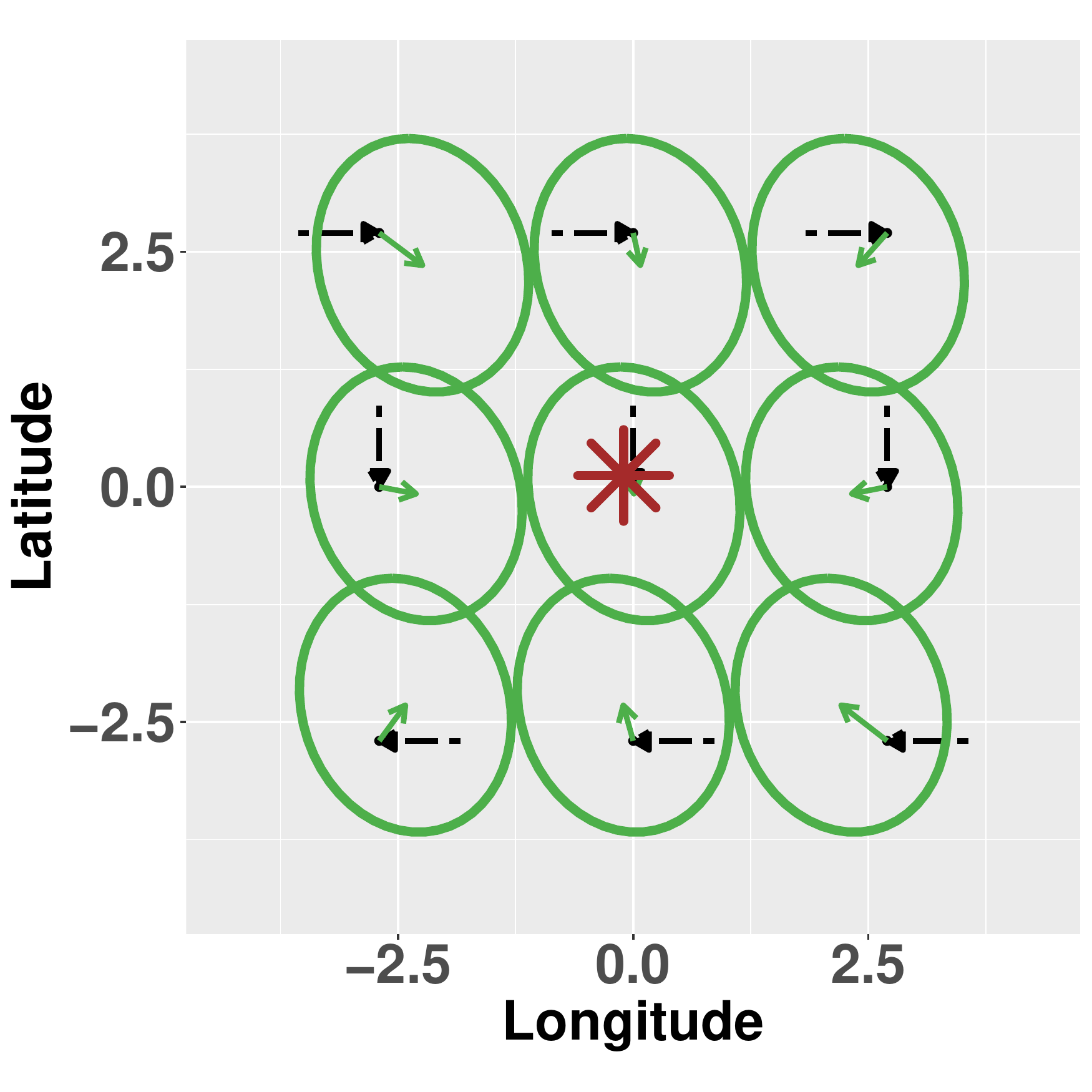}}}\\
  {\subfloat[Sherlock - $\text{B}_{21}$]{\includegraphics[trim = 0 20 0 20,scale=0.27]{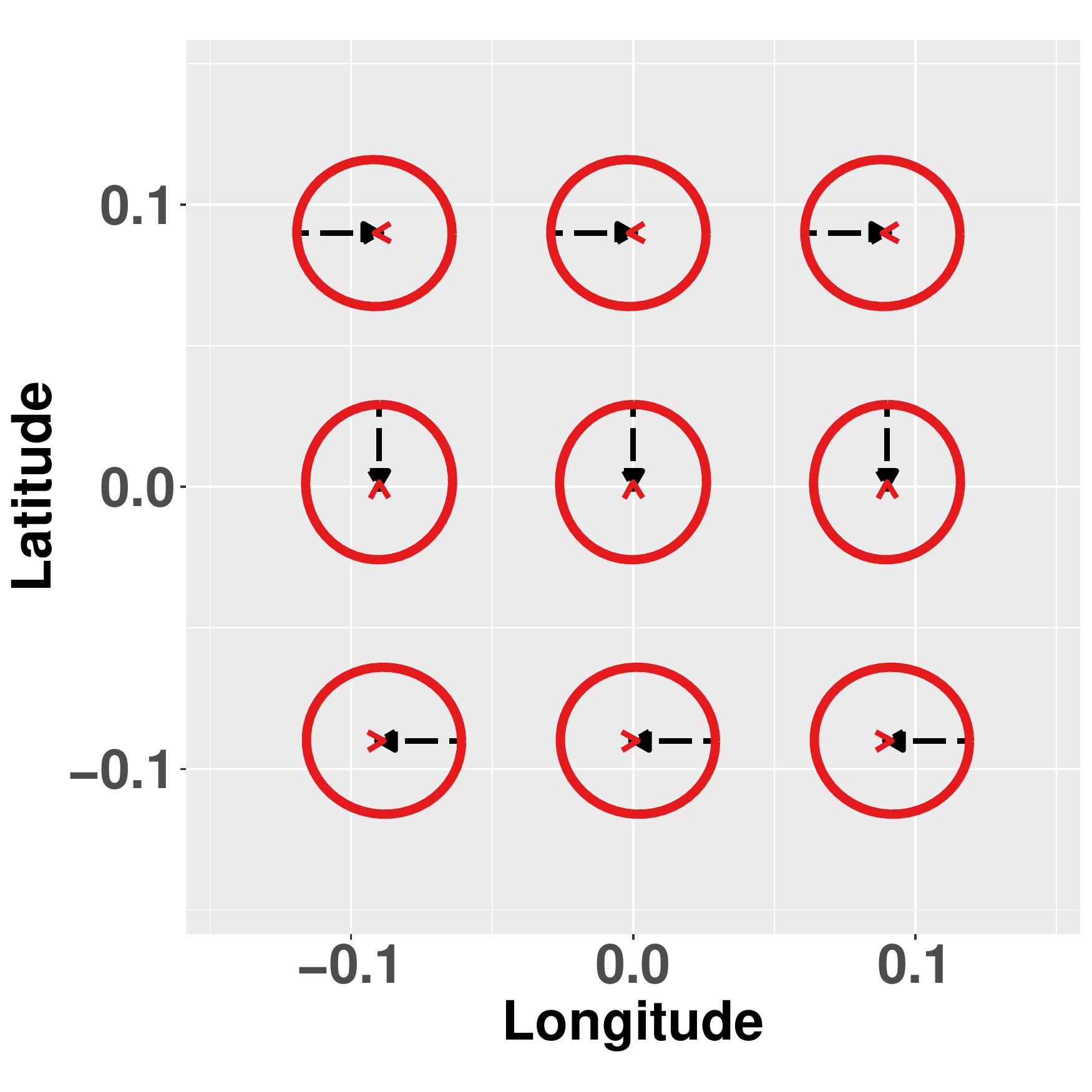}}}
  {\subfloat[Sherlock - $\text{B}_{22}$]{\includegraphics[trim = 0 20 0 20,scale=0.27]{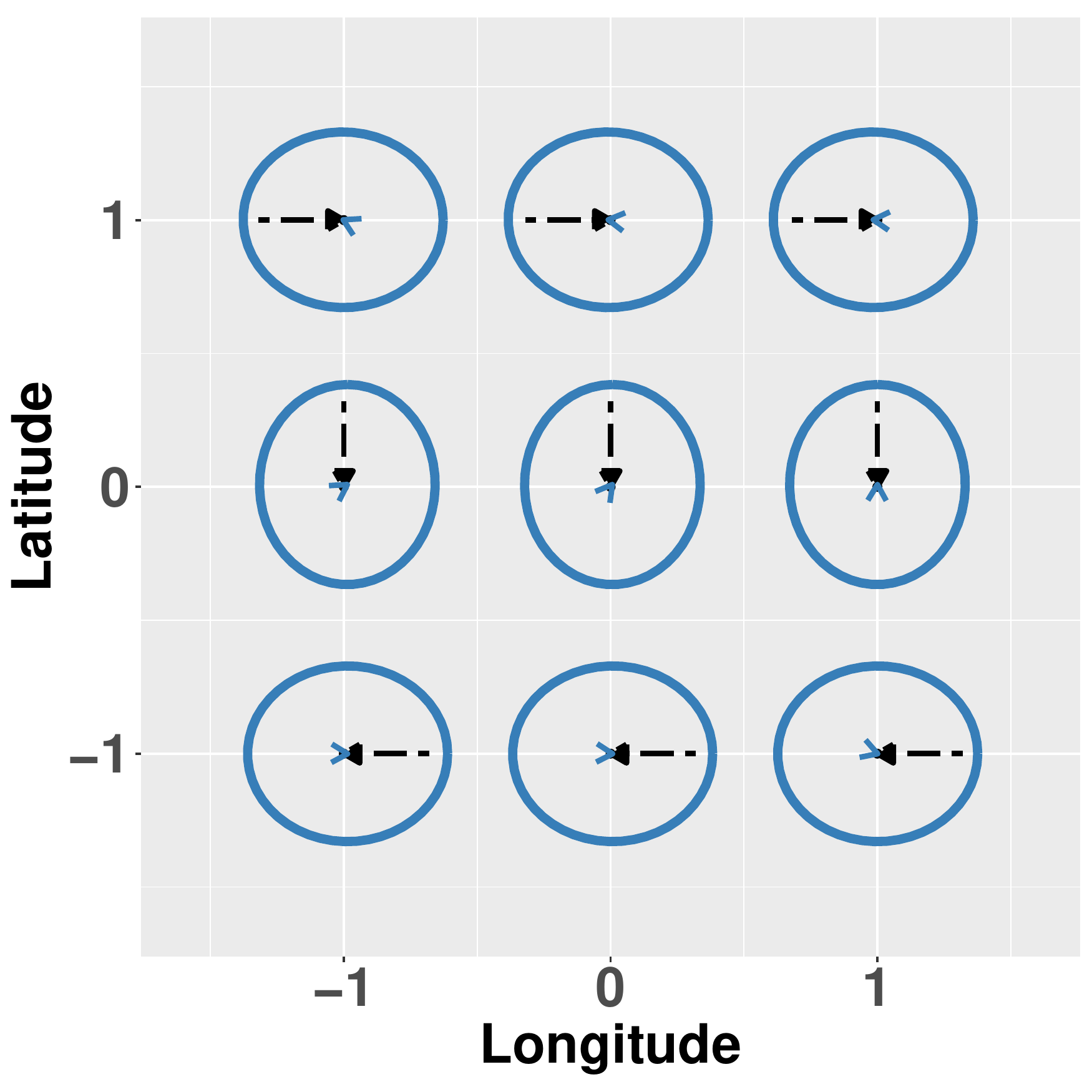}}}
  {\subfloat[Sherlock - $\text{B}_{23}$]{\includegraphics[trim = 0 20 0 20,scale=0.27]{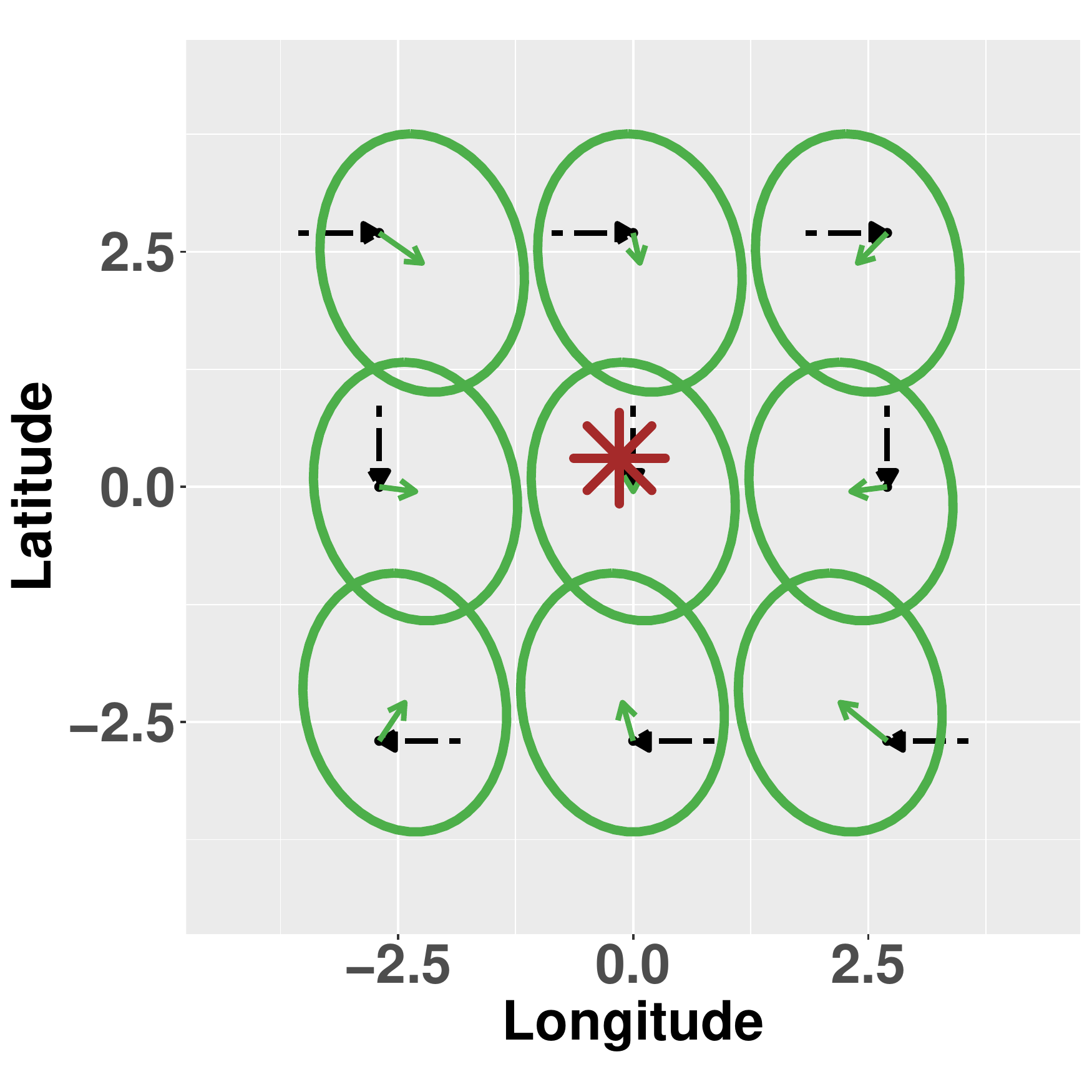}}}\\
	{\subfloat[Alvin - $\text{B}_{31}$]{\includegraphics[trim = 0 20 0 20,scale=0.27]{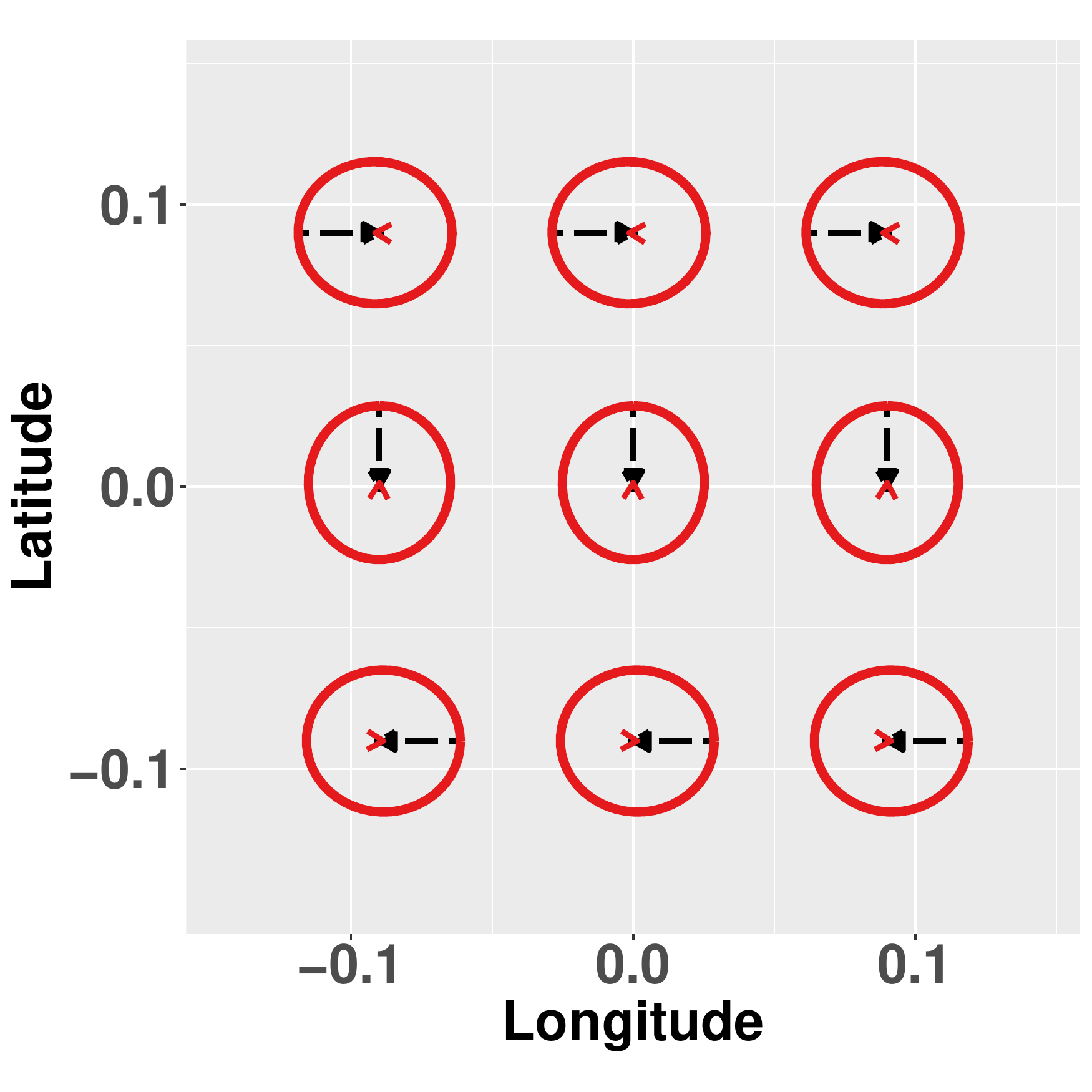}}}
	{\subfloat[Alvin - $\text{B}_{32}$]{\includegraphics[trim = 0 20 0 20,scale=0.27]{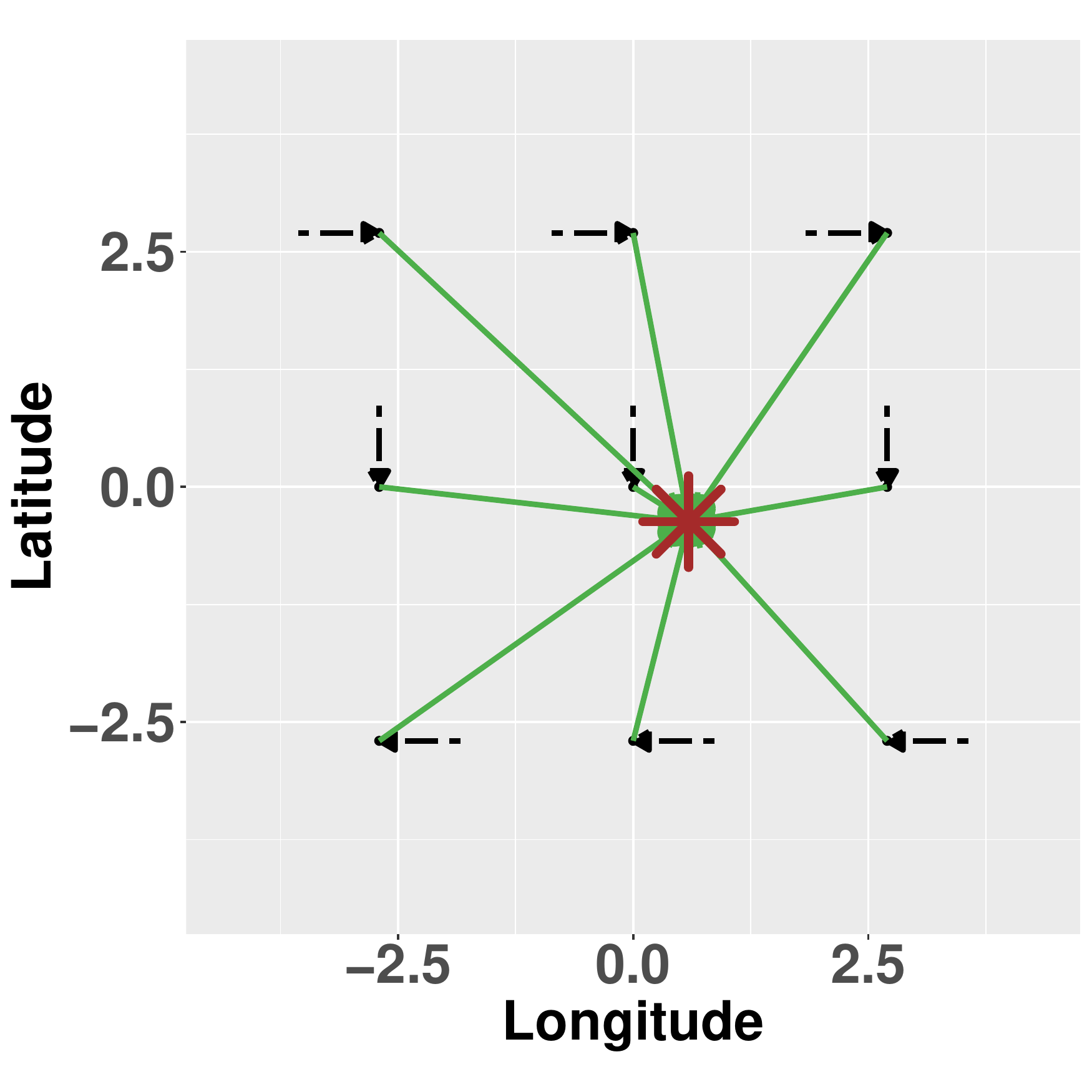}}}
	{\subfloat[Alvin - $\text{B}_{33}$]{\includegraphics[trim = 0 20 0 20,scale=0.27]{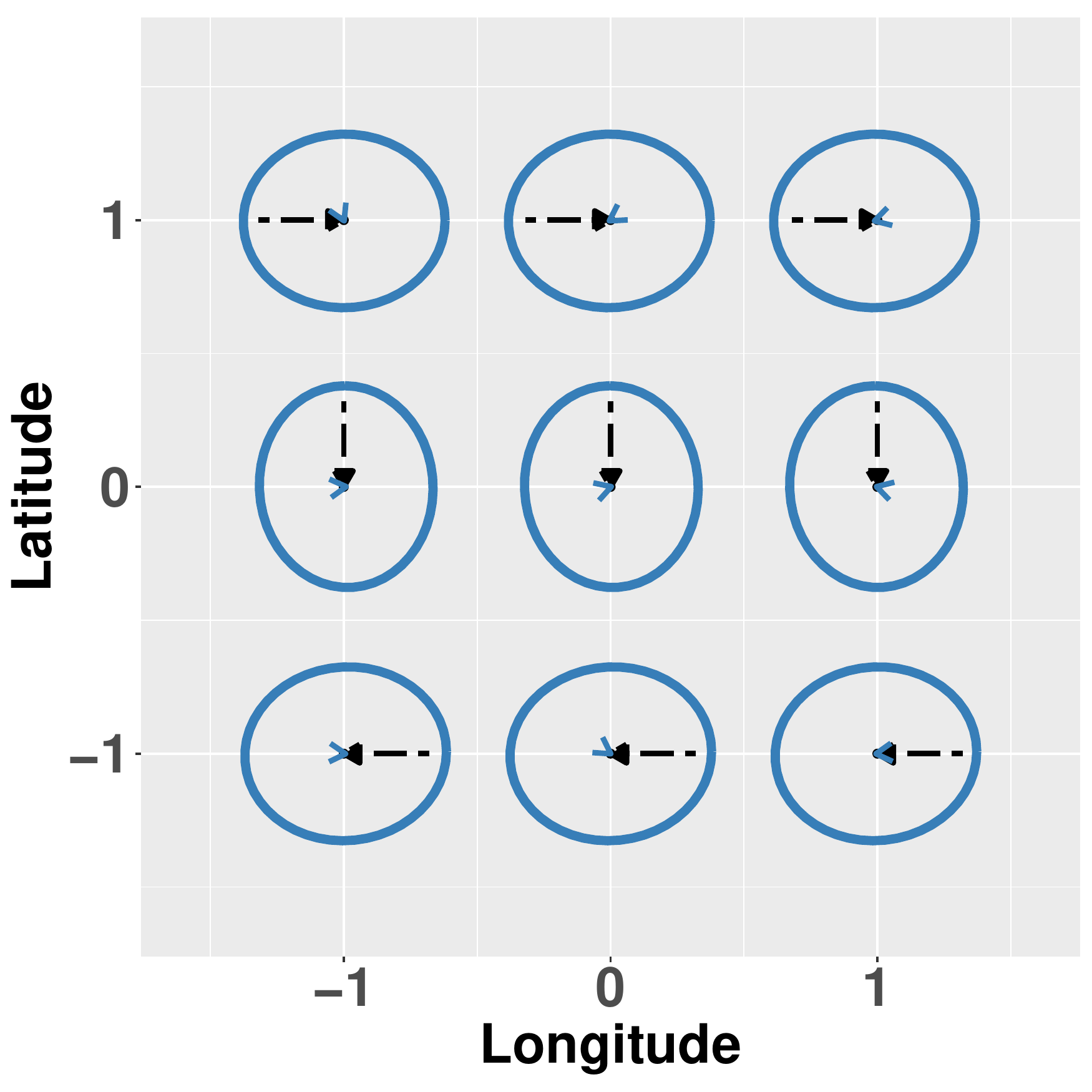}}}
  \caption{M2 -  Graphical representation of the conditional distribution of $\mathbf{s}_{j,t_{j,i+1}}$ for the first  three dogs,  for different possible values of $\mathbf{s}_{j,t_{j,i}}$ and  previous directions.
	The images has been obtained by using the posterior values that maximize the data likelihood of each animal, given the representative clusterization $\hat{z}_{j, t_{j,i}}$.
	The dashed arrow represents the movement between  $\mathbf{s}_{j,t_{j,i-1}}$ and $\mathbf{s}_{j,t_{j,i}}$. The solid arrow is $\protect\vec{\mathbf{F}}_{j,t_{j,i}}$,  while the ellipse is an area containing  95\% of the probability mass of  the  conditional  distribution of $\mathbf{s}_{j,t_{j,i+1}}$. The asterisk represents the  attractor,  and it is only shown for behaviors that have posterior values of
	${\rho}_{j,k}<0.9$ and ${\tau}_{j,k}>0.1$. }\label{I-newfig:a2_1Single}
\end{figure}

\begin{figure}[t!]
  \centering
  {\subfloat[Rosie - $\text{B}_{41}$]{\includegraphics[trim = 0 20 0 20,scale=0.27]{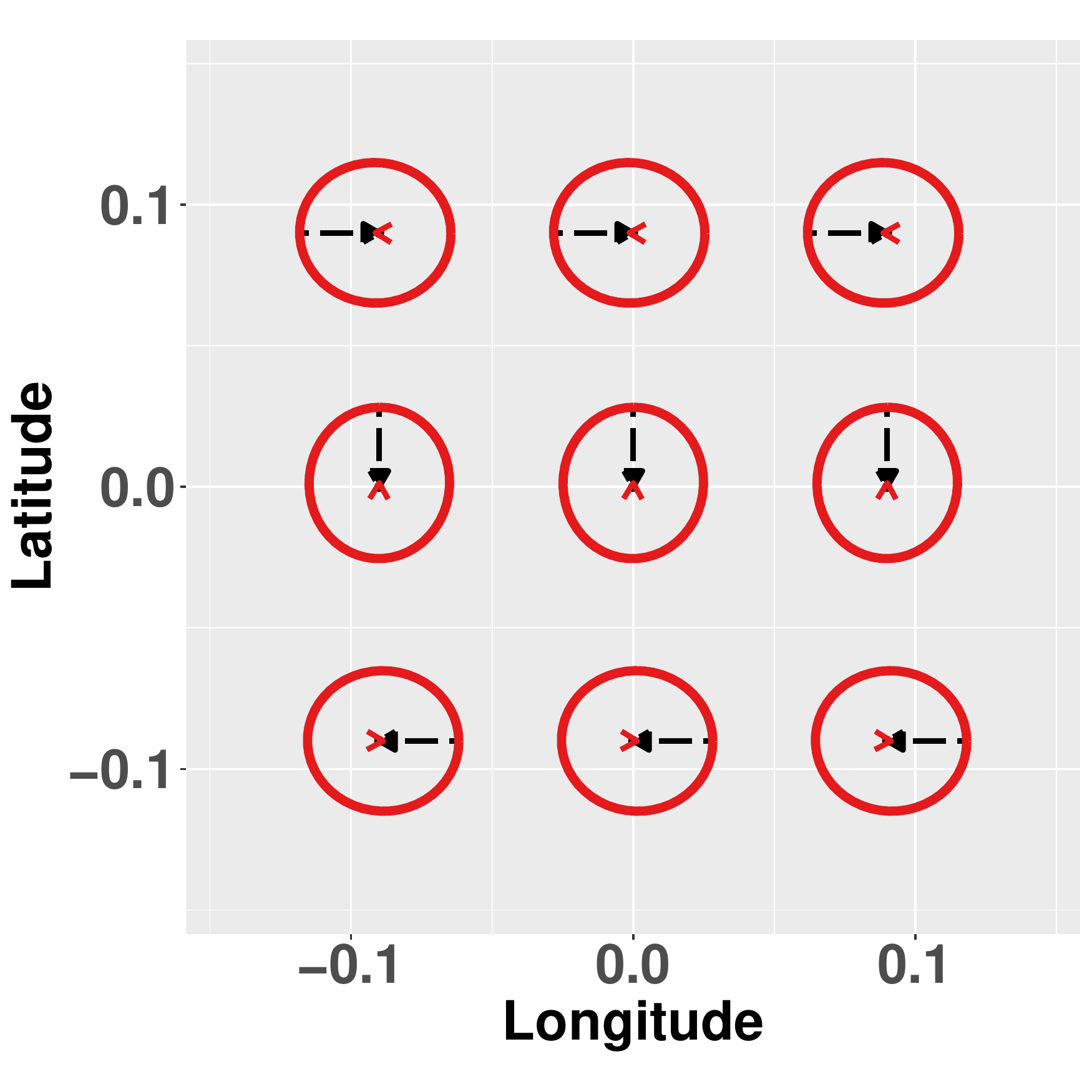}}}
  {\subfloat[Rosie - $\text{B}_{42}$]{\includegraphics[trim = 0 20 0 20,scale=0.27]{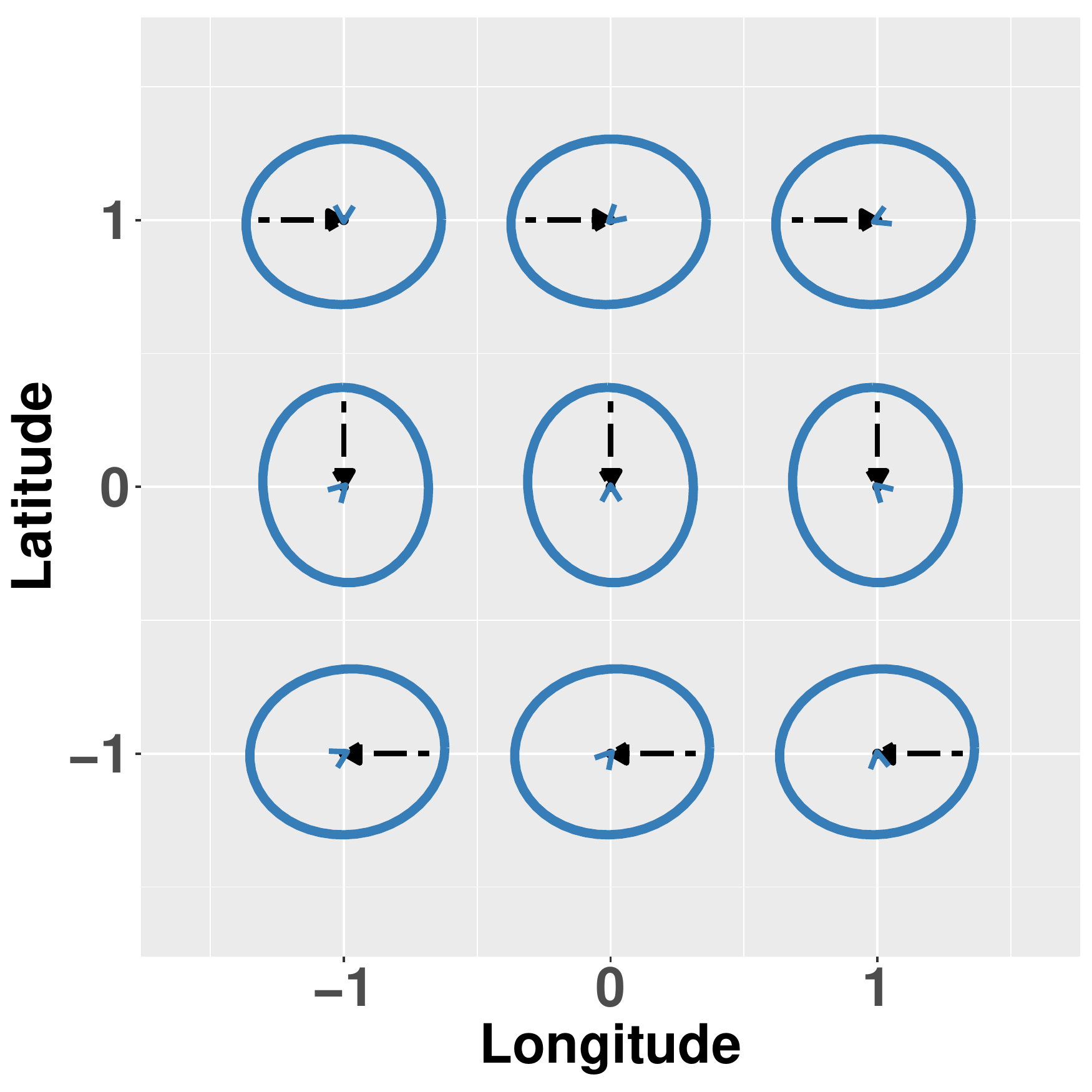}}}\\
  %{\subfloat[Rosie - $\text{B}_{43}$]{\includegraphics[trim = 0 20 0 20,scale=0.27]{Single4_3Mov}}}
	{\subfloat[Bear - $\text{B}_{51}$]{\includegraphics[trim = 0 20 0 20,scale=0.27]{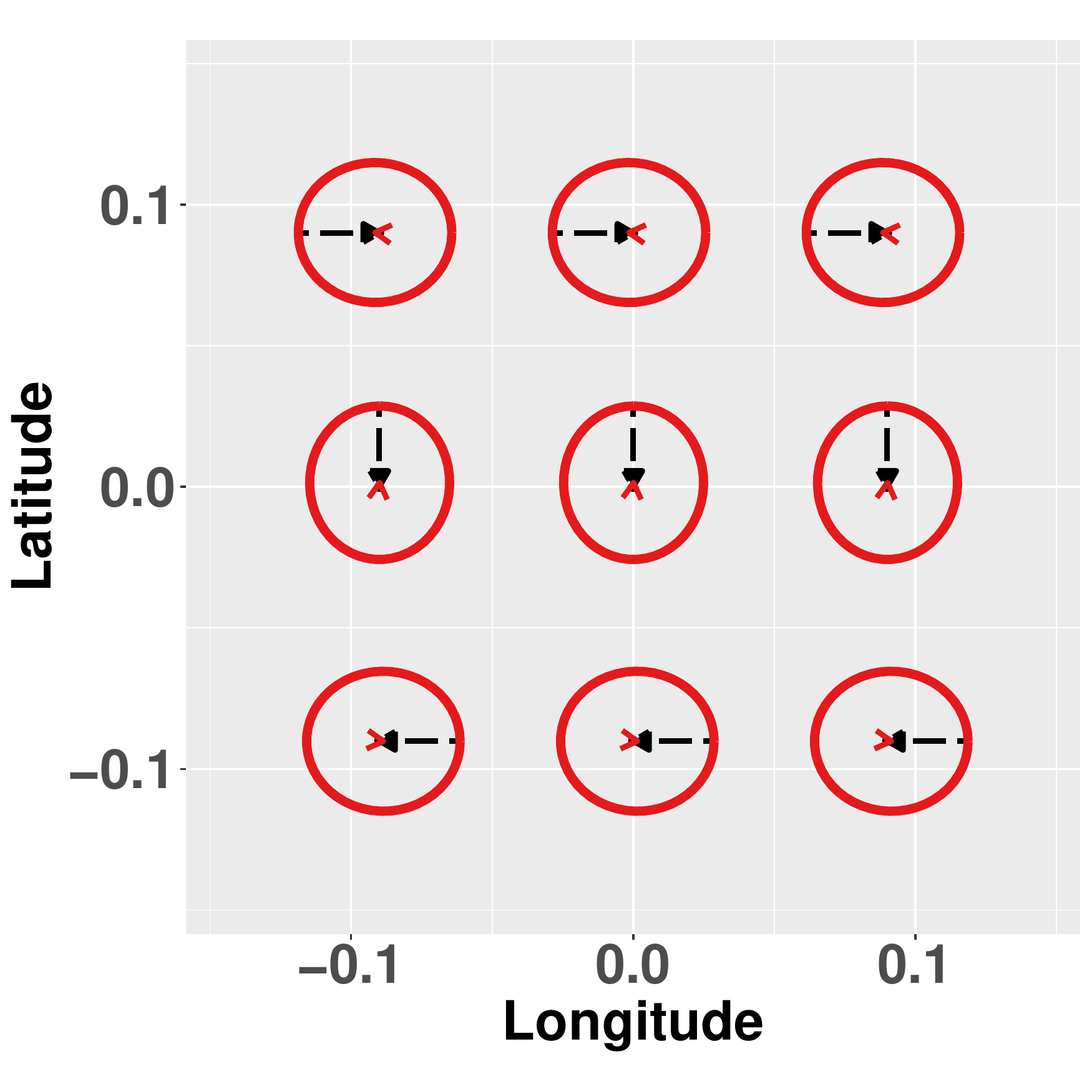}}}
	{\subfloat[Bear - $\text{B}_{52}$]{\includegraphics[trim = 0 20 0 20,scale=0.27]{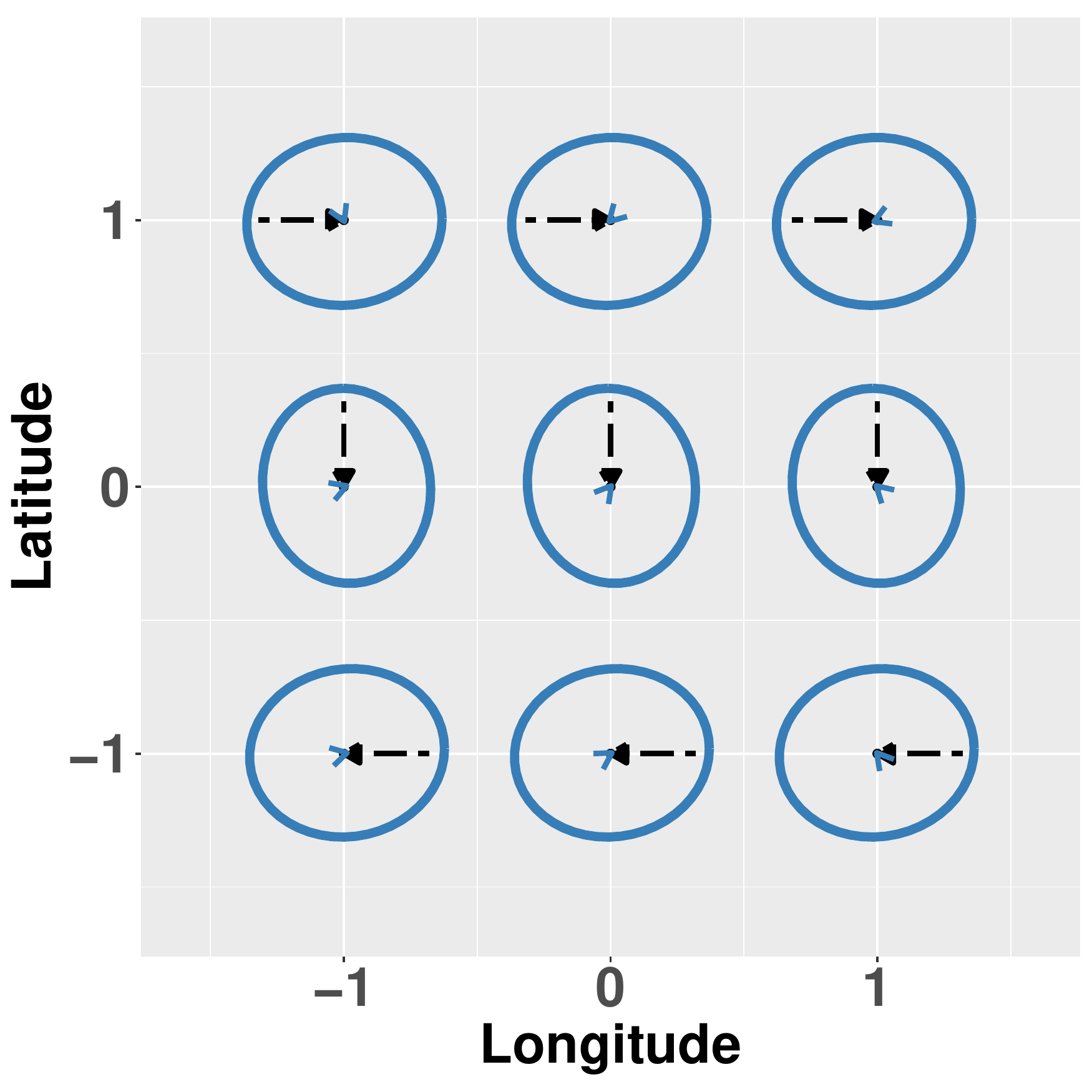}}}
	{\subfloat[Bear - $\text{B}_{53}$]{\includegraphics[trim = 0 20 0 20,scale=0.27]{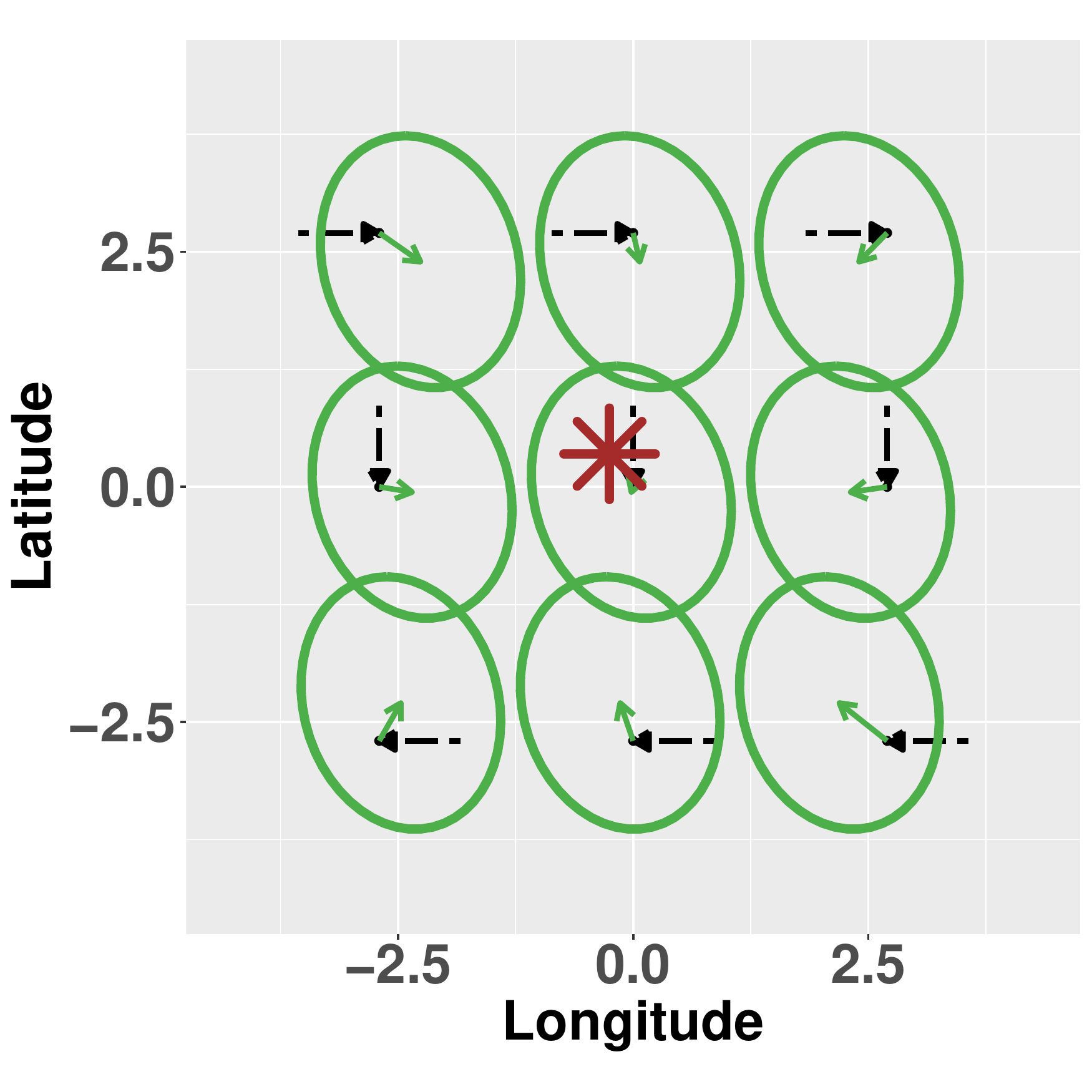}}}\\
	{\subfloat[Lucy - $\text{B}_{61}$]{\includegraphics[trim = 0 20 0 20,scale=0.27]{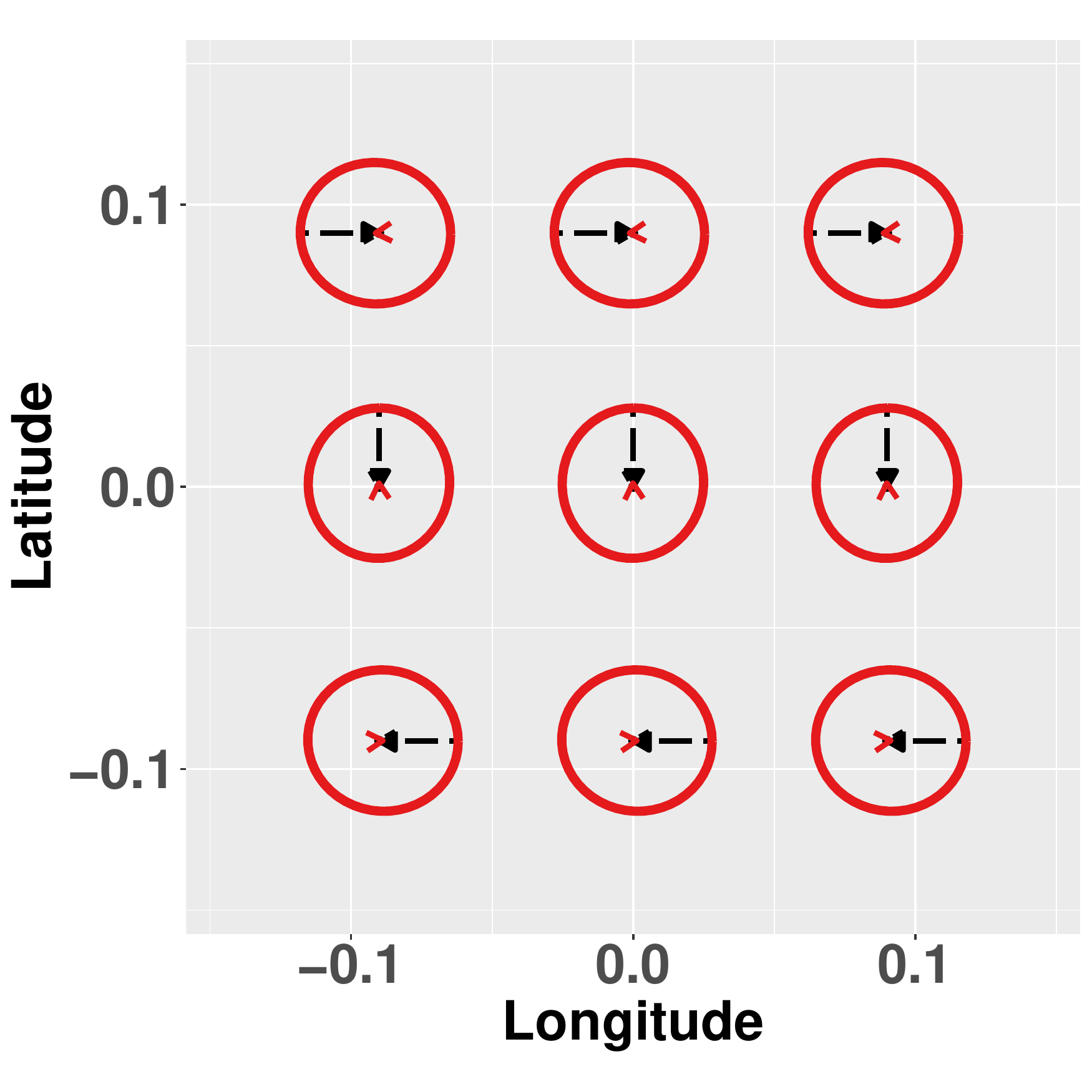}}}
	{\subfloat[Lucy - $\text{B}_{62}$]{\includegraphics[trim = 0 20 0 20,scale=0.27]{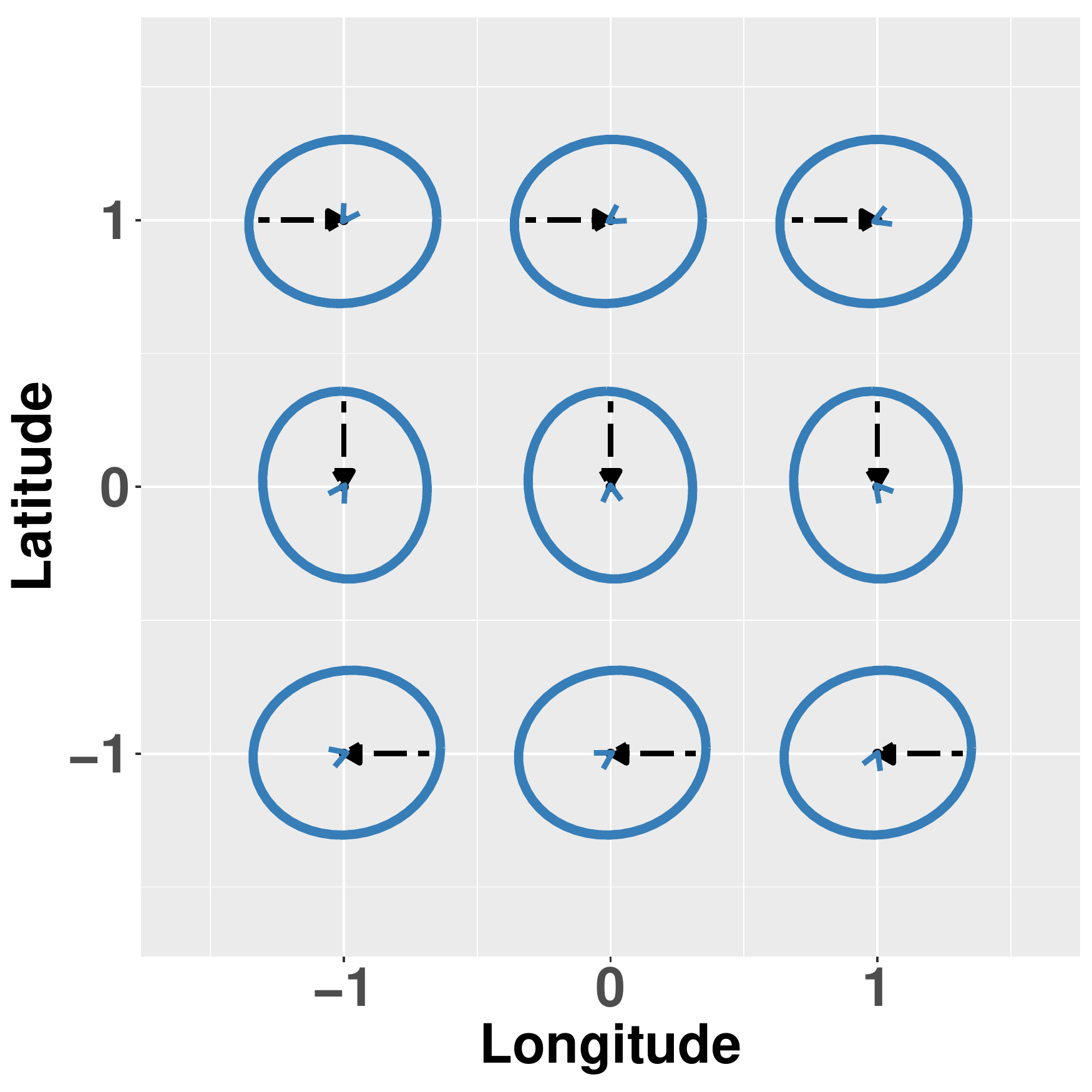}}}
	{\subfloat[Lucy - $\text{B}_{63}$]{\includegraphics[trim = 0 20 0 20,scale=0.27]{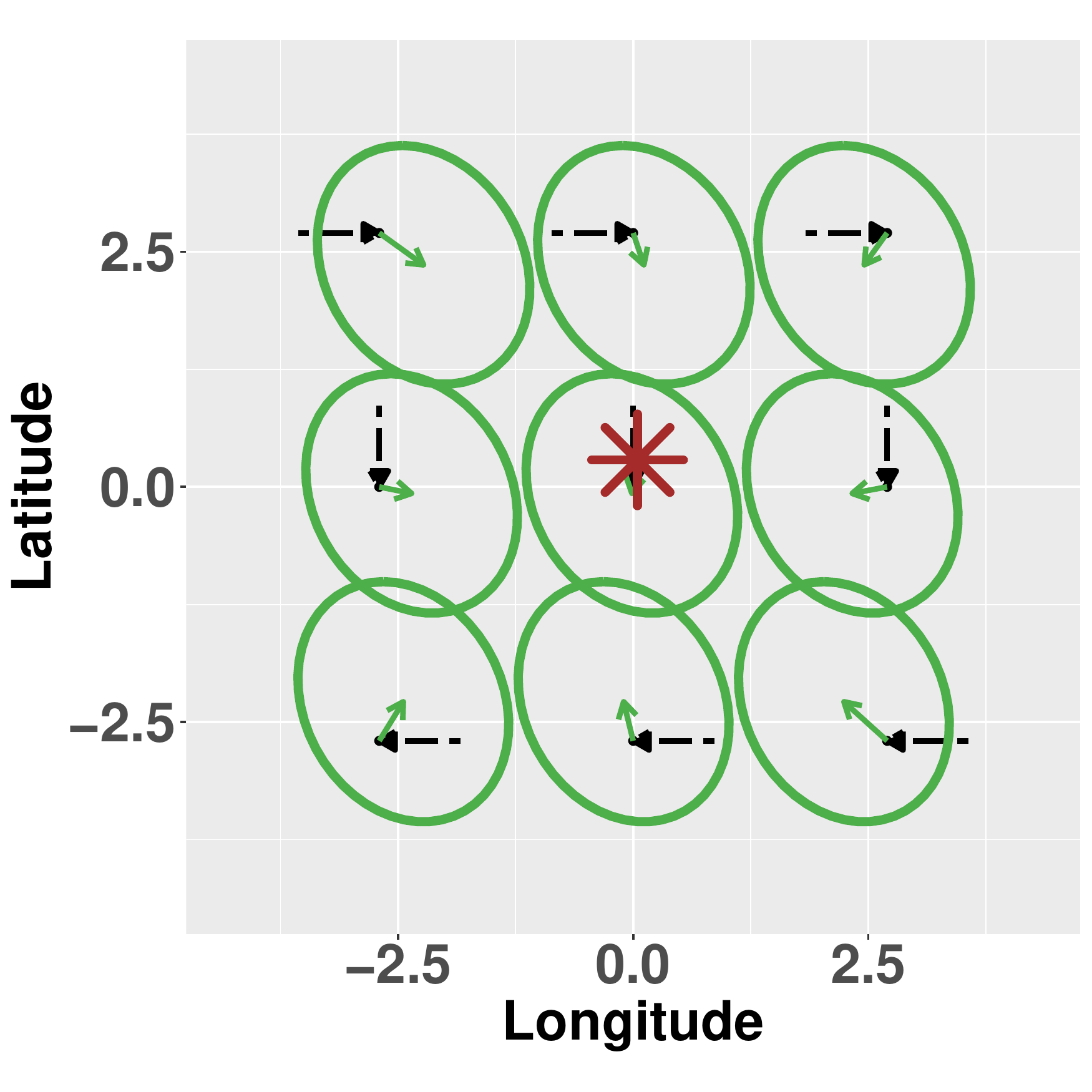}}}
  \caption{M2 - Graphical representation of the conditional distribution of $\mathbf{s}_{j,t_{j,i+1}}$ for the last three dogs,  for different possible values of $\mathbf{s}_{j,t_{j,i}}$ and  previous directions.
	The images has been obtained by using the posterior values that maximize the data likelihood of each animal, given the representative clusterization $\hat{z}_{j, t_{j,i}}$.
	The dashed arrow represents the movement between  $\mathbf{s}_{j,t_{j,i-1}}$ and $\mathbf{s}_{j,t_{j,i}}$. The solid arrow is $\protect\vec{\mathbf{F}}_{j,t_{j,i}}$,  while the ellipse is an area containing  95\% of the probability mass of  the  conditional  distribution of $\mathbf{s}_{j,t_{j,i+1}}$. The asterisk represents the  attractor,  and it is only shown for behaviors that have posterior values of
	${\rho}_{j,k}<0.9$ and ${\tau}_{j,k}>0.1$.
	}\label{I-newfig:a2_new_2Single}
\end{figure}

\begin{table}[t!]
\caption{\label{I-tab:anim1Standard}  M3 -  Woody: posterior means and CIs of the model parameters (j=1).}
%%\scriptsize
%\scriptsize
  \centering
  \fbox{
\begin{tabular}{c|cccccc}
 & k=1& k=2 & k=3  \\
 \hline
 $\mu_{j,k,1}$   &  2.827  &  -0.157  &  -0.301  \\
 (CI)  & (-4.211 9.069) & (-0.529 0.214) & (-2.109 1.54)  \\
 $\mu_{j,k,2}$   &  0.103  &  0.131  &  0.459  \\
 (CI)  & (-6.173 6.606) & (-0.231 0.506) & (-2.069 2.062)  \\
 $\eta_{j,k,1}$   &  -0.085  &  0.035  &  0.055  \\
 (CI)  & (-8.901 8.846) & (0.017 0.054) & (-8.647 8.947)  \\
 $\eta_{j,k,2}$   &  -0.027  &  0.002  &  0.13  \\
 (CI)  & (-8.622 8.429) & (-0.013 0.018) & (-8.406 8.445)  \\
 $\tau_{j,k}$   &  0  &  0.726  &  0.631  \\
 (CI)  & (0 0.001) & (0.356 0.984) & (0.074 0.981)  \\
 $\rho_{j,k}$   &  0  &  0.937  &  0.001  \\
 (CI)  & (0 0) & (0.877 0.96) & (0 0.013)  \\
 $\boldsymbol{\Sigma}_{j,k,1,1}$   &  0.001  &  0.108  &  0.875  \\
 (CI)  & (0.001 0.001) & (0.096 0.118) & (0.327 2.289)  \\
 $\boldsymbol{\Sigma}_{j,k,1,2}$   &  0  &  -0.004  &  -0.867  \\
 (CI)  & (0 0) & (-0.009 0.001) & (-2.264 -0.277)  \\
 $\boldsymbol{\Sigma}_{j,k,2,2}$   &  0.001  &  0.085  &  1.455  \\
 (CI)  & (0.001 0.001) & (0.077 0.094) & (0.607 3.361)  \\
 ${\pi}_{j,1,k}$   &  0.814  &  0.186  &  0  \\
 (CI)  & (0.799 0.828) & (0.172 0.201) & (0 0)  \\
 ${\pi}_{j,2,k}$   &  0.362  &  0.628  &  0.009  \\
 (CI)  & (0.337 0.388) & (0.602 0.654) & (0.003 0.023)  \\
 ${\pi}_{j,3,k}$   &  0.001  &  0.94  &  0.05  \\
 (CI)  & (0 0.001) & (0.653 1) & (0 0.322)  \\
 $n_{j,k}$   &  3203  &  1587  &  10  \\
\end{tabular}
}
\end{table}

\begin{table}[t!]
   \caption{\label{I-tab:anim2Standard} M3 -  Sherlock: posterior means and CIs of the model parameters (j=2).}
%\scriptsize
  \centering
  \fbox{
\begin{tabular}{c|cccccc}
 & k=1& k=2 & k=3& k=4& k=5  \\
 \hline
 $\mu_{j,k,1}$   &  -0.042  &  0.012  &  0.089  &  -2.476  &  2.245  \\
 (CI)  & (-8.78 8.892) & (-0.325 0.352) & (-0.181 0.348) & (-5.019 -1.158) & (-7.673 10.247)  \\
 $\mu_{j,k,2}$   &  -0.128  &  -0.14  &  -0.927  &  3.947  &  0.257  \\
 (CI)  & (-8.41 8.689) & (-0.548 0.281) & (-1.272 -0.596) & (1.81 7.873) & (-7.609 7.576)  \\
 $\eta_{j,k,1}$   &  -0.003  &  0.538  &  -0.097  &  0.087  &  -0.028  \\
 (CI)  & (-0.004 -0.002) & (-5.408 5.995) & (-8.929 8.571) & (-8.56 8.862) & (-8.682 9.341)  \\
 $\eta_{j,k,2}$   &  0  &  0.371  &  -0.082  &  0.235  &  -0.017  \\
 (CI)  & (-0.001 0.001) & (-5.128 5.73) & (-8.536 8.629) & (-8.33 8.733) & (-8.791 8.584)  \\
 $\tau_{j,k}$   &  0.5  &  0.048  &  0.813  &  0.359  &  0.46  \\
 (CI)  & (0.027 0.975) & (0.034 0.063) & (0.674 0.949) & (0.159 0.629) & (0.043 0.965)  \\
 $\rho_{j,k}$   &  1  &  0.07  &  0.004  &  0.004  &  0.292  \\
 (CI)  & (1 1) & (0 0.175) & (0 0.046) & (0 0.062) & (0 1)  \\
 $\boldsymbol{\Sigma}_{j,k,1,1}$   &  0.001  &  0.067  &  0.183  &  0.16  &  2.052  \\
 (CI)  & (0.001 0.001) & (0.06 0.074) & (0.094 0.365) & (0.074 0.352) & (0.133 9.781)  \\
 $\boldsymbol{\Sigma}_{j,k,1,2}$   &  0  &  0.01  &  0.061  &  -0.021  &  0.28  \\
 (CI)  & (0 0) & (0.005 0.016) & (-0.022 0.185) & (-0.161 0.107) & (-3.659 3.496)  \\
 $\boldsymbol{\Sigma}_{j,k,2,2}$   &  0.001  &  0.097  &  0.207  &  0.25  &  2.274  \\
 (CI)  & (0.001 0.001) & (0.085 0.11) & (0.106 0.407) & (0.109 0.552) & (0.117 10.195)  \\
 ${\pi}_{j,1,k}$   &  0.842  &  0.157  &  0  &  0  &  0  \\
 (CI)  & (0.829 0.855) & (0.144 0.17) & (0 0) & (0 0.001) & (0 0.002)  \\
 ${\pi}_{j,2,k}$   &  0.408  &  0.569  &  0.013  &  0.01  &  0  \\
 (CI)  & (0.375 0.44) & (0.538 0.6) & (0.005 0.029) & (0.004 0.017) & (0 0)  \\
 ${\pi}_{j,3,k}$   &  0.368  &  0.45  &  0.171  &  0  &  0  \\
 (CI)  & (0.141 0.638) & (0.151 0.768) & (0 0.423) & (0 0) & (0 0)  \\
 ${\pi}_{j,4,k}$   &  0.001  &  0.724  &  0  &  0.26  &  0  \\
 (CI)  & (0 0.002) & (0.475 0.929) & (0 0.001) & (0.056 0.505) & (0 0)  \\
 ${\pi}_{j,5,k}$   &  0.007  &  0.012  &  0.35  &  0.008  &  0.397  \\
 (CI)  & (0 0.022) & (0 0.08) & (0 0.998) & (0 0.019) & (0 1)  \\
 $n_{j,k}$   &  3466  &  1301  &  16  &  14  &  3  \\
\end{tabular}
}
\end{table}

\begin{table}[t!]
  \caption{\label{I-tab:anim3Standard}  M3 - Alvin: posterior means and CIs of the model parameters (j=3).}
%\scriptsize
  \centering
  \fbox{
\begin{tabular}{c|cccccc}
 & k=1& k=2 & k=3& k=4  \\
 \hline
 $\mu_{j,k,1}$   &  0.048  &  0.602  &  0.863  &  -1.208  \\
 (CI)  & (-8.603 8.615) & (0.591 0.621) & (0.665 1.149) & (-5.366 4.64)  \\
 $\mu_{j,k,2}$   &  -0.202  &  -0.362  &  -0.211  &  2.213  \\
 (CI)  & (-9.292 8.204) & (-0.371 -0.35) & (-0.423 0.024) & (-4.438 5.592)  \\
 $\eta_{j,k,1}$   &  -0.004  &  0.069  &  0.412  &  0.017  \\
 (CI)  & (-0.005 -0.003) & (-8.273 8.831) & (-8.391 8.73) & (-8.527 8.38)  \\
 $\eta_{j,k,2}$   &  0  &  -0.022  &  -0.241  &  -0.077  \\
 (CI)  & (-0.001 0.001) & (-8.691 8.888) & (-8.45 8.582) & (-8.723 8.379)  \\
 $\tau_{j,k}$   &  0.496  &  0.916  &  0.238  &  0.683  \\
 (CI)  & (0.024 0.969) & (0.827 0.96) & (0.14 0.318) & (0.126 0.915)  \\
 $\rho_{j,k}$   &  1  &  0  &  0.004  &  0.095  \\
 (CI)  & (1 1) & (0 0) & (0 0.05) & (0 1)  \\
 $\boldsymbol{\Sigma}_{j,k,1,1}$   &  0.001  &  0.011  &  0.104  &  0.883  \\
 (CI)  & (0.001 0.001) & (0.008 0.019) & (0.079 0.158) & (0.065 3.261)  \\
 $\boldsymbol{\Sigma}_{j,k,1,2}$   &  0  &  0.002  &  0.005  &  0.098  \\
 (CI)  & (0 0) & (0 0.006) & (-0.013 0.022) & (-0.82 0.765)  \\
 $\boldsymbol{\Sigma}_{j,k,2,2}$   &  0  &  0.008  &  0.141  &  0.639  \\
 (CI)  & (0 0) & (0.005 0.015) & (0.106 0.216) & (0.068 3.051)  \\
 ${\pi}_{j,1,k}$   &  0.899  &  0.069  &  0.032  &  0.001  \\
 (CI)  & (0.887 0.909) & (0.058 0.084) & (0.015 0.043) & (0 0.002)  \\
 ${\pi}_{j,2,k}$   &  0.527  &  0.41  &  0.058  &  0.004  \\
 (CI)  & (0.458 0.589) & (0.346 0.493) & (0.031 0.086) & (0 0.011)  \\
 ${\pi}_{j,3,k}$   &  0.152  &  0.436  &  0.41  &  0  \\
 (CI)  & (0.095 0.209) & (0.366 0.515) & (0.335 0.485) & (0 0.003)  \\
 ${\pi}_{j,4,k}$   &  0.282  &  0.003  &  0.146  &  0.493  \\
 (CI)  & (0 0.651) & (0 0.014) & (0 0.469) & (0.002 0.999)  \\
 $n_{j,k}$   &  3907  &  627  &  253  &  13  \\
\end{tabular}
}
\end{table}

\begin{table}[t!]
  \caption{\label{I-tab:anim4Standard}  M3 - Rosie: posterior means and CIs of the model parameters (j=4).}
%\scriptsize
  \centering
  \fbox{
\begin{tabular}{c|ccccc}
 & k=1& k=2 & k=3  \\
 \hline
 $\mu_{j,k,1}$   &  0  &  -0.222  &  -0.005  \\
 (CI)  & (-8.879 8.897) & (-2.847 2.427) & (-8.67 8.758)  \\
 $\mu_{j,k,2}$   &  -0.008  &  0.987  &  0.034  \\
 (CI)  & (-9.01 8.879) & (-1.65 4.187) & (-8.876 8.487)  \\
 $\eta_{j,k,1}$   &  -0.002  &  -0.284  &  0.167  \\
 (CI)  & (-0.002 -0.001) & (-8.681 8.557) & (-8.144 8.953)  \\
 $\eta_{j,k,2}$   &  0  &  0.116  &  0.25  \\
 (CI)  & (-0.001 0.001) & (-8.305 8.793) & (-8.742 9.179)  \\
 $\tau_{j,k}$   &  0.502  &  0.007  &  0.501  \\
 (CI)  & (0.028 0.974) & (0.001 0.015) & (0.026 0.973)  \\
 $\rho_{j,k}$   &  1  &  0.002  &  0.487  \\
 (CI)  & (1 1) & (0 0.019) & (0 1)  \\
 $\boldsymbol{\Sigma}_{j,k,1,1}$   &  0.001  &  0.031  &  8.024  \\
 (CI)  & (0 0.001) & (0.028 0.034) & (0.14 14.776)  \\
 $\boldsymbol{\Sigma}_{j,k,1,2}$   &  0  &  0  &  -3.989  \\
 (CI)  & (0 0) & (-0.002 0.002) & (-5.117 5.942)  \\
 $\boldsymbol{\Sigma}_{j,k,2,2}$   &  0  &  0.033  &  6.494  \\
 (CI)  & (0 0) & (0.03 0.037) & (0.133 17.609)  \\
 ${\pi}_{j,1,k}$   &  0.884  &  0.116  &  0  \\
 (CI)  & (0.872 0.895) & (0.105 0.128) & (0 0)  \\
 ${\pi}_{j,2,k}$   &  0.449  &  0.549  &  0  \\
 (CI)  & (0.41 0.49) & (0.508 0.589) & (0 0)  \\
 ${\pi}_{j,3,k}$   &  0.031  &  0.014  &  0.595  \\
 (CI)  & (0 0.567) & (0 0.069) & (0 1)  \\
 $n_{j,k}$   &  3841  &  957  &  2  \\
\end{tabular}
}
\end{table}

\begin{table}[t!]
  \caption{ \label{I-tab:anim5Standard}  M3 - Bear: posterior means and CIs of the model parameters (j=5).}
%\scriptsize
\centering
\fbox{
\begin{tabular}{c|ccc}
 & k=1& k=2 & k=3  \\
 \hline
 $\mu_{j,k,1}$   &  -0.09  &  -0.245  &  -0.762  \\
 (CI)  & (-8.953 9.085) & (-0.636 0.153) & (-4.482 3.091)  \\
 $\mu_{j,k,2}$   &  0  &  0.252  &  0.97  \\
 (CI)  & (-8.605 8.655) & (-0.21 0.722) & (-3.735 4.985)  \\
 $\eta_{j,k,1}$   &  -0.001  &  0.209  &  0.517  \\
 (CI)  & (-0.002 0) & (-8.325 8.759) & (-8.878 9.856)  \\
 $\eta_{j,k,2}$   &  0  &  -0.069  &  -0.105  \\
 (CI)  & (-0.001 0.001) & (-8.79 8.872) & (-8.846 8.431)  \\
 $\tau_{j,k}$   &  0.517  &  0.033  &  0.313  \\
 (CI)  & (0.028 0.977) & (0.023 0.044) & (0.007 0.879)  \\
 $\rho_{j,k}$   &  1  &  0  &  0.013  \\
 (CI)  & (1 1) & (0 0) & (0 0.103)  \\
 $\boldsymbol{\Sigma}_{j,k,1,1}$   &  0.001  &  0.055  &  0.567  \\
 (CI)  & (0.001 0.001) & (0.049 0.061) & (0.271 1.203)  \\
 $\boldsymbol{\Sigma}_{j,k,1,2}$   &  0  &  -0.005  &  -0.361  \\
 (CI)  & (0 0) & (-0.01 -0.001) & (-1.013 0.021)  \\
 $\boldsymbol{\Sigma}_{j,k,2,2}$   &  0.001  &  0.079  &  1.131  \\
 (CI)  & (0.001 0.001) & (0.067 0.089) & (0.506 2.469)  \\
 ${\pi}_{j,1,k}$   &  0.831  &  0.169  &  0  \\
 (CI)  & (0.817 0.844) & (0.156 0.183) & (0 0)  \\
 ${\pi}_{j,2,k}$   &  0.383  &  0.607  &  0.01  \\
 (CI)  & (0.356 0.411) & (0.576 0.635) & (0.003 0.024)  \\
 ${\pi}_{j,3,k}$   &  0.001  &  0.571  &  0.42  \\
 (CI)  & (0 0.006) & (0.331 0.817) & (0.171 0.655)  \\
 $n_{j,k}$   &  3337  &  1447  &  16  \\
\end{tabular}
}
\end{table}

\begin{table}[t!]
  \caption{\label{I-tab:anim6Standard} M3 - Lucy: posterior means and CIs of the model parameters (j=6).}
%\scriptsize
  \centering
\fbox{
\begin{tabular}{c|cccc}
 & k=1& k=2 & k=3  \\
 \hline
 $\mu_{j,k,1}$   &  1.986  &  -0.105  &  -0.409  \\
 (CI)  & (-4.873 8.395) & (-0.438 0.259) & (-6.908 7)  \\
 $\mu_{j,k,2}$   &  1.196  &  0.005  &  0.72  \\
 (CI)  & (-5.206 7.673) & (-0.413 0.404) & (-6.704 7.169)  \\
 $\eta_{j,k,1}$   &  -2.121  &  -0.042  &  0.629  \\
 (CI)  & (-10.102 7.19) & (-8.624 8.723) & (-9.071 11.768)  \\
 $\eta_{j,k,2}$   &  0.466  &  0.012  &  0.006  \\
 (CI)  & (-7.098 7.046) & (-8.756 8.734) & (-8.964 8.851)  \\
 $\tau_{j,k}$   &  0  &  0.038  &  0.523  \\
 (CI)  & (0 0.001) & (0.028 0.047) & (0.032 0.957)  \\
 $\rho_{j,k}$   &  0.001  &  0  &  0.235  \\
 (CI)  & (0 0.007) & (0 0) & (0 1)  \\
 $\boldsymbol{\Sigma}_{j,k,1,1}$   &  0.001  &  0.068  &  2.064  \\
 (CI)  & (0.001 0.001) & (0.06 0.075) & (0.17 8.422)  \\
 $\boldsymbol{\Sigma}_{j,k,1,2}$   &  0  &  0  &  -0.084  \\
 (CI)  & (0 0) & (-0.006 0.005) & (-2.633 2.692)  \\
 $\boldsymbol{\Sigma}_{j,k,2,2}$   &  0.001  &  0.086  &  2.821  \\
 (CI)  & (0.001 0.001) & (0.076 0.097) & (0.179 9.906)  \\
 ${\pi}_{j,1,k}$   &  0.795  &  0.205  &  0  \\
 (CI)  & (0.779 0.812) & (0.188 0.221) & (0 0)  \\
 ${\pi}_{j,2,k}$   &  0.365  &  0.626  &  0.005  \\
 (CI)  & (0.34 0.39) & (0.6 0.652) & (0 0.017)  \\
 ${\pi}_{j,3,k}$   &  0.008  &  0.397  &  0.408  \\
 (CI)  & (0 0.05) & (0 0.998) & (0 1)  \\
 $n_{j,k}$   &  3082  &  1704  &  14  \\
\end{tabular}
}
\end{table}

\begin{figure}[t!]
  \centering
  {\subfloat[Woody - $\text{B}_{11}$]{\includegraphics[trim = 0 20 0 20,scale=0.27]{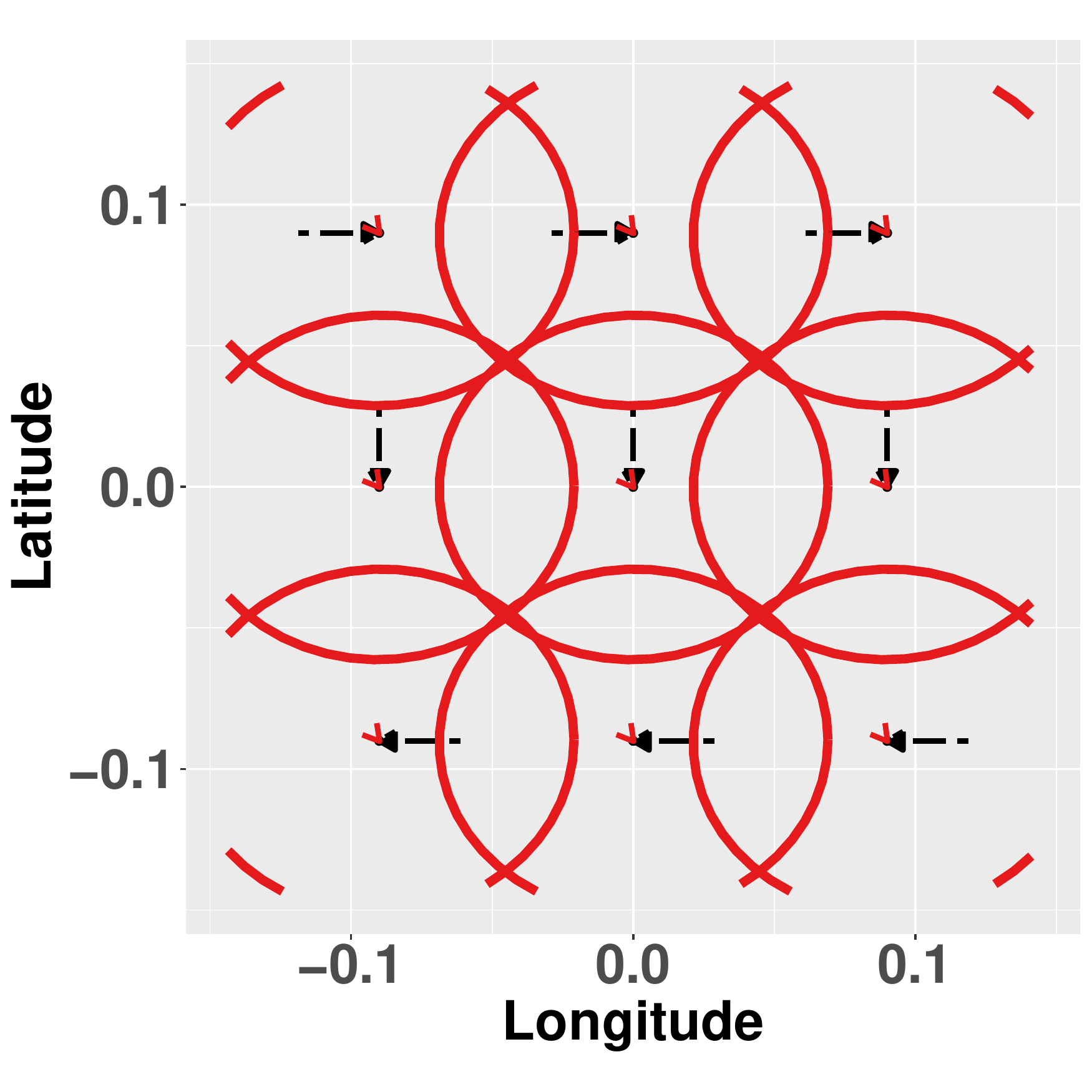}}}
  {\subfloat[Woody - $\text{B}_{12}$]{\includegraphics[trim = 0 20 0 20,scale=0.27]{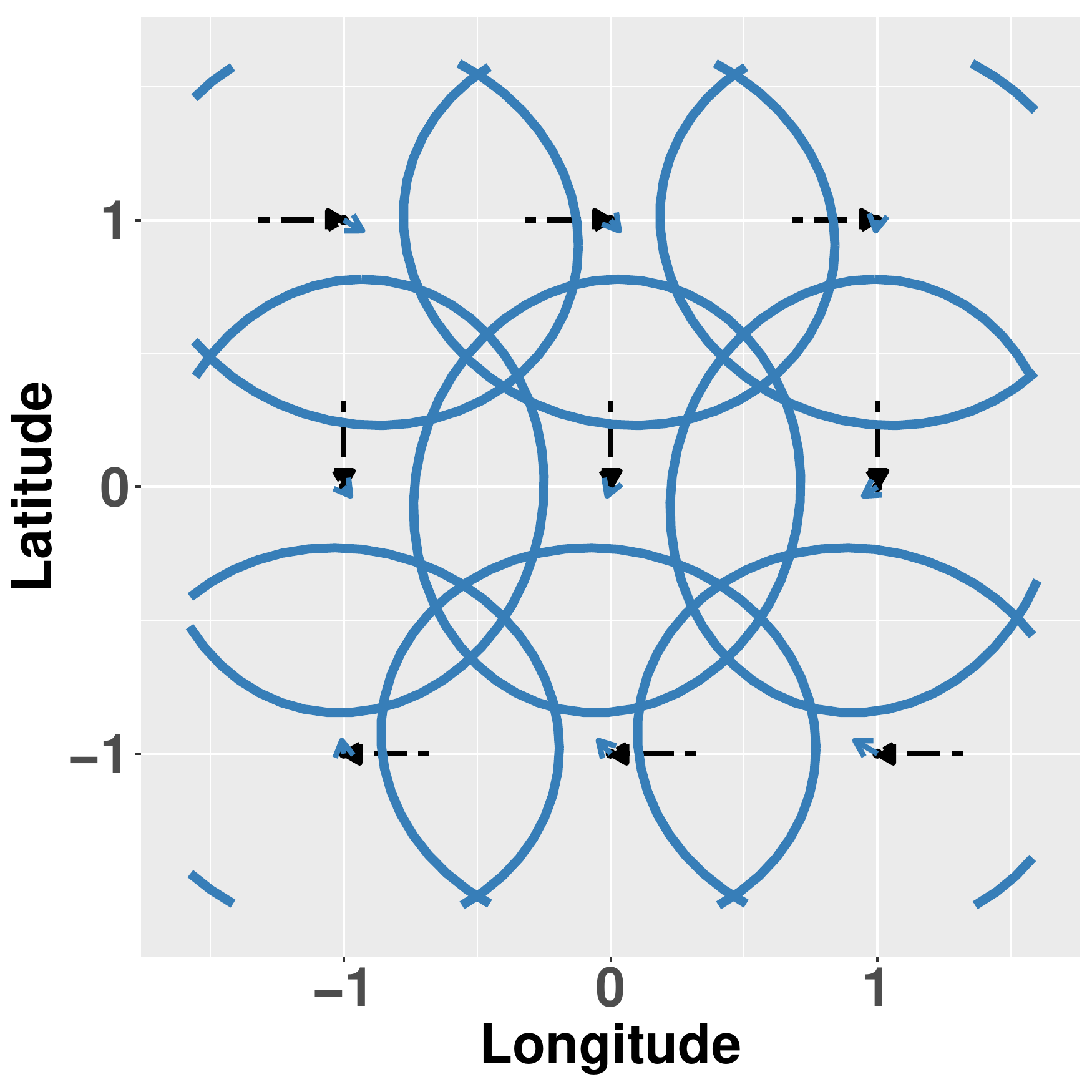}}}\\
  %{\subfloat[Woody - $\text{B}_{13}$]{\includegraphics[trim = 0 20 0 20,scale=0.27]{Standard1_3Mov}}}\\
  {\subfloat[Sherlock - $\text{B}_{21}$]{\includegraphics[trim = 0 20 0 20,scale=0.27]{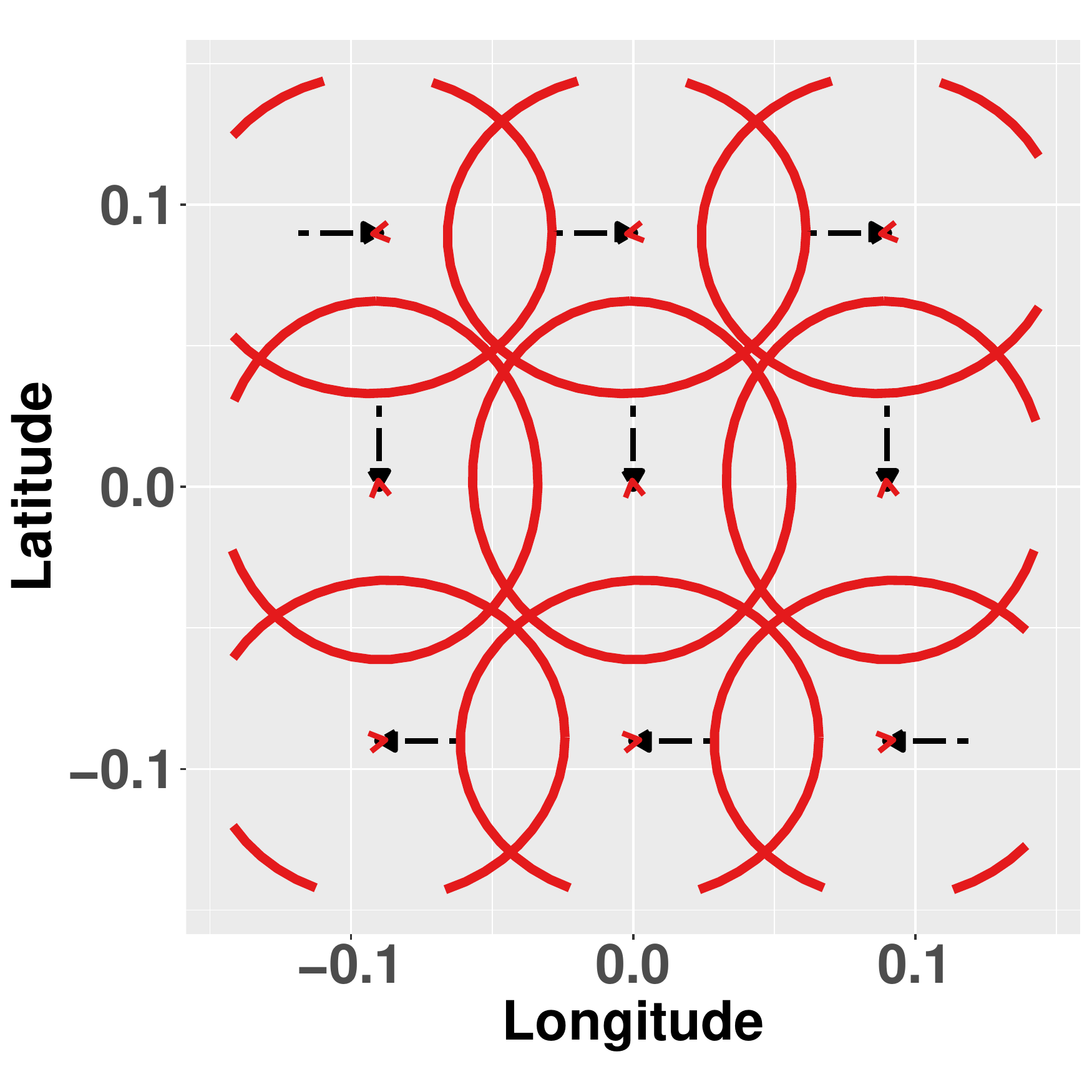}}}
  {\subfloat[Sherlock - $\text{B}_{22}$]{\includegraphics[trim = 0 20 0 20,scale=0.27]{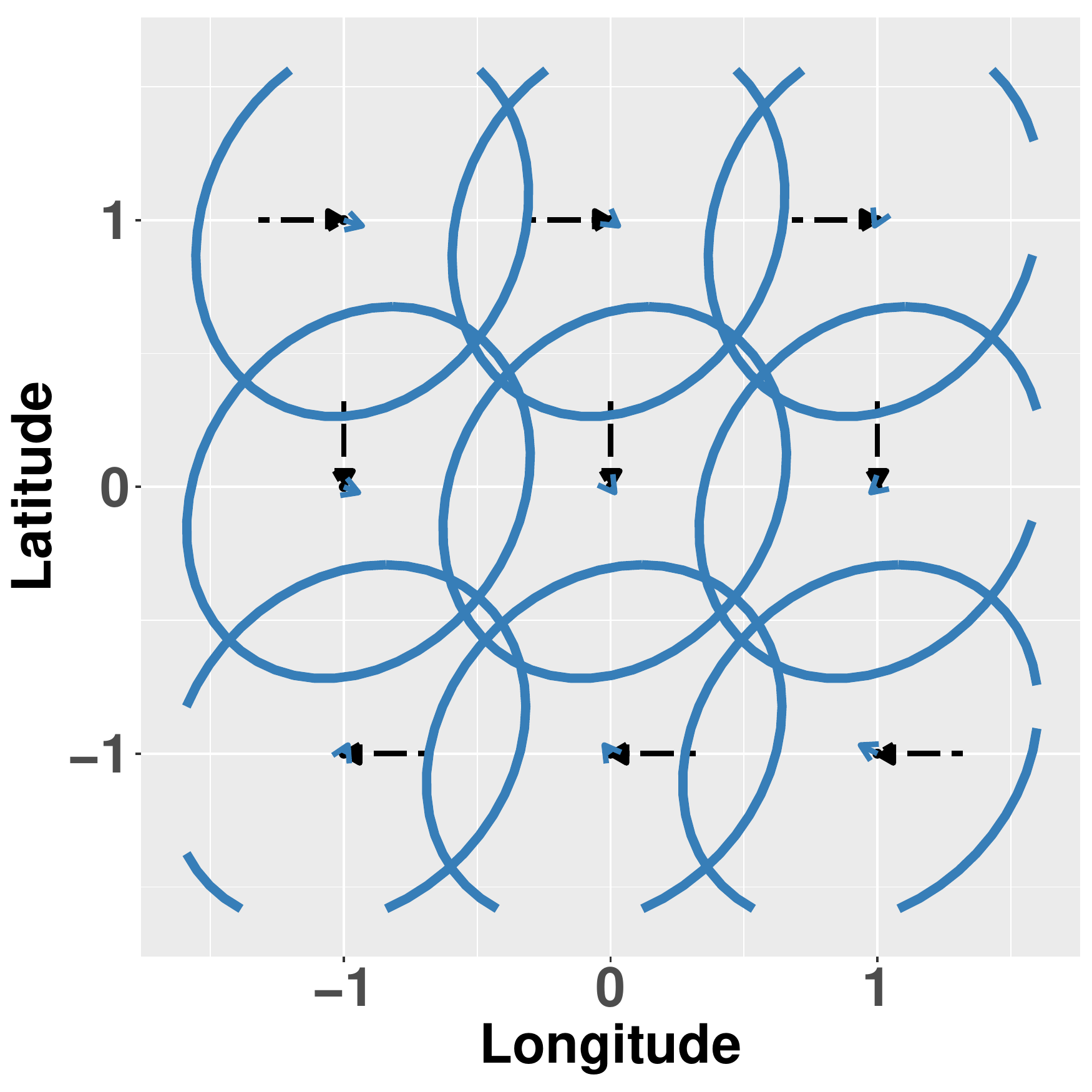}}}\\
  %{\subfloat[Sherlock - $\text{B}_{23}$]{\includegraphics[trim = 0 20 0 20,scale=0.27]{Standard2_3Mov}}}\\
	{\subfloat[Alvin - $\text{B}_{31}$]{\includegraphics[trim = 0 20 0 20,scale=0.27]{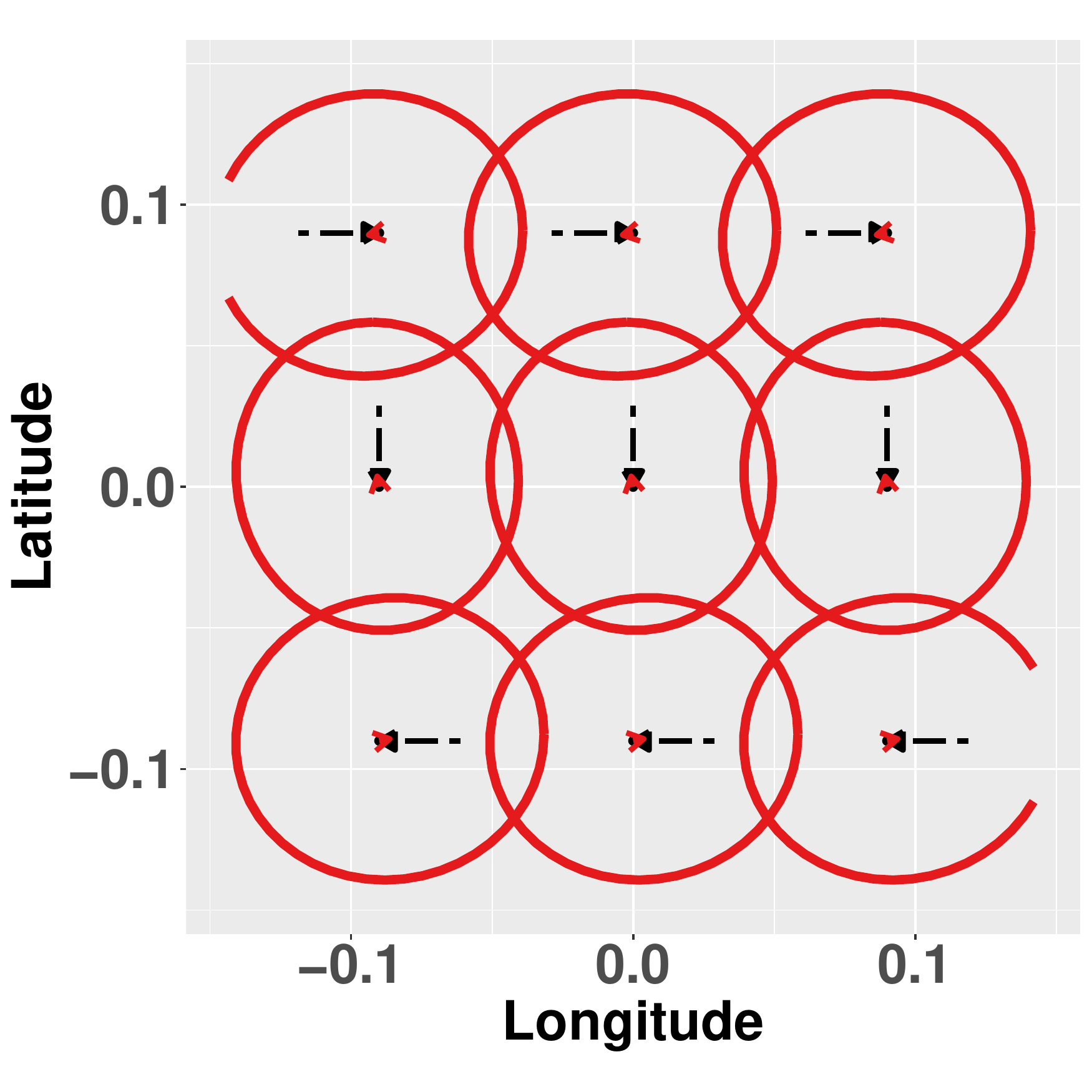}}}
	{\subfloat[Alvin - $\text{B}_{32}$]{\includegraphics[trim = 0 20 0 20,scale=0.27]{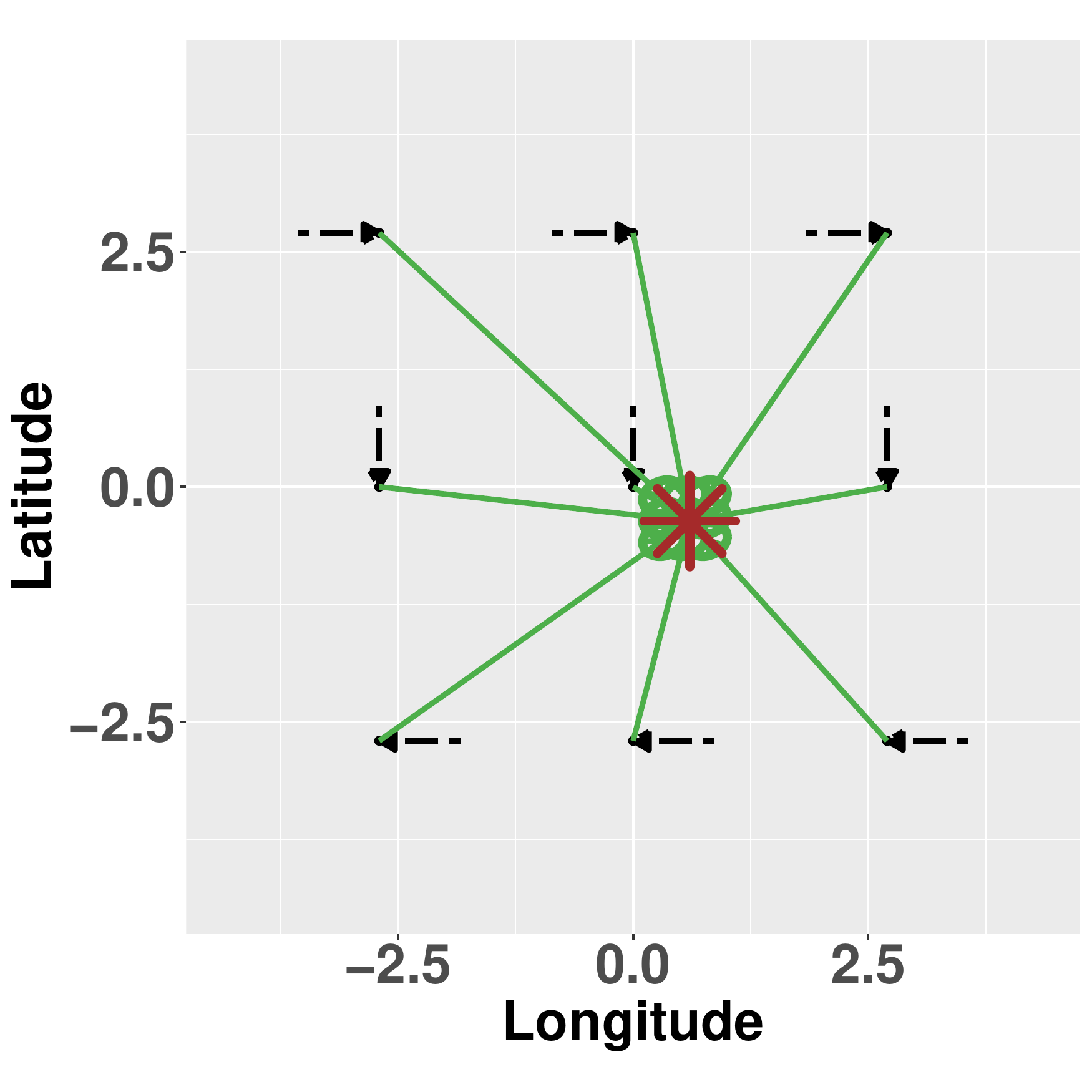}}}
	{\subfloat[Alvin - $\text{B}_{33}$]{\includegraphics[trim = 0 20 0 20,scale=0.27]{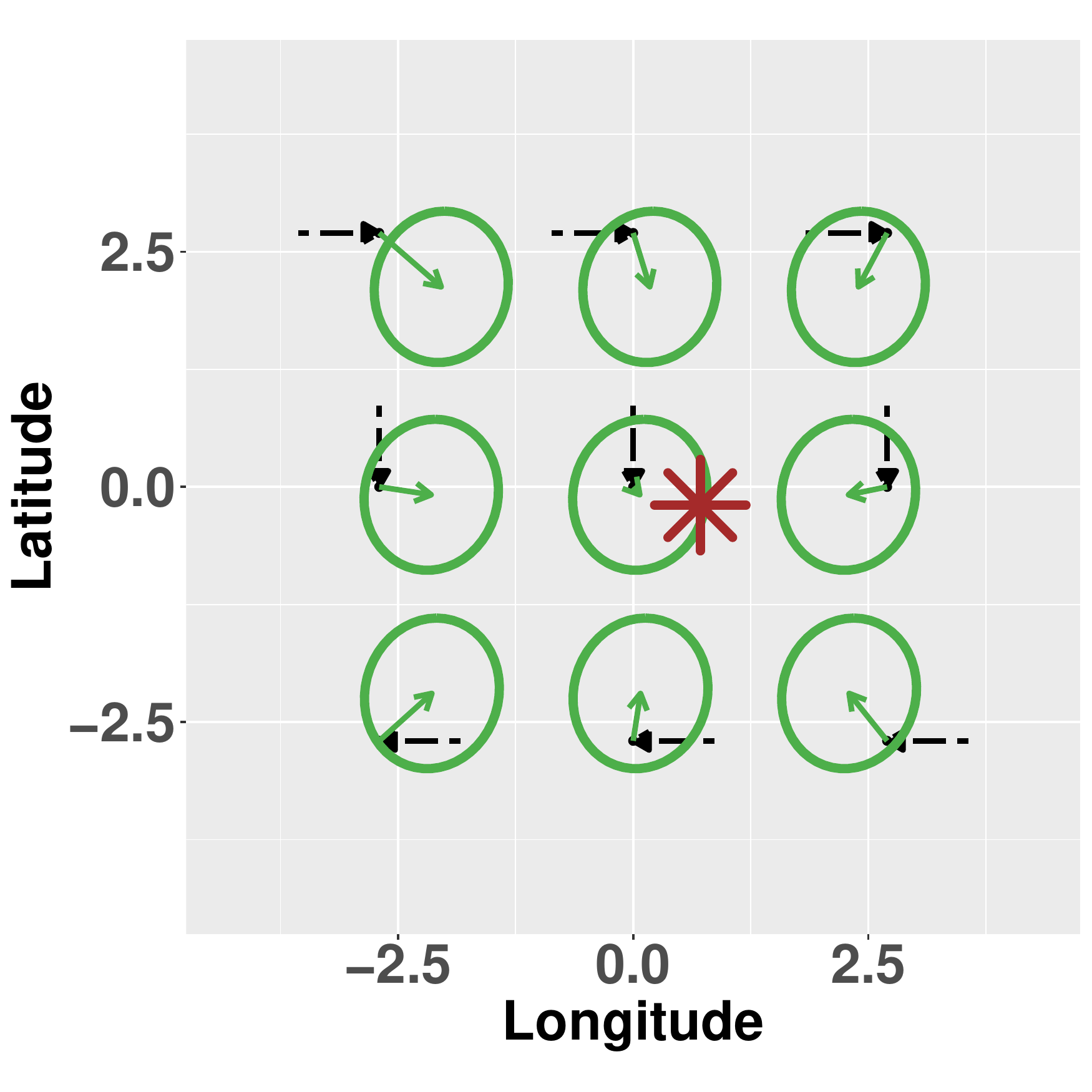}}}
	%{\subfloat[Alvin - $\text{B}_{34}$]{\includegraphics[trim = 0 20 0 20,scale=0.27]{Standard3_4Mov}}}
  \caption{M3 -  Graphical representation of the conditional distribution of $\mathbf{s}_{j,t_{j,i+1}}$ for the first three dogs,  for different possible values of $\mathbf{s}_{j,t_{j,i}}$ and  previous directions.
	The images has been obtained by using the posterior values that maximize the data likelihood of each animal, given the representative clusterization $\hat{z}_{j, t_{j,i}}$.
	The dashed arrow represents the movement between  $\mathbf{s}_{j,t_{j,i-1}}$ and $\mathbf{s}_{j,t_{j,i}}$. The solid arrow is $\protect\vec{\mathbf{F}}_{j,t_{j,i}}$,  while the ellipse is an area containing  95\% of the probability mass of  the  conditional  distribution of $\mathbf{s}_{j,t_{j,i+1}}$. The asterisk represents the  attractor,  and it is only shown for behaviors that have posterior values of
	${\rho}_{j,k}<0.9$ and ${\tau}_{j,k}>0.1$. }\label{I-newfig:a2_1Standard}
\end{figure}

\begin{figure}[t!]
  \centering
  {\subfloat[Rosie - $\text{B}_{41}$]{\includegraphics[trim = 0 20 0 20,scale=0.27]{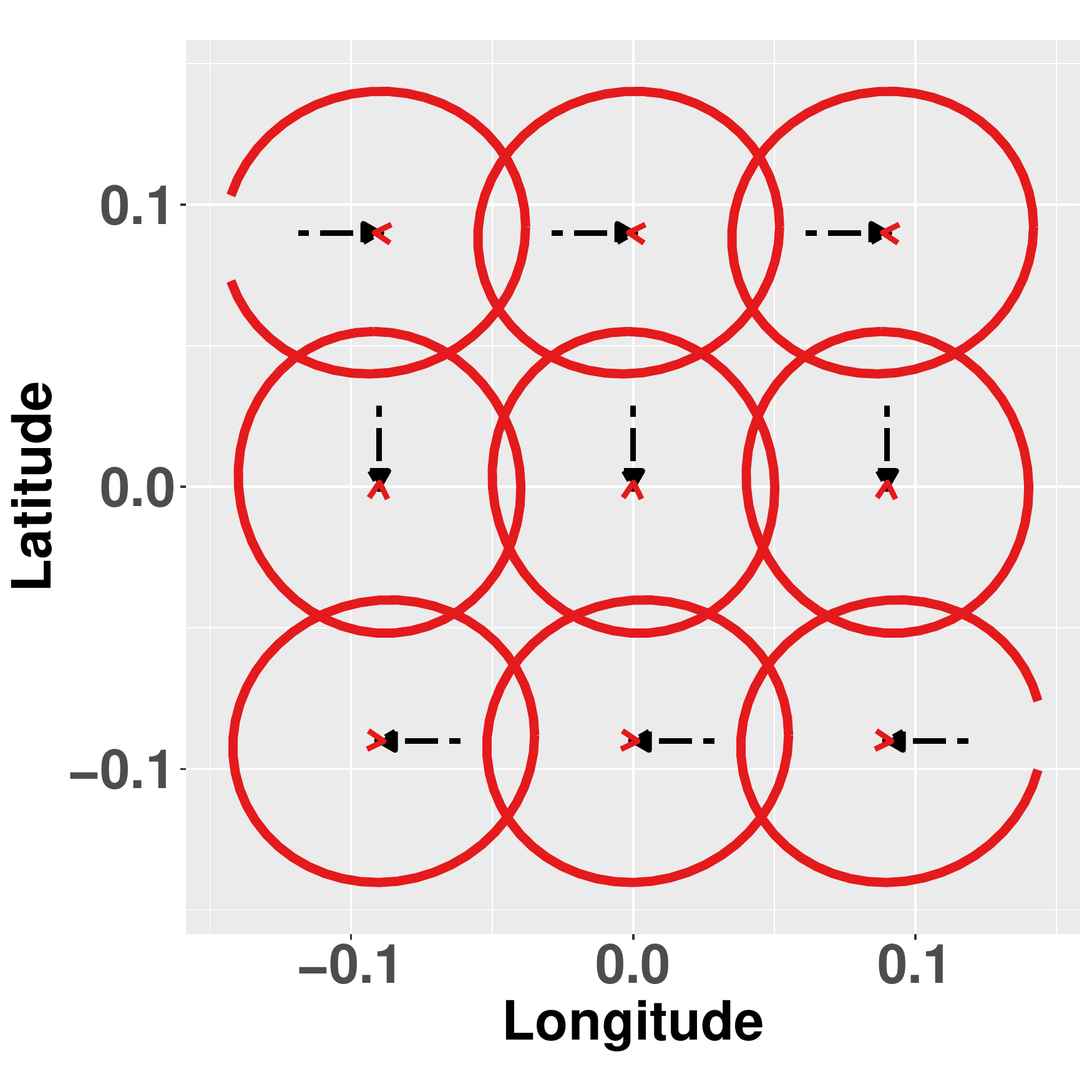}}}
  {\subfloat[Rosie - $\text{B}_{42}$]{\includegraphics[trim = 0 20 0 20,scale=0.27]{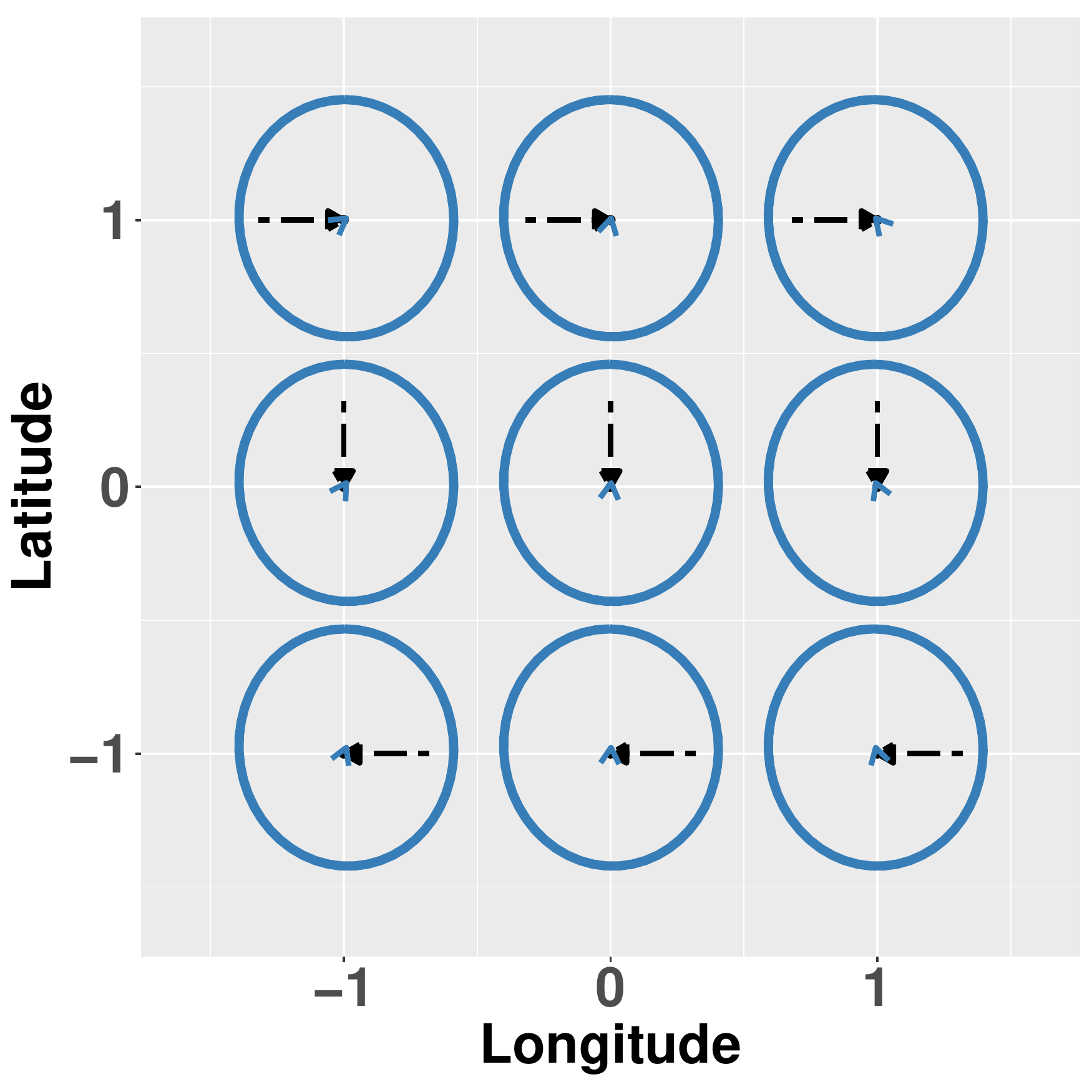}}}\\
  %{\subfloat[Rosie - $\text{B}_{43}$]{\includegraphics[trim = 0 20 0 20,scale=0.27]{Standard4_3Mov}}}
	{\subfloat[Bear - $\text{B}_{51}$]{\includegraphics[trim = 0 20 0 20,scale=0.27]{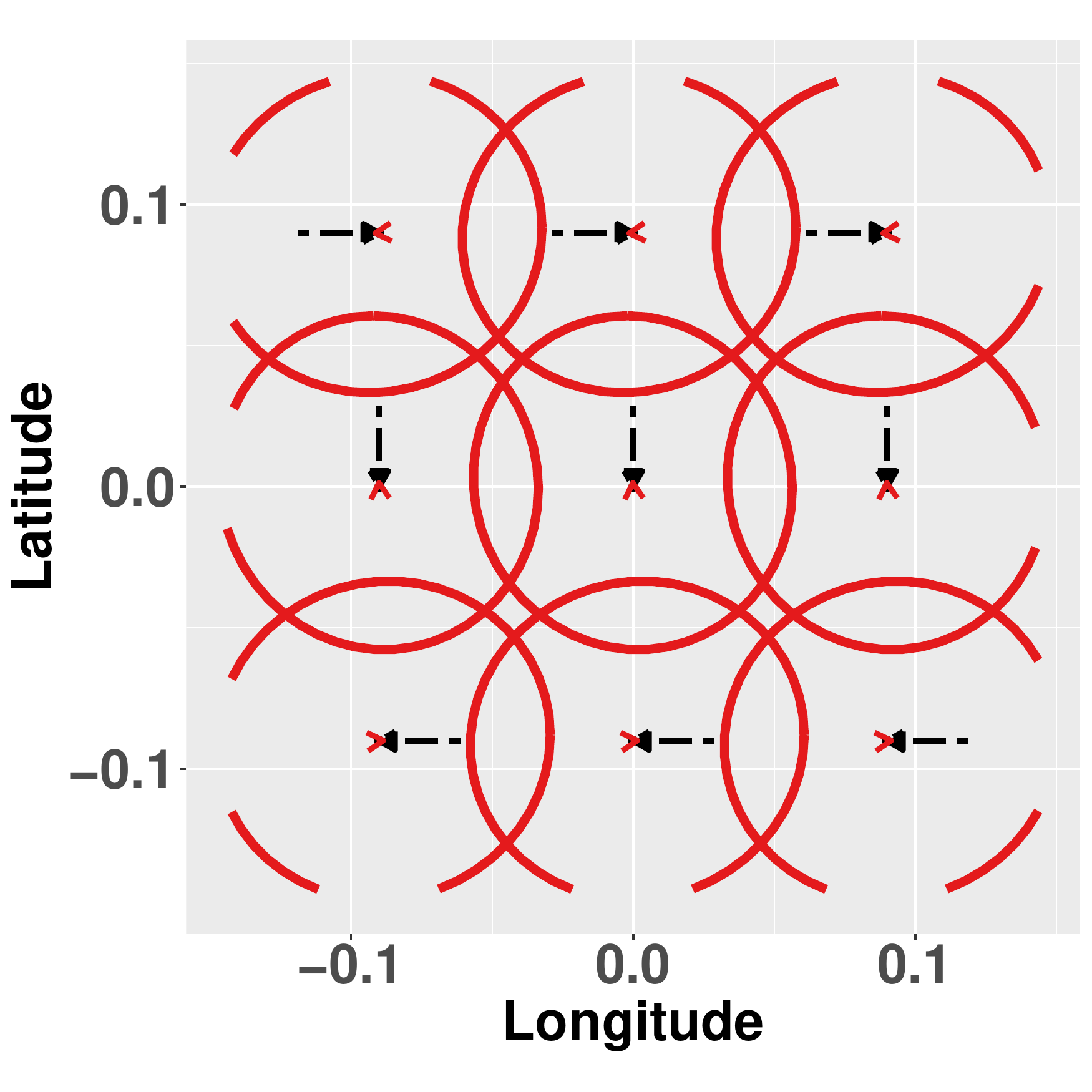}}}
	{\subfloat[Bear - $\text{B}_{52}$]{\includegraphics[trim = 0 20 0 20,scale=0.27]{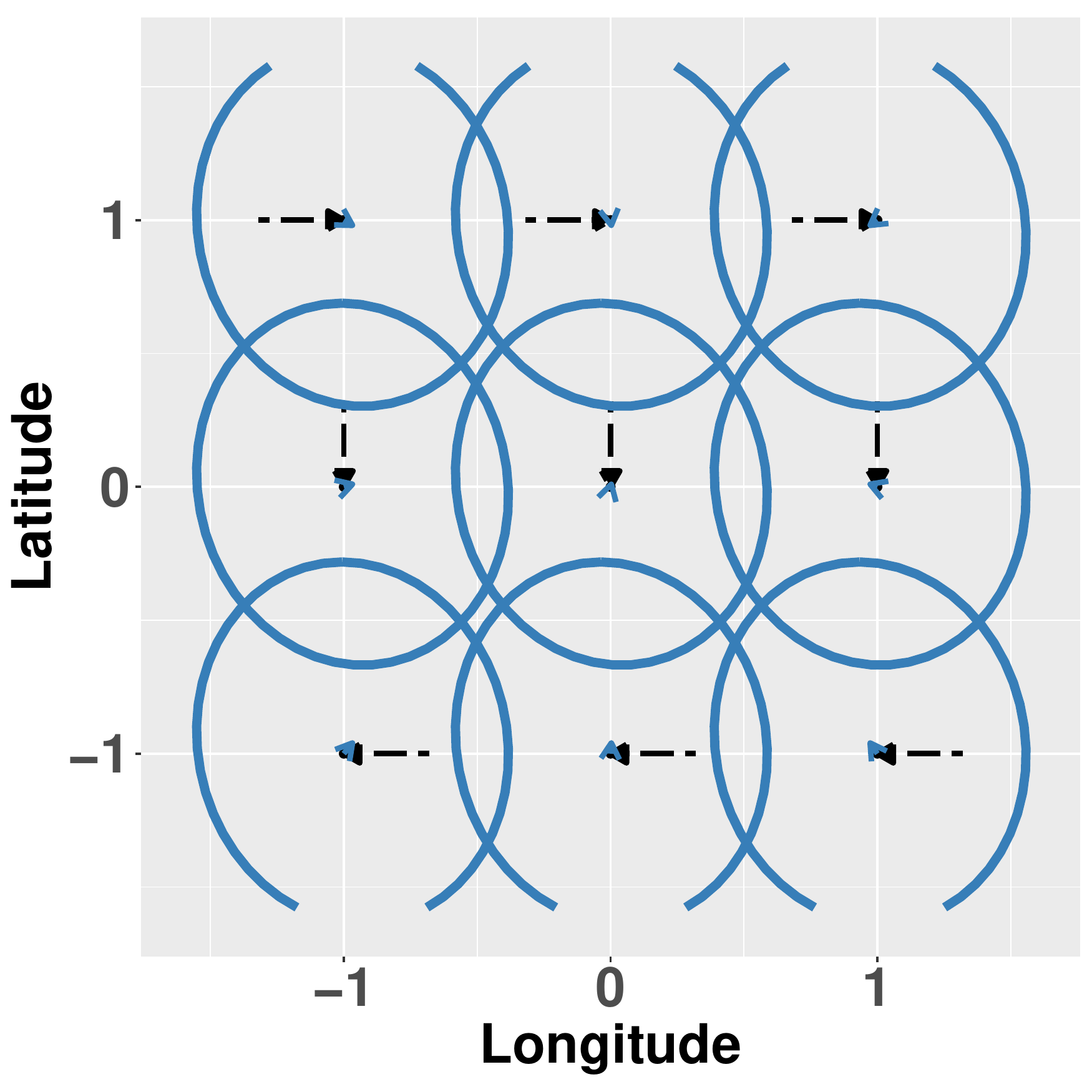}}}\\
	%{\subfloat[Bear - $\text{B}_{53}$]{\includegraphics[trim = 0 20 0 20,scale=0.27]{Standard5_3Mov}}}
	{\subfloat[Lucy - $\text{B}_{61}$]{\includegraphics[trim = 0 20 0 20,scale=0.27]{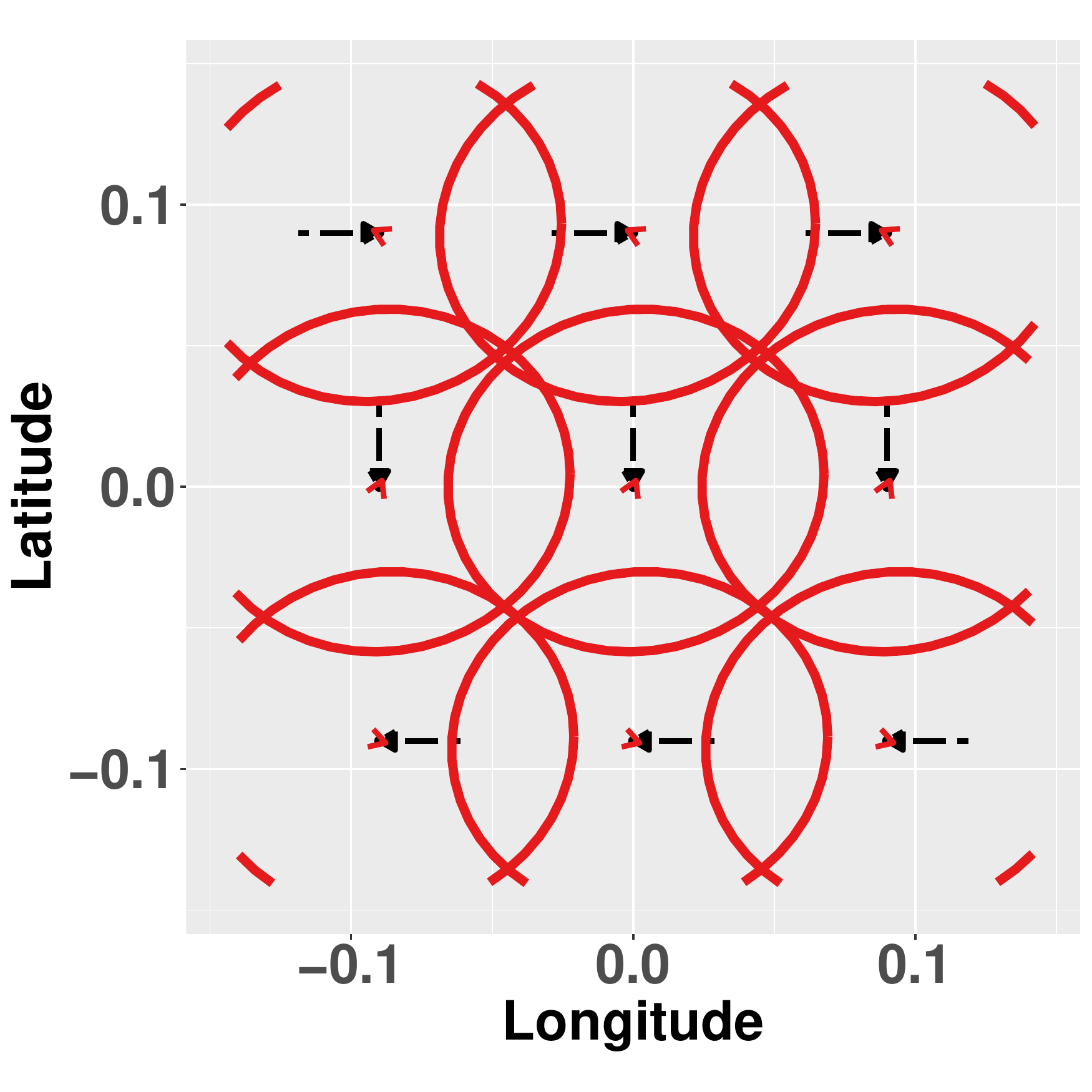}}}
	{\subfloat[Lucy - $\text{B}_{62}$]{\includegraphics[trim = 0 20 0 20,scale=0.27]{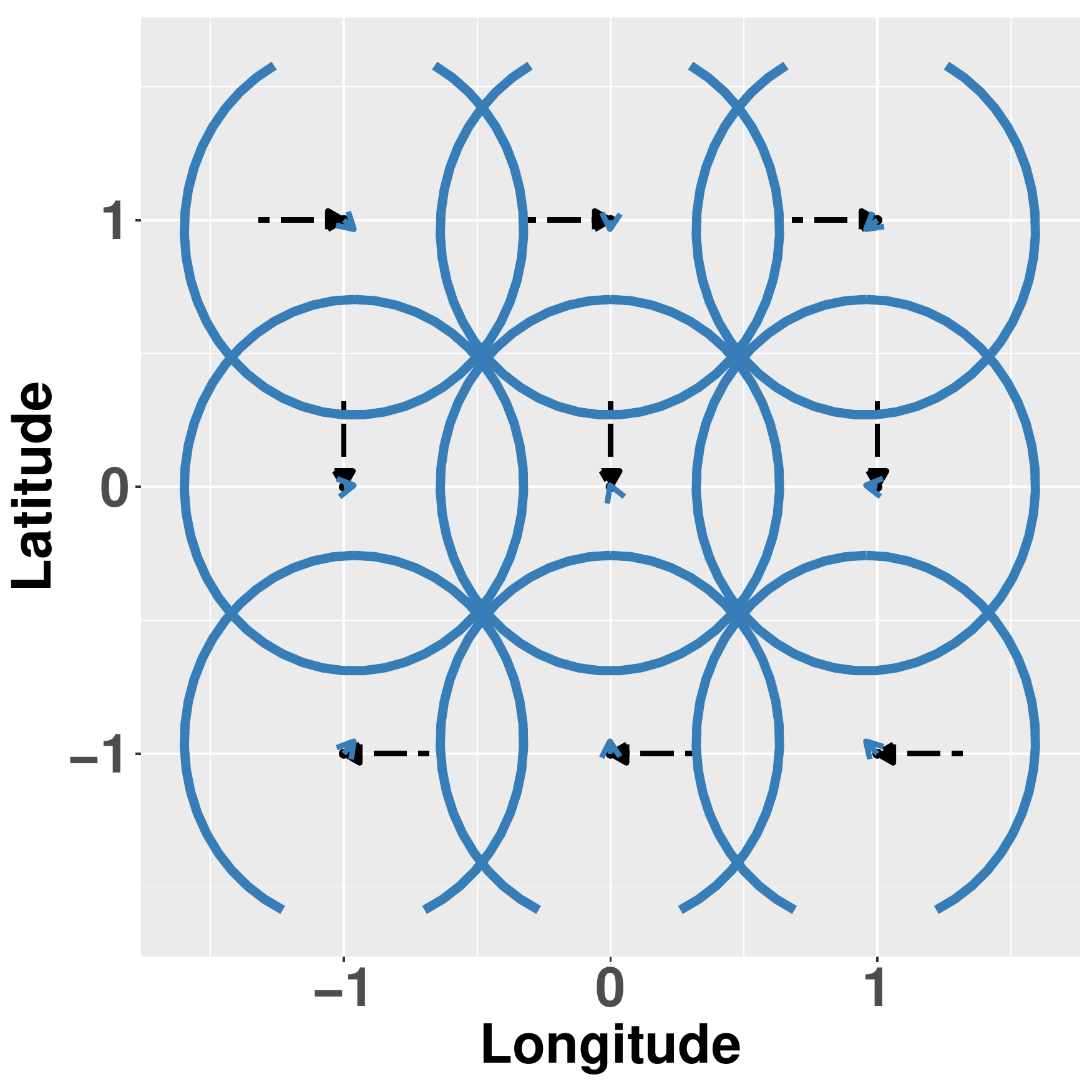}}}
	%{\subfloat[Lucy - $\text{B}_{63}$]{\includegraphics[trim = 0 20 0 20,scale=0.27]{Standard6_3Mov}}}
  \caption{M3 - Graphical representation of the conditional distribution of $\mathbf{s}_{j,t_{j,i+1}}$ for the last three dogs,  for different possible values of $\mathbf{s}_{j,t_{j,i}}$ and  previous directions.
	The images has been obtained by using the posterior values that maximize the data likelihood of each animal, given the representative clusterization $\hat{z}_{j, t_{j,i}}$.
	The dashed arrow represents the movement between  $\mathbf{s}_{j,t_{j,i-1}}$ and $\mathbf{s}_{j,t_{j,i}}$. The solid arrow is $\protect\vec{\mathbf{F}}_{j,t_{j,i}}$,  while the ellipse is an area containing  95\% of the probability mass of  the  conditional  distribution of $\mathbf{s}_{j,t_{j,i+1}}$.
	}\label{I-newfig:a2_new_2Standard}
\end{figure}

In this section, we show the posterior means and CIs of the model parameters of our proposal (M1) (Tables \ref{I-tab:anim1}-\ref{I-tab:anim6}), model M2 (Tables \ref{I-tab:anim1single}-\ref{I-tab:anim6single}), and model M3 (Tables \ref{I-tab:anim1Standard}-\ref{I-tab:anim1Standard}). We also show the graphical description of the behaviors found for M2 and M3 in
Figures \ref{I-newfig:a2_1Single}-\ref{I-newfig:a2_new_2Single} and
Figures \ref{I-newfig:a2_1Standard}-\ref{I-newfig:a2_new_2Standard}, respectively. As in the main manuscript, we   only discuss behaviors with  $n_{j,k}>100$, and in order to distinguish between the results of different models, we use the superscript  to indicate  which model we are referring to.

\paragraph{M2 description}
The  model M2 finds the same number of behaviors as M1, and similar parameters. The major difference between  the two is in the ordering of the behaviors of the third dog, since the boundary-patrolling behavior of M2 ($B_{31}^{M2}$) is the one with the largest number of temporal points ($n_{3,1}^{M2} = 3537 $), while  it is the third  ($n_{3,3}^{M1}=  608$) in M1. The behavior   with  the strongest attraction to the central tendency   is $B_{32}^{M2}$ (the second) in M2, with only $n_{3,2}^{M2} = 818$  temporal-points, while it is  $B_{31}^{M1}$ (the first) in M1. It is interesting to note that all the   boundary-patrolling behaviors of the 6 dogs are described by the same set of parameters, i.e., they are the same behaviors, and the same is also true for  the explore-behaviors.
On the other hand, the attending-livestock behaviors, $B_{13}^{M2}$, $B_{23}^{M2}$, and $B_{53}^{M2}$,  are the same, while
$B_{63}^{M2}$ has its own set of parameters, with a slightly different  attractor.

A possible explanation for why, unlike  our proposal, the boundary-patrolling behavior  of the third dog has the highest number of temporal points, can be found in the way the model is formalized.  All the animals, in M2, share the same set of parameters, and 5 of them have the same  boundary-patrolling behavior as the first one. This means that  the expected value of the probability of remaining in this behavior, or of moving to this behavior from any other, is also very high  for the third dog, which may  lead to an overestimation of the number of temporal-points that belong to this behavior.
This problem is not present in M1 because, even though all the  boundary-patrolling behaviors are very similar, they  only share the  covariance matrix $\boldsymbol{\Sigma}_{j,k}^{M1}$, but all the other parameters are different, see, for example, parameters $\rho_{j,k}^{M1}$ and $\tau_{j,k}^{M1}$ in Figure \ref{newfig:a2_44} (c)   and (e) of the Manuscript. This means that there is only the probability of that specific covariance matrix to be higher.
%This can be also be seen by the fact that $B_{31}^{M1}$ and $B_{33}^{M1}$ have the same value of $\boldsymbol{\Sigma}_{j,k}$, see Figure
%\ref{I-newfig:a2_44} of the Manuscript.
It is also interesting to note that the CIs of the spatial attractors are wider in M2,  in the behaviors where they are relevant, than for M1, especially for dog 3.

\paragraph{M3 description}

The results of M3 are very different from those obtained from the other two models, since  dog 3 has 3 behaviors, while the others only have 2.  The first behavior of the 6 dogs  is similar to the boundary-patrolling-behavior of M1 and M2, but the ellipses  are much wider, i.e., the variances in $\boldsymbol{\Sigma}_{j,k}^{M3}$ are larger. With the exception of dog 2, the second behaviors   lack directional persistence and  attractors. On the other hand,   $B_{32}^{M3}$ and $B_{33}^{M3}$ each have  an attractor, located in a similar place  to the one in  $B_{31}^{M1}$, and the attraction is strong in $B_{32}^{M3}$ and weak in $B_{33}^{M3}$.  The CI of the spatial attractor is wider than the one in   $B_{31}^{M1}$.

The difference between the results of M3 and those of M2 and M1 can be explained by   the independence  between animals since, as already pointed out by \cite{Jonsen2016}, the inference and description of behaviors improve if multiple animals are considered.  M3 fails especially in the description of the dogs in the cohesive group,
which could point out the benefits of a joint modeling. This is further highlighted by the  values of the   information criteria shown in Table
\ref{newtab:cv} in the Manuscript, which indicate that M3 always has  the worst goodness-of-fit.

\end{appendices}
\bibliographystyle{ba}
\bibliography{all}

\end{document}